\documentclass{aa}  

\usepackage{graphicx}
\usepackage{txfonts}
\usepackage{verbatim}
\usepackage{todonotes}
\usepackage{booktabs}
\usepackage[switch]{lineno}
\usepackage{graphicx} 
\usepackage{color, soul}
\usepackage{placeins}
\usepackage{tabularx}
\usepackage{natbib}
\bibpunct{(}{)}{;}{a}{}{,}
\usepackage{hyperref}
\begin{document}

   \title{DBNets2.0: simulation-based inference for planet-induced dust substructures in protoplanetary discs}
\author{A. Ruzza\inst{1}, G. Lodato\inst{1}, G. P. Rosotti\inst{1} \and P. J. Armitage \inst{2,3}}

\institute{\inst{1}Università degli Studi di Milano, Dipartimento di Fisica, via Celoria 16, 20133 Milano, Italy. \\
\inst{2}Center for Computational Astrophysics, Flatiron Institute, 162 Fifth Avenue, New York, NY 10010, USA \\
\inst{3}Department of Physics and Astronomy, Stony Brook University, Stony Brook, NY 11794, USA\\
\email{alessandro.ruzza@unimi.it}}

\date{Received 06/03/2025; accepted 10/06/2025}
\titlerunning{DBNets2.0 SBI for dusty discs}

    \abstract{
Dust substructures observed in protoplanetary discs can be interpreted as signatures of embedded young planets whose detection and characterization would provide a better understanding of planet formation. Traditional techniques used to link the morphology of these substructures to the properties of putative embedded planets present several limitations that the use of deep learning methods has partly overcome. In our previous work, we used these new techniques to develop DBNets, a tool exploiting an ensemble of Convolutional Neural Networks (CNNs) to estimate the mass of putative planets in disc dust substructures. This inference problem, however, is degenerate as planets of different masses could produce the same rings and gaps if other physical disc properties were different. In this paper, we address this issue by improving our tool to estimate three other disc properties in addition to the planet mass: the disc $\alpha$-viscosity, the disc scale height and the dust Stokes number. For a given dust continuum observation, the full joint posterior for these four properties is inferred, exposing the existing degeneracies and enabling the integration of external constraints to improve the planet mass estimates. In addition to this new feature, we also addressed a few minor issues with our previous tool which reduced its accuracy depending on the observations' resolution or in the case of peculiar disc morphologies.
  The new pipeline involves a CNN that summarizes the input images in a set of summary statistics, followed by an ensemble of normalizing flows that model the inferred posterior for the target properties.
  We tested our pipeline on a dedicated set of synthetic observations using the TARP test and standard metrics to demonstrate that our estimates are good approximations of the actual posteriors. Additionally, we use the results obtained on the test set to study the presence and shape of degeneracies between pairs of parameters. 
  Finally, we apply the developed pipeline to a set of 49 gaps in 34 protoplanetary discs' continuum observations. The results show typically low values of $\alpha$-viscosity, disc scale heights, and planet masses, with 83\% of them being lower than 1M$_J$. These low masses are consistent with the non-detections of these putative planets in direct imaging surveys.
  Our tool is publicly available.
}


   \maketitle
%

\section{Introduction}

Continuum observations of protoplanetary discs' dust emission often show  annular substructures in the form of gaps and rings (e.g. \citealt{Isella2016RingedALMA, Andrews2018TheOverview, vanTerwisga2018V1094Star, Huang2020Large-scaleLup}).
Although other mechanisms have been proposed (e.g. \citealt{Hawley2001GlobalDisks, Barge2017GapsDust,   Dullemond2018Dust-drivenDisks, Hu2019NonidealRings, Bae2023StructuredDisks}), a promising explanation for their origin is the presence of embedded planets gravitationally interacting with the disc material \citep{Dipierro2015OnTau, Rosotti2016TheObservations, Zhang2018TheInterpretation}. In the specific case of the PDS70 transition disc, two planets in the disc cavity have been detected through observations of accretion tracers \citep{Wagner2018MagellanDisk, Haffert2019Two70, Zhou2021HubbleB}, direct imaging in the infrared \citep{Keppler2019HighlyALMA, Christiaens2019EvidenceB, Mesa2019VLT/SPHEREPDS70}, and, for PDS70c, in the millimeter and submillimeter \citep{Isella2019DetectionProtoplanets, Benisty2021APDS70c}. Furthermore, evidence has been presented for candidate planets in other systems with dust substructures, including in AS209 \citep{Fedele2023Kinematics209}, AB Aur \citep{Currie2022ImagesAurigae} and HD169142 \citep{Hammond2023ConfirmationB}. Setting aside these cases, however, systematic infrared surveys of discs with substructure have to date had poor success in directly detecting planets, being only able to put upper limits on the mass and temperature of putative embedded planets (e.g. \citealt{Reggiani2016TheOrbits, Nielsen2019TheAu, Vigan2021TheSHINE, Asensio-Torres2021Perturbers:Disks}).

In the scenario of planet-disc interaction, the morphology of observed substructures is determined by the physical properties of the planet and surrounding disc, and can be used to provide an estimate of these quantities. The collation and systematic analysis of substructures thus constitutes a useful tool for studying the population of young planets \citep{Lodato2019TheDiscs,Zhang2018TheInterpretation}, which is difficult to probe with the more standard techniques used for exoplanet detection (e.g. transit, radial velocity). Additionally, it can also support observational surveys by testing the consistency of (non-)detections and possibly suggesting promising candidates for new observations.

Several works investigated the dependence of substructure  morphology on the disc and planet properties (e.g. \citealt{Rosotti2016TheObservations, Kanagawa2016MassWidth, Dipierro2017AnDiscs}), proposing empirical relations that link morphological features, like the gap widths and depths, with the mass of the putative planet and, for example, the disc $\alpha$-viscosity and scale height.
Nevertheless, these formulae are limited in accuracy and precision as they rely only on a few properties of the observed substructures while actual observations can present a richer morphology with asymmetries or other local features. In some selected cases, more accurate analyses have been carried out by using an ensemble of fine-tuned hydrodynamical and radiative transfer models to produce synthetic observations that match data as close as possible (e.g. \citealt{Fedele2018ALMA209, Clarke2018High-resolutionAu,Toci2020PlanetData, Veronesi2020IsPlanet}). Computational and time costs are the main downsides of this approach, and these hinder a systematic study of all disc observations. Furthermore, both empirical formulae and case-specific modelling typically lack a proper statistical formalization of their uncertainties. 

Deep learning methods have been proposed to overcome these issues and provide a fast, accurate and reliable way to estimate the properties of discs and planets that would produce the observed dust substructures.
\cite{Auddy2020ADisks} and  \cite{Auddy2022UsingDisks} implemented respectively feed-forward and Bayesian neural networks to estimate the planet mass from the observed gap width and other disc properties, overcoming the rigid functional forms used in empirical formulae. \cite{Auddy2021DPNNet-2.0Gaps}, \cite{Zhang2022PGNets:Discs} and \cite{Ruzza2024DBNets:Discs} improved on these works by using, instead, convolutional neural networks (CNNs) that directly take as input dust continuum observations so that the entire substructure morphology is taken into account in the inference process. 
\cite{Mao2024Disk2Planet:Systems} proposed a different approach by directly fitting the disc's gas density map to the hydrodynamical models with an evolutionary algorithm tailored for complex optimization problems (Covariance Matrix Adaptation Evolution Strategy). This was made feasible by the use of an emulator \citep{Mao2023PPDONet:Systems} to substitute for the expensive hydrodynamical simulations.
In all these works, the techniques employed outperformed the more traditional methods in estimating the mass of the putative planets. 
In our previous work \citep{Ruzza2024DBNets:Discs} we developed a tool, DBNets, that takes as input a dust continuum observation of a protoplanetary disc and, through an ensemble of CNNs, provides an estimate for the mass of putative planets in the observed substructures. We focused on equipping our tool with a robust uncertainty quantification method and we extensively tested how the results were affected by non-idealities that might be present in actual data. Nevertheless, DBNets still faced some limitations that set our goals for the further developments that we present here.


\section{Improvements to DBNets and paper outline}

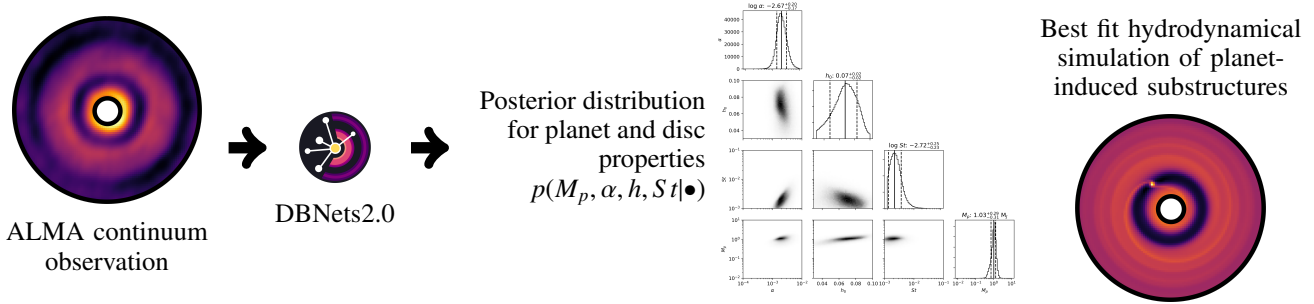
\begin{figure*}[t]
    \centering
    \begin{tikzpicture}
        \node (alma) at (0,0.5) {\includegraphics[width=2.5cm]{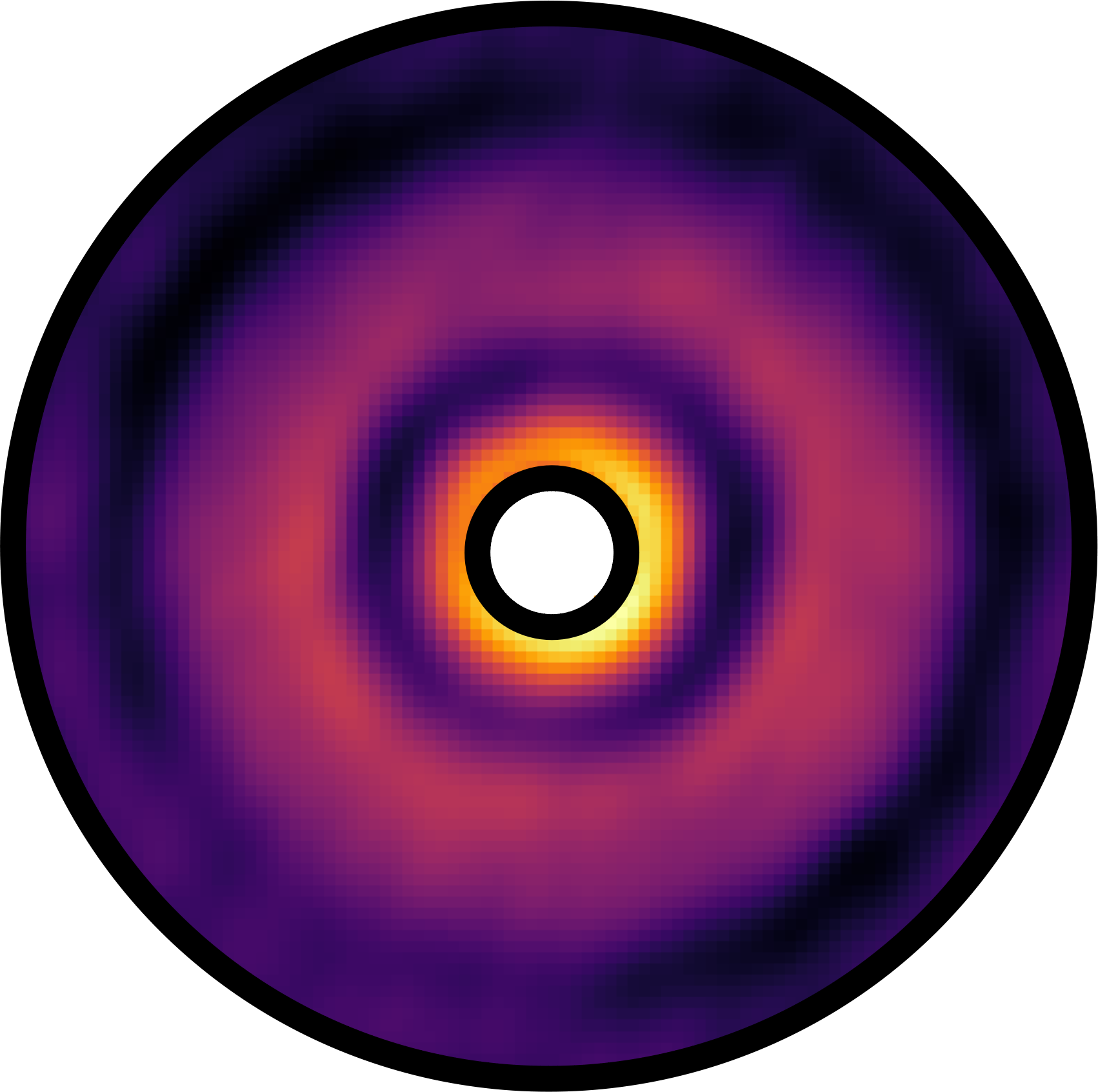}};
        \node[below=0cm of alma, align=center, text width=3.5cm] {ALMA continuum \\ observation};
        
        \node (dbnets) at (3,0) {\includegraphics[width=1cm]{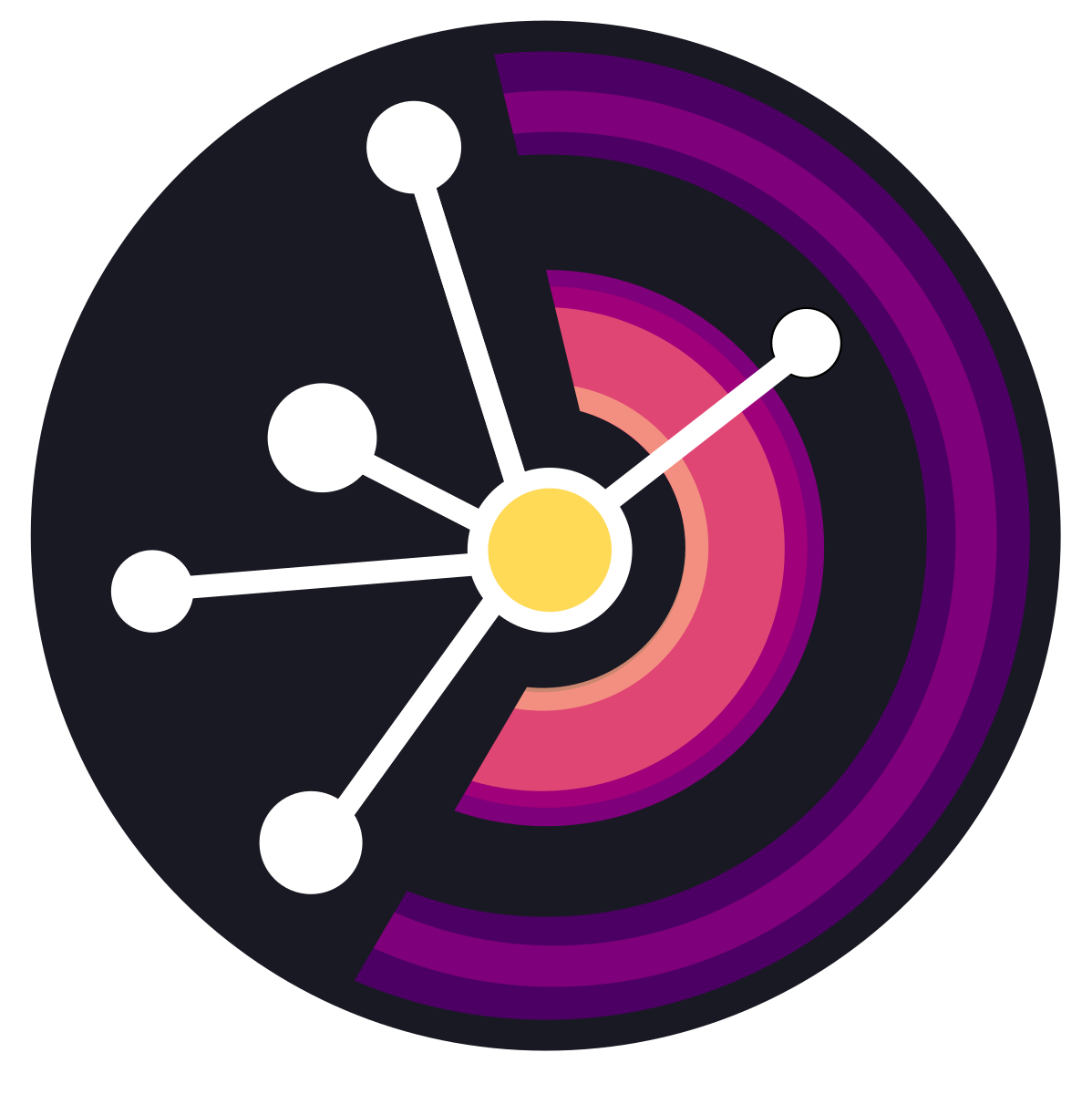}};
        \node[below=0cm of dbnets] {DBNets2.0};
        
        \node (posterior) at (10,0) {\includegraphics[width=4cm]{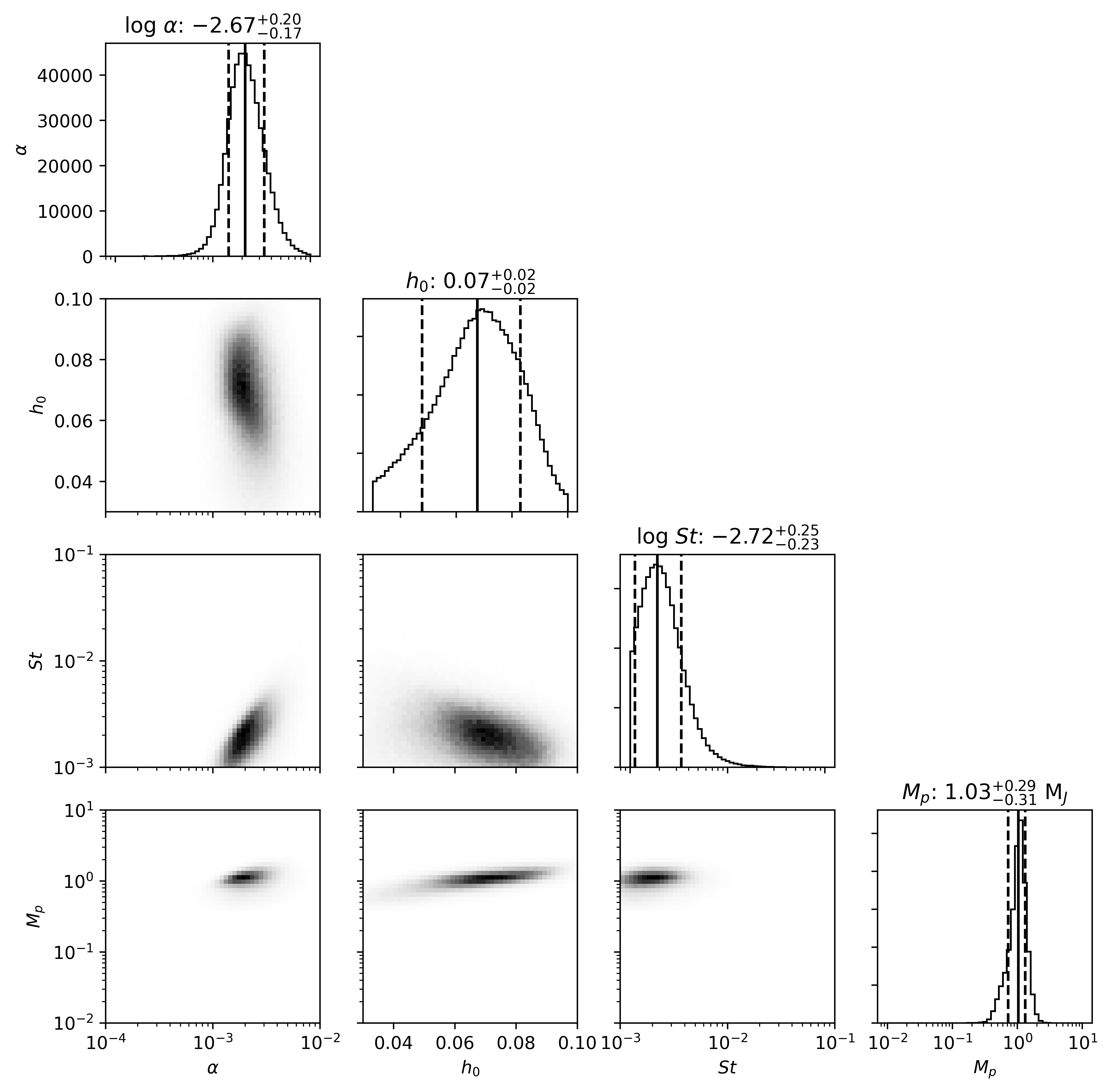}};
        \node[anchor=east, align=right] at (8, 0) { Posterior distribution \\ for planet and disc \\ properties \\ $p(M_p, \alpha, h, St|\bullet)$};
        
        \node (bestfit) at (14,-0.8) {\includegraphics[width=2.5cm]{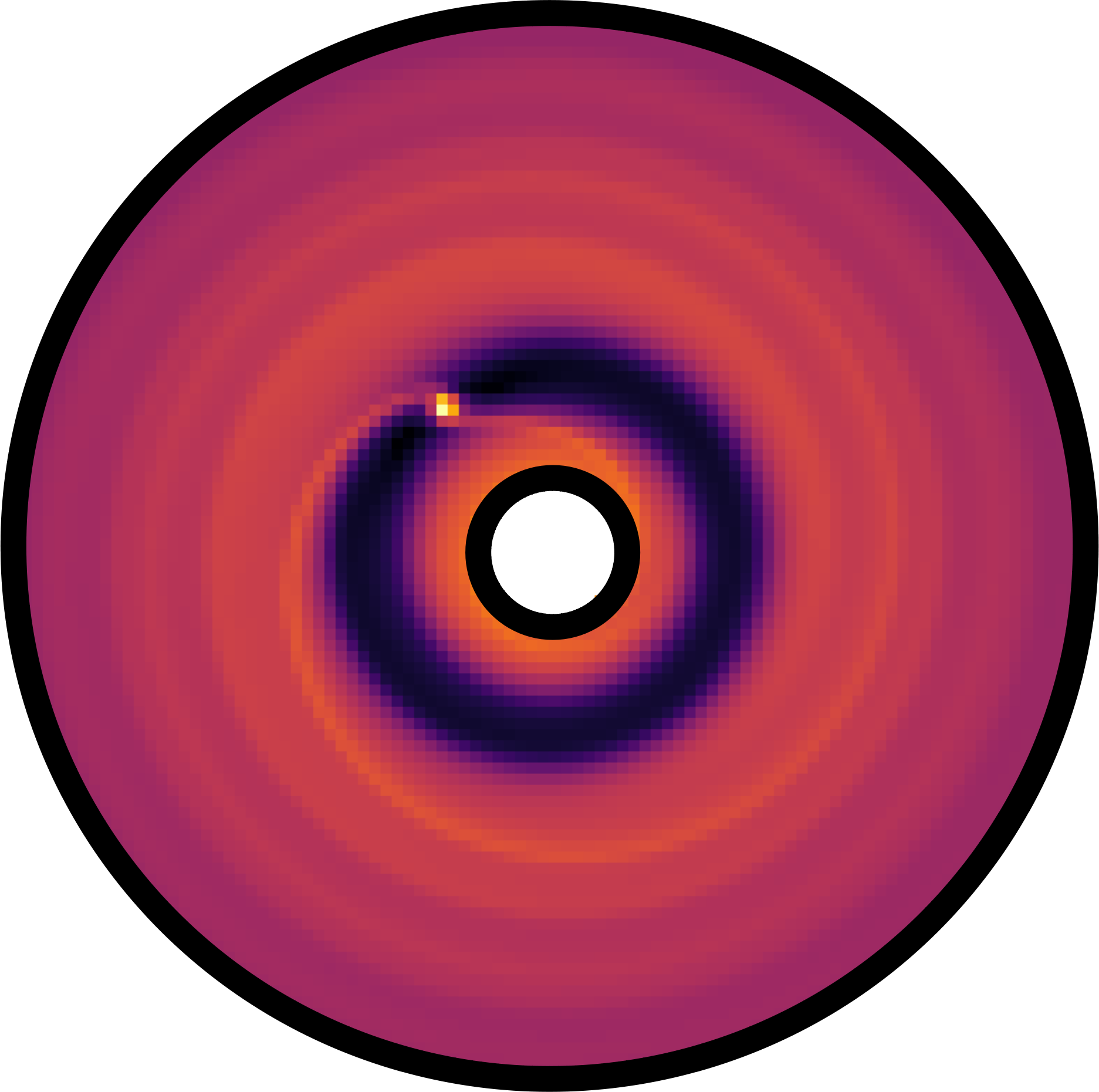}}; 
        \node[above=0cm of bestfit, align=center, text width=3.5cm] {Best fit hydrodynamical simulation of planet-induced substructures};
        
        \draw[thick, -{>}, line width=1.5mm] (1.6, 0) -- (2.1, 0);
        \draw[thick, -{>}, line width=1.5mm] (4, 0) -- (4.5, 0);   

    \end{tikzpicture}
    \caption{Schematic of DBNets2.0 pipeline and objective.}
    \label{fig:fig1}
\end{figure*}

The morphology of planet-induced dust substructures is determined not only by the planet's mass but also by other disc properties which make the inference problem degenerate. In DBNets \citep{Ruzza2024DBNets:Discs} we only provided estimates for the planet mass accounting for the degeneracy between different disc properties in the estimated uncertainties. The major improvement introduced in this work is the possibility to expose and explore these degeneracies through direct inference, given an observation, of the full joint posterior for the planet mass and 3 additional disc parameters: the disc scale height, $\alpha$-viscosity and dust Stokes number. Figure \ref{fig:fig1} provides an overview of this pipeline's general idea and objective.
Inferring the four-dimensional posterior not only enables a systematic quantification of the relation between these parameters but could also potentially provide a way to reduce the uncertainties of our constraints by setting proper priors, informed by other works, on some or all of the target properties.
Additionally, a statistical formalisation of the results allows us to properly validate them through rigorous testing.

Another limitation of DBNets was its limited flexibility as it could only infer posteriors for the planet's mass as mixtures of 50 Gaussians. This limitation is overcome in this work as the techniques adopted can effectively approximate posteriors described by virtually any functional form.

We also addressed some minor issues 
that reduced the generalization capabilities of our tool.
Specifically, in \cite{Ruzza2024DBNets:Discs} we showed that the inference results were affected by the resolution of the input observation with the best accuracy obtainable only when the beam size was close to that used to convolve the synthetic observations used for training. We also noticed, through further testing, that the position of the outer disc boundary could, in some cases, affect the outcome of DBNets inference, often returning results that did not pass the rejection criterion that we established. This was especially happening when the outer boundary appeared to be close to the gap edge.
Both these limitations have now been removed.

The paper is organized as follows. In Sect. \ref{sec:methods} we outline the dataset of synthetic observations used in this work and the inference pipeline adopted. In Sect. \ref{sec:evmethods} we explain the metrics and tests used to benchmark our tool. Results on our test set will be presented in Sect. \ref{sec:results}. We also applied our final pipeline to a large set of actual observations. Results of this survey will be shown in Sect. \ref{sec:real_obs}. Finally, in Sect. \ref{sec:discussion}, we discuss some of the key features of this tool and present our conclusions in Sect. \ref{sec:conclusions}.

\section{DBNets2.0 pipeline}
\label{sec:methods}

DBNets2.0 aims at inferring the full posterior $p(\theta|x, b)$ for four disc and planet properties ($\theta$, introduced in Sect. \ref{sec:dataset}) given a disc dust continuum observation $(x)$ and its resolution $(b)$.
The pipeline employed, similar to \cite{Lemos2023SimBIG:Clustering}, consists of two main components: 1) a CNN used for feature extraction and compression of the input data and 2)~an ensemble of normalizing flows for neural posterior estimation (NPE) trained on the summary statistics extracted in the previous step.

Normalizing flows provide the flexibility to represent any arbitrary posterior in a form that can be directly evaluated providing an accessible interface for modifying the priors, computing marginalized posteriors and evaluating the likelihood of the
training dataset. However, normalizing flows are difficult to train
directly on a high dimensional feature space. Additionally, the inductive bias of CNNs makes them more effective in extracting information from structured data types such as images. For these reasons, as has been done in other works (e.g.  \citealt{Lemos2023SimBIG:Clustering}), we supply our pipeline with a CNN that extracts and compresses the main features of the input data in a set of lower-dimensional summary statistics on which the NPE algorithm is trained.

In section \ref{sec:dataset} we introduce the dataset of mock observations used to train, test and evaluate our methods. We describe in detail the first component of our pipeline (feature extraction CNN) in section \ref{sec:CNN} and the simulation-based inference (SBI) method implemented in section \ref{sec:SBI}.

\subsection{Dataset}
\label{sec:dataset}
We use the same dataset of hydrodynamical simulations presented in \cite{Ruzza2024DBNets:Discs}. These are two dimensional simulations of protoplanetary discs with one embedded planet, run with the mesh code FARGO3D \citep{Benitez-Llambay2016FARGO3D:CODE}. We adopted a locally isothermal equation of state for the gas component while dust is simulated as a pressureless fluid subject to the gas drag \citep{Benitez-Llambay2019AsymptoticallyFARGO3D}. We neglected dust feedback on the gas, self-gravity, planet migration and accretion. 

The original dataset contains 1000 simulations with differing values for six disc and planet properties randomly sampled with a Latin Hypercube Sampling (LHS) algorithm to provide the best coverage of the parameter space. These properties are: the disc $\alpha$-viscosity ($\alpha$), aspect ratio at the planet position ($h$), Stokes number of the dust-gas interaction (St), slope of the power-law profile for the disc surface density ($\sigma$), flaring index ($\beta$) and planet/star mass ratio ($M_p$). Both $\alpha$ and St are considered constants across the entire disc. For definitions and physical meanings of these properties we refer the reader to our previous work \citep{Ruzza2024DBNets:Discs}. 
To simplify the problem, we chose to explicitly infer with our pipeline only four of these properties ($\alpha$, $h$, $St$, $M_p$) assuming an implicit marginalization of the posteriors over all possible $\sigma$ and $\beta$.
For convenience, we report in Table \ref{tab:par_table} the ranges where these four main properties were sampled.

Supplementing the 1000 simulations that were used in \cite{Ruzza2024DBNets:Discs}, we added 300 additional simulations that we used as the test set. For each simulation, we consider the snapshots after 500, 1000 and 1500 orbits of the embedded planet corresponding respectively to $1.7 \times 10^5$~yr, $3.5 \times 10^5$~yr and $5.3 \times 10^5$~yr for an orbit at 50~au around a solar mass star. 

\begin{table}
        \centering
        \caption{Dynamical properties sampled in the simulation's dataset.}
 \label{tab:par_table}
        \begin{tabular}{lccr} 
                \toprule
                Property & Symbol & Values & Type of sampling\\
                \midrule
                $\alpha$-viscosity & $\alpha$ & $10^{-4} - 10^{-2}$ & log \\
                Stokes number & St & $10^{-3} - 10^{-1}$ & log\\
                Aspect ratio & $h_0$ & $0.03 - 0.1$ & lin \\
        \begin{tabular}[x]{@{}c@{}} Planet/star \\ mass ratio \\ \end{tabular} & $M_p$ & $10^{-5} - 10^{-2}$ & log \\
                \bottomrule
        \end{tabular}
 
\end{table}

From each simulated map of the dust density ($\Sigma_d$) we compute the expected brightness temperature ($T_s$), as done in \cite{Ruzza2024DBNets:Discs}, through
\begin{equation}
    T_s = T_d \left[1-\exp(-\kappa \Sigma_d)\right]
\end{equation}
where we set the disc temperature $T_d$ consistent with the disc aspect ratios ($h$) of the hydrodynamical simulations. Assuming vertical hydrostatic equilibrium $h = c_s/v_k \propto \sqrt{Tr}$ (where $r$ is the radial coordinate).
We compute the opacity $\kappa$ with the same model and assumptions used in \cite{Ruzza2024DBNets:Discs}.

As done in our previous work, we remove all synthetic observations that do not exhibit visible substructures. The final size of our dataset is 2151 synthetic observations used for training and validation, with an additional 534 used for testing. These numbers are the result of the original choice of running 1000+300 hydrodynamical simulations, which is a trade-off between good coverage of the parameter space and computational costs informed by similar works that have been able to achieve good performance of their deep learning models with datasets of similar sizes. For instance, \cite{Auddy2021DPNNet-2.0Gaps} used 1200 independent simulations selecting 4 snapshots from each of them, while \cite{Zhang2022PGNets:Discs} used 6240 synthetic observations obtained from only 195 hydrodynamical simulations.

Each synthetic observation in our dataset is finally resized to 128x128 pixels and standardised
by subtracting the mean value of its pixels and dividing by their
standard deviation. 
Regarding the target values, a logarithmic transformation is first applied to the parameters that were uniformly sampled in their log space (see Table \ref{tab:par_table}). Then, all targets ($\theta$) are normalised with a linear transformation to bring their values between -1 and 1.

\subsection{First step: the CNN}
\label{sec:CNN}

Our pipeline takes as input the dust continuum observation of a protoplanetary disc showing substructure.
The first step is aimed at compressing the information content extracted from the input data into a set of summary statistics. For this purpose, we chose to employ a convolutional neural network (CNN) that produces as output a first guess for the four target parameters that characterize the input disc. We implemented the CNN using Keras \citep{Cholletetal.2015Keras} with the TensorFlow backend \cite{Abadi2015TensorFlow:Systems}. Training is performed by minimizing the mean squared error between the true ($\theta$) and estimated ($\hat \theta$) parameters for each one of the $N$ elements in the training dataset $\mathcal{T}$, that is:
\begin{equation}
    \text{mse} = \frac{1}{N} \sum_{i\in \mathcal T} \Vert\theta_i-\hat \theta_i\Vert^2.
    \label{eq:mse}
\end{equation}
Our summary statistics for a given observation is a set of 1500 samples of $\hat \theta$ obtained by enabling dropout layers during inference. This approach, also known as Monte Carlo dropout, has been proposed to approximate a Bayesian sampling of the model uncertainty \citep{Gal2015DropoutLearning, Kendall2017WhatVision}, although subsequent studies have raised concerns about its limitations and statistical justification \citep{Loic2021IsBayesian}. We thus only employ this method to easily capture model variability in our summary statistics. We adopt the same dropout rate used for training. 

The architecture of the CNN we used is composed of three main parts: 1) augmentation layers, 2) residual convolutional blocks and 3) dense layers.  
The first block of layers performs image augmentation during training by, in order, randomly translating each input image up to 1\% of its dimension, applying a random rotation, adding Gaussian noise of mean 0 and variance 0.1, randomly masking the disc to simulate a different outer boundary and, finally, convolving the input with a Gaussian beam of random semi-major axis up to $0.2r_p$, where $r_p$ is the planet orbital radius. We introduced the last two layers to make our tool less sensitive to, respectively, the position of the disc outer boundary and the resolution of the observation provided as input. The dimension of the Gaussian beam convolved to the input image is saved and run through a dedicated dense layer whose output is concatenated to the flattened output of the convolutional residual blocks. At inference time, it is thus necessary to provide as input to the CNN both the dust continuum observation and its observational beam size.
The second part of the CNN contains the convolutional and maxpooling layers commonly found in convolutional networks. They are organized in groups of three convolutional layers followed by one pooling layer with the introduction of a residual connection that has been proven to facilitate training \citep{He2015DeepRecognition}. We called these `residual convolutional blocks'.
In the third part of the CNN, a set of dense layers gradually compresses the information down to four real values interpreted as estimates of the target disc and planet properties. Connections between these layers are randomly dropped with a 20\% rate. The $\tanh$ activation function is finally applied to the CNN outputs to limit their values between -1 and 1, the same domain of the normalized targets.

We used the WeightAndBias framework to track our experiments and perform a Bayesian hyperparameters optimization varying the number and complexity of residual convolutional blocks, the number and dimension of dense layers, the learning rate, batch size and the normalization method of the input data. The architecture presented here is the optimal configuration that resulted from this optimization process. We observed that standardizing the input images, instead of normalizing them, and using simpler architectures were the keys to achieving the best results and avoiding overfitting.
The optimized CNN configuration used for the following steps was trained for 3000 epochs with the Adam optimizer, a batch size of 64 items and learning rate $5\times10^{-5}$.

\subsection{Second step: NPE with Normalizing Flows}
\label{sec:SBI}
In the second part of our pipeline we run simulation based inference (SBI) of the disc and planet properties in Table \ref{tab:par_table} using the summary statistics ${\hat \theta(x)}$, returned by the CNN in the previous step, as input. We implement masked autoregressive flows (MAF) to perform neural posterior estimation. The goal is thus inferring the joint posterior $p(\theta | \hat \theta(x), b)$ where $x$ denotes the specific disc observation under examination.
MAF are a special class of normalising flows (NF), a class of generative models that work by learning an appropriate set of change of variables that transforms a simple distribution into the complex one that is targeted. The different flavours of NF available, such as MAF, differ in how the change of variables is constructed, making them more or less flexible and easy to train.

We implemented this method using the Python package sbi \citep{Tejero-Cantero2020Sbi:Inference}, which provides ready-to-use implementations of most SBI algorithms. We train the model for a minimum of 100 epochs and until the loss function on the validation set is observed to no longer improve. Training is performed minimising the negative log-likelihood of the training dataset, that is:
\begin{equation}
    \mathcal{L} = - \sum_{i\in\mathcal{T}} \log \hat p(\theta_i | \hat \theta_i(x_i))
\end{equation}
where we denote as $\hat p(\cdot)$ the NF estimate for the target posterior.
The addition of this SBI method to our pipeline enables the inference of virtually any posterior without the limitation of Gaussian (or Gaussian mixtures) approximations. We also found, as discussed in Sect. \ref{sec:results}, that this is a necessary step to estimate accurate posteriors. Furthermore, the access to the full joint posterior allows us to understand degeneracies and correlation between the different disc and planet properties as well as change priors to integrate constraints that may be already available from other analyses.

\subsection{Usage and result interpretation guidelines}

DBNets2.0 is an SBI tool designed to fit the morphology of substructures observed in the dust continuum emission of protoplanetary discs, inferring the posterior distribution for model parameters corresponding to disc and planet properties. The model is inherently defined by the training dataset, which encodes the assumptions made to sample the parameter space and the physics assumptions made running and postprocessing the simulations (see Sect. \ref{sec:dataset}). The returned posteriors are conditioned on this model and represent the disc and planet properties that could produce, under the model assumptions, the observed disc morphology. In particular, the inferred distributions cannot account for scenarios which are not included in our model, e.g. presence of other planets. 

The suitability of our model for describing dust continuum observations must be carefully pondered in each case, considering all the information available for the system under examination. In support of this analysis, we design a new metric called ``confidence score" that quantifies, with a value between 0 and 1, how our underlying model can explain the observed substructures. We integrated the computation of this metric into our code, ensuring it is automatically calculated and returned with every DBNets2.0 posterior estimate. The definition, calibration and testing of this metric are thoroughly discussed in Appendix \ref{app:conf_score}. Based on our testing, we recommend rejecting any DBNets2.0 estimate associated with a confidence score lower than 0.6. Scores above this threshold can be treated as a continuous metric of the model's fit quality. Note, however, that even though a high confidence score indicates that our model can well describe the observed substructures, it cannot exclude different models and scenarios.

\section{Evaluation methods}
\label{sec:evmethods}
\subsection{Train-validation-test dataset split}
The dataset used in this work collects, before filtering, 3900 synthetic observations of protoplanetary discs obtained from 1300 hydrodynamical simulations, considering three snapshots for each of them. To perform cross-validation and testing of our pipeline we implemented the following split.
We selected 900 synthetic observations
uniformly in the parameter space used only for validating the full pipeline (test set). To avoid introducing any bias, the results on these data are never used to optimise any parameter or hyperparameter. Instead, we use the remaining 3000, appropriately split into training and validation sets, for training and tuning each model.

We train the CNN used in the first part of our pipeline (Sect. \ref{sec:CNN}) using 5-fold cross-validation for preventing overlearning and optimising the CNN architecture and hyperparameters. This means that the 3000 synthetic observations are split in five different ways into a training set, accounting for 80\% of these data, and a validation set containing the remaining 20\%. For each fold, the training set is used to train the CNN while the results on the validation set are used to early stop the training and to optimise the hyperparameters. The use of five folds maximises the use of the data for training and allows us to prevent the performed optimisations from being fine-tuned for a specific split.
In the second part of our pipeline, to train and test the normalizing flows, we collated a new dataset of $\{\hat \theta_i, \theta_i\}$ pairs obtained from the original dataset $\{x_i, \theta_i\}$. The summary statistics $\hat \theta_i$ are obtained, for each element, using the CNN trained on the fold where the respective $i^{\text{th}}$ simulation was contained in the validation set. We then train the models performing a single 80-20\% split of the data into a training and a validation set.

Finally, we use the test set to check our final pipeline and evaluate its performance. The results shown in the following sections were obtained on these data.

\subsection{Evaluation tests and metrics}

To evaluate our pipeline, we perform two types of tests on the test set. First, we assess the accuracy of the full joint posterior $p(\alpha, h_0, \text{St}^{-1}, M_p | x)$ using the Test of Accuracy with Random Points (\textbf{TARP}; \citealt{Lemos2023Sampling-BasedInference}).
\textbf{TARP} provides a necessary and sufficient condition for checking the accuracy of estimated posteriors. It is based on computing the coverage probability of some credibility regions of the inferred posteriors constructed around randomly sampled points. The expected coverage probability (ECP) should equal the credibility level ($\alpha_\text{TARP}$) of the considered region, if, and only if, the estimation $\hat p(\theta|x)$ well represents the actual target posterior $p(\theta|x)$. We refer the reader to \cite{Lemos2023Sampling-BasedInference} for further details. For the purpose of interpreting the plots in this paper, it is worth knowing that a TARP curve where the ECP is systematically lower that $\alpha_\text{TARP}$ indicates a biased posterior, an ECP that tends towards 0.5 for all values of $\alpha_\text{TARP}$ is indicative of an overconfident posterior while an underconfident posterior is characterized by ECPs significantly lower than $\alpha_\text{TARP}$ for $\alpha_\text{TARP}<0.5$ and higher for $\alpha_\text{TARP}>0.5$. To quantitatively evaluate the results of a TARP tests we provide two metrics that we call `ks-pval' and `atc'. The former is the p-value for a two sample Kolmogorov-Smirnov test where the null hypothesis is that the distributions of ECP and $\alpha_\text{TARP}$ are identical. If they were, it would also indicate a perfect match between the inferred and target posteriors and the p-value would be close to 1. The null hypothesis is generally rejected if the p-value is below 0.05. The latter metric (`atc') is defined as the area of the difference between the ECP and $\alpha_\text{TARP}$ curves for $\alpha_\text{TARP}$ values larger than 0.5. This number should be close to 0.
Lower values indicate underconfident or biased distributions while positive values indicate overconfident distributions.
These metrics are meant to be used to compare different results. For example, in the future, we might release minor updates to our public tool that could improve the accuracy of the pipeline. In this case, these metrics will be recomputed and provided to allow a proper comparison with the previous versions.
We perform the same TARP test on the 1D posteriors for each target property marginalizing over the inferred 4D posteriors with uniform priors in the parameter space where the models were trained.

Finally, in order to also evaluate the precision of our estimates and quantify the biases and under or overconfidence revealed by the TARP tests, we consider the medians of the inferred distributions as the best estimates and compute, for each target property, the root mean squared error (rmse) and r2-score metrics. These are defined as
\begin{equation}
    \text{rmse} 
    = \sqrt{\frac 1 N \sum_{i\in \mathcal T}\Vert \theta_i-\hat \theta_i \Vert^2}
\end{equation}
and
\begin{equation}
    \text{r2-score} = 1 - \frac{\sum_{i\in \mathcal{T}} (\theta_i - \hat \theta_i)}{\sum_{i\in \mathcal{T}} (\theta_i - \bar \theta)},
\end{equation}
where $\theta$ and $\hat \theta$ indicate, respectively, the target and estimated values of the property under exam for each one of the $N$ elements in the test set $\mathcal T$, while $\bar \theta$ is the mean of the target values $\sum_{i\in\mathcal T} \theta_i /N$.
We point out that, when computed on normalized values, the rmse can be interpreted as the root mean square of the relative errors (of the non-normalized property) with respect to the mean value of the parameter space, explored with our datasets, for that property (listed in Table. \ref{tab:par_table}).
We also compute the standard deviation ($\sigma$) of the inferred 1D marginalized distributions and compare them with the rmse to check whether the inferred uncertainties are well representative of the typical errors.

\section{Results}
\label{sec:results}
In this section, we present our testing of the entire pipeline on the test set. A brief discussion regarding the performance of the CNN alone can be found in Appendix \ref{app:appa}.
To evaluate how the tool is affected by the observations' resolution, we performed our tests convolving the input images of the test set with Gaussian beams of four different sizes: 0, 0.1, 0.15 and 0.2 $r_p$. We present both aggregated and separated results with regard to the synthetic image resolution.

We first test the full joint posterior inferred with DBNets2.0, then we evaluate the resulting single-parameter distributions obtained by marginilizing the joint. Finally, we check and discuss the correlations between pairs of variables encoded in the inferred posteriors. To further test our pipeline, we also performed posterior predictive checks (PPC) for three elements of our test set, which consist of a comparison between the input synthetic images and new simulations run with DBNets2.0 best estimates for the disc and planet properties. If the inference is correct, a synthetic observation generated from the inferred properties should well reproduce the actual data. We discuss the results in appendix \ref{app:ppc}.
\subsection{Full joint posterior}
\label{sec:results4d}
\begin{figure}
    \centering
    \includegraphics[width=\linewidth]{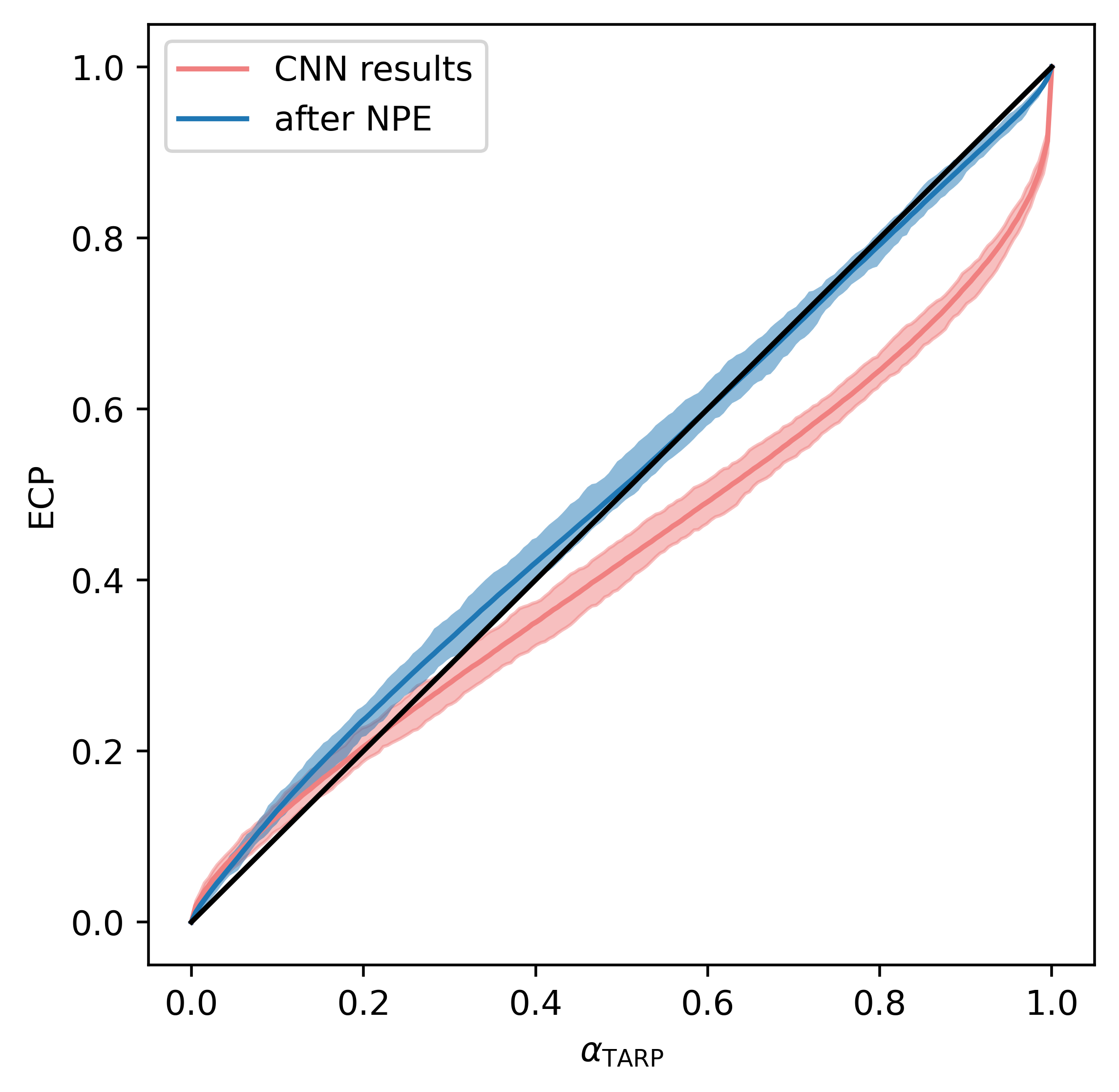}
    \caption{TARP curves computed on the test set using the entire pipeline (blue line) and only the feature extracting CNN (red line). The shaded areas mark the curves uncertainty evaluated bootstrapping the test set.}
    \label{fig:tarpall}
\end{figure}

\begin{figure}
    \centering
    \includegraphics[width=\linewidth]{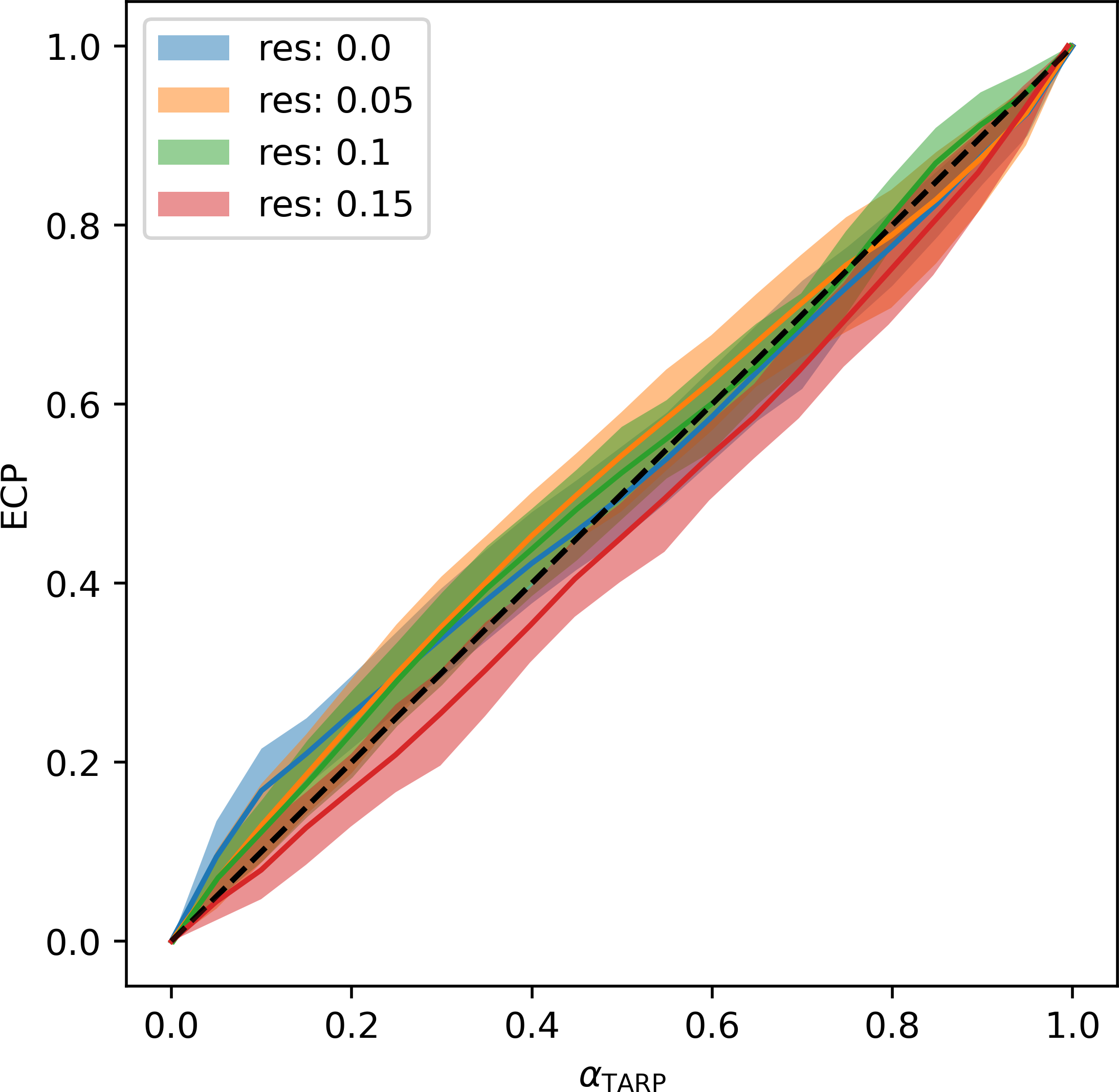}
    \caption{TARP curves computed, using the entire pipeline, on the test set with the input images convolved with gaussian beams of different sizes. The shaded areas mark the curves uncertainty evaluated bootstrapping the test data.}
    \label{fig:tarpallres}
\end{figure}

Figure \ref{fig:tarpall} shows the result of the TARP test on the full 4D posteriors $p(M_p, \text{\emph{St}}, \alpha, h | x)$ inferred by our tool. The shaded area indicates the curve uncertainty computed by bootstrapping the simulations in the test set.
The curve of the expected coverage shows good agreement with the target diagonal indicating that our posteriors are well calibrated (ks-pval: 0.999, atc: $-0.86$). We also reported in the same figure the same test performed on the summary statistics returned by the CNN, assuming them to be samples from the target posterior (ks-pval: 0.009, atc: $-13.27$). The results highlight how Bayesian dropout on the CNN alone is not sufficient, in our pipeline, for providing an accurate estimate of the full four-dimensional posterior.
Figure \ref{fig:tarpallres} shows instead the TARP curves computed separating input images of different observational resolutions. The results show a fair agreement with the target, within the uncertainties, independently from the observation's resolution.
 \subsection{Single parameter predictions}

\begin{figure}
    \centering
    \includegraphics[width=\linewidth]{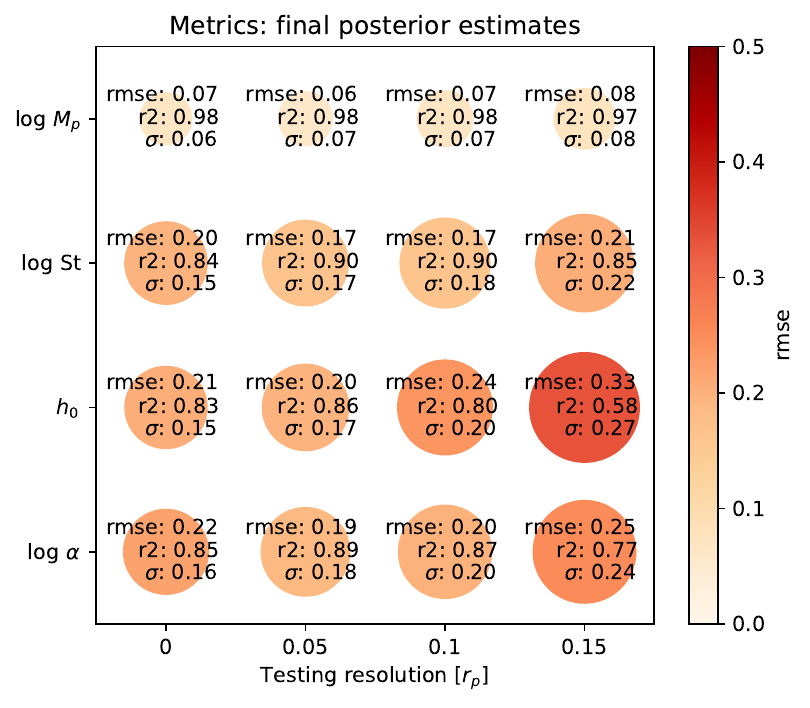}
    \caption{Metrics computed on the test set for the 4 different inferred parameters and at varying resolution of the input images. The rmse and r2-score are computed using the median of the inferred distributions as best estimates. The $\sigma$ indicates the mean standard deviation of the inferred distributions.}
    \label{fig:metrics_sing}
\end{figure}
\begin{figure*}
    \centering
    \includegraphics[width=0.48\linewidth]{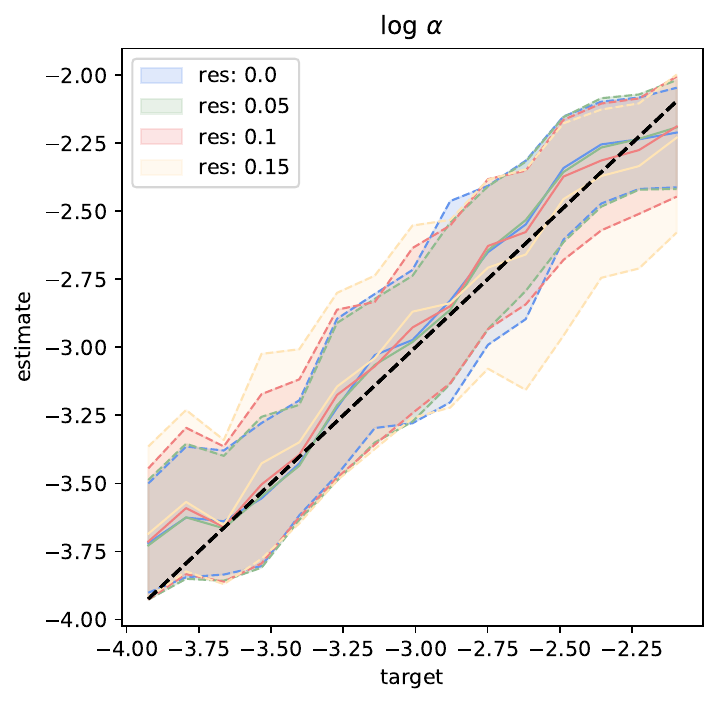}
    \includegraphics[width=0.48\linewidth]{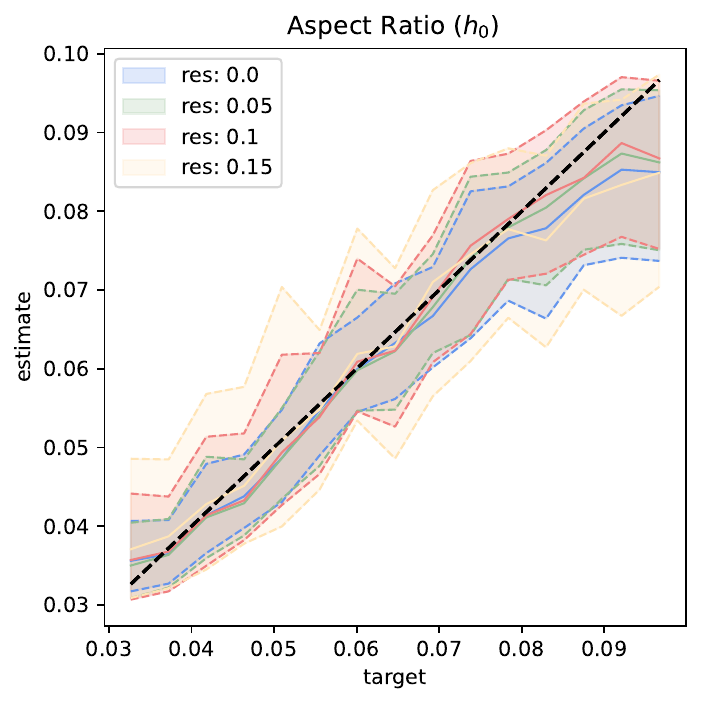}
    \includegraphics[width=0.48\linewidth]{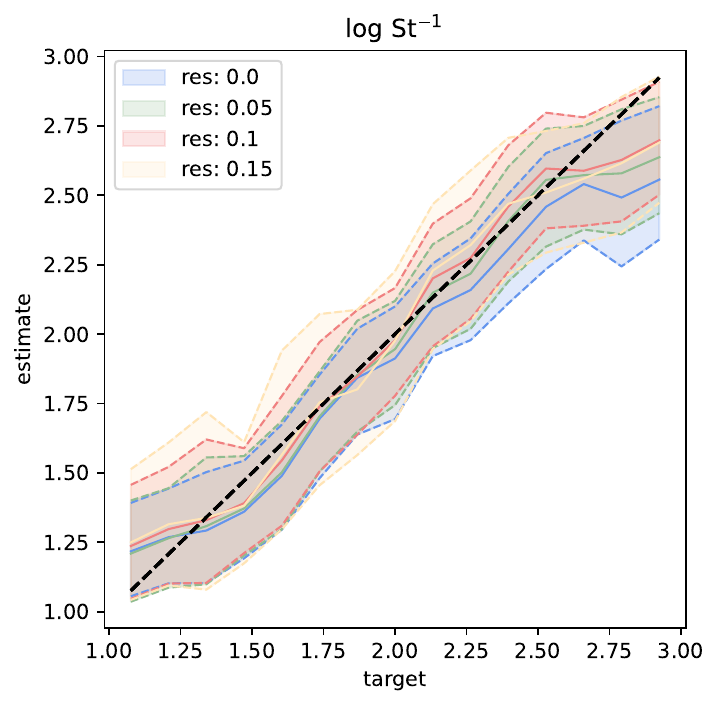}
    \includegraphics[width=0.48\linewidth]{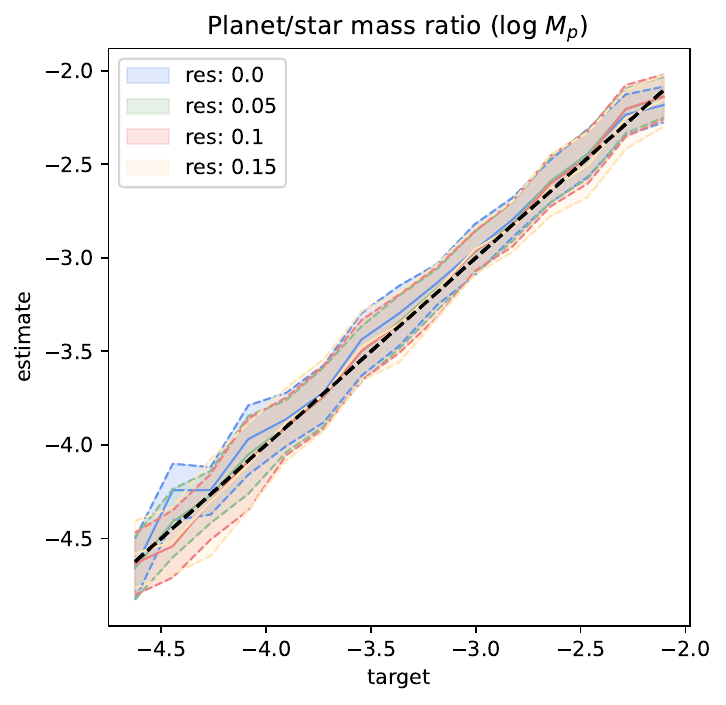}
    \caption{Results on the test set for each inferred property. For each targeted disc or planet property, the plots show the correlation between DBNets2.0 estimates and target values by plotting the median of the inferred distributions and the region between the $16^\text{th}$ and $84^\text{th}$ percentiles. In all cases, different colours refer to the results obtained on synthetic images of different resolutions. }
    
    \label{fig:1ddist}
\end{figure*}
The accuracy of the inferred posteriors evaluated in the previous section does not imply anything regarding their precision. In this section we are interested in evaluating both the accuracy and precision of single parameter estimates, that can be obtained by marginalizing the full joint posterior over the other disc or planet properties. 
In this context, for this marginalization, we assume uniform priors in the tool's scope of the parameter space.
We call ``best estimate'' of a given parameter $\theta$ the median of the respective posterior $p(\theta|x)$.

In Fig. \ref{fig:metrics_sing} we report metrics computed on the best estimates for each inferred property and dimension of the Gaussian beam used for convolving the input image. As noted in Sect. \ref{sec:results4d}, we still observe that the results are not strongly affected by the resolution of the input images although a slight worsening of precision is noted for the lower resolution tested. The average standard deviation of the 1D inferred posteriors ($\sigma$ in Fig. \ref{fig:metrics_sing}) is typically close to the rmse with respect to the target values, indicating that, in general, the estimated uncertainties well represent the actual errors.

In our pipeline, all the target properties are treated in the same way and are all normalized to take values between -1 and 1. Therefore, the fact that, according to these metrics, the planet mass can be inferred with higher precision with respect to the other properties is indicative that the morphology of dust substructures is primarily controlled by the planet mass. The disc scale height, instead, appears to be the more difficult property to infer showing also a sensitive drop in accuracy with the lowering of the observation resolution.
For comparison with our previous work, taking into account the different normalization of the target data, the rmse of DBNets \citep{Ruzza2024DBNets:Discs} estimates for the planet mass was 0.07 while the r2-score amounted to 97\%. Hence, limited to the scope and objective of our previous work, the new pipeline achieves the same accuracy.

The target-estimate plots in Fig. \ref{fig:1ddist} show, for each inferred property, the distribution of samples extracted from the marginalized inferred posteriors on the test set. After binning, median, $16^\text{th}$ and $84^\text{th}$ percentiles of these distributions are plotted as a function of the target values.
 We note that all the inferred parameters exhibit a good agreement with the targets which are always within the uncertainties. A systematic over and under-estimation at, respectively, the lower and higher end of the explored range of values is observed for all of the inferred properties with the exception of the planet's mass.
 We also performed TARP tests for these marginalized posteriors. Results confirming their accuracy are shown in detail in Appendix \ref{app:tarp_single}.
Although the TARP curves highlight, in some cases, the presence of slight biases, these are usually of small entities as can be seen in the target-estimate plots.

\subsection{Correlations between pairs of variables in the inferred marginalized posterios}

\label{sec:results_corr}
\begin{figure*}
    \centering
    \includegraphics[width=\linewidth]{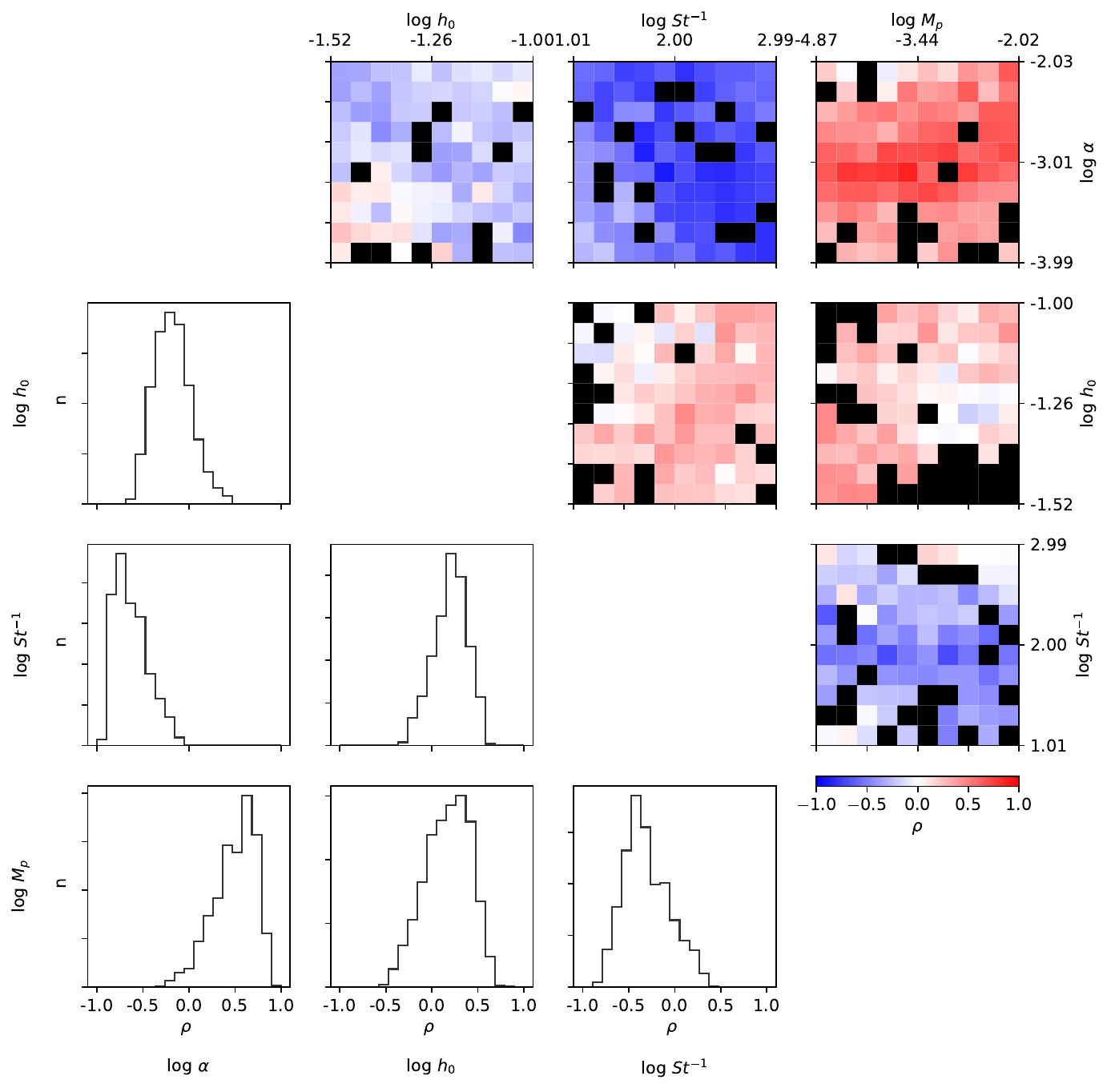}
    \caption{Pearson correlation coefficients between pairs of inferred properties computed for each element of the test set using 5000 samples from the inferred posterior. The bottom left corner shows for each pair of target properties the distribution of values of the relative Pearson correlation coefficients over the entire test set. The upper right corner similarly shows the same distribution but as a function of the two parameters considered. Bins of the 2D histograms with no data are marked in black.}
    \label{fig:pearson}
\end{figure*}

One of the key advantages provided by the availability of the full joint posterior is the possibility of looking at correlations between disc and planet properties inferred from substructures. These emerge because usually there is not a unique combination of disc and planet properties that determines the observed substructures but several of them can result in similar disc morphologies.
To understand which are the main properties involved in this degeneracy we computed, using 5000 samples from the inferred posterior for each item in our test set, the Pearson correlation coefficient between all pairs of inferred variables.
The results are shown in Fig. \ref{fig:pearson}. The bottom left corner of this figure shows the distribution of the computed Pearson coefficients for each pair of variables. The higher correlation is observed between log~St and $\log \alpha$ and between $\log M_p $ and $ \log \alpha$. Other pairs of variables still show distributions of Pearson coefficients that peak away from zero but at lower values. An interpretation of these results is given in Sect. \ref{sec:deg_study}. We also observe that these distributions are typically wide indicating that correlations in the inferred posteriors are not consistent for all items in our test set or across the entire parameter space. This variability is shown in the top right corner. We note that correlation coefficients close to 0 or in the opposite direction to the general trend are usually segregated to the boundaries of the parameter space. This may be the result of a selection effect due to the low density of simulations in those regions. Like every deep learning model trained on large data the user should be aware that the models might be less accurate in the extreme cases at the boundaries of the parameter space covered by the training data.

\section{Results on actual observations}
\label{sec:real_obs}

\begin{figure*}
    \centering
    \includegraphics[width=\linewidth]{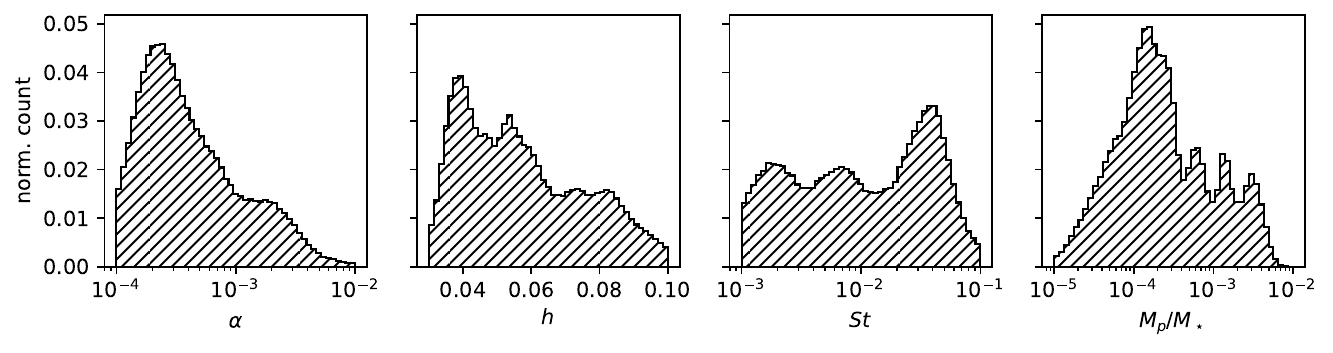}
    \caption{Distribution of the inferred properties for the 49 actual gap observations considered. To construct these histograms we combine 5000 samples extracted from each inferred posterior. }
    \label{fig:all_samples_real}
\end{figure*}

\begin{figure}
    \centering
    \includegraphics[width=\linewidth]{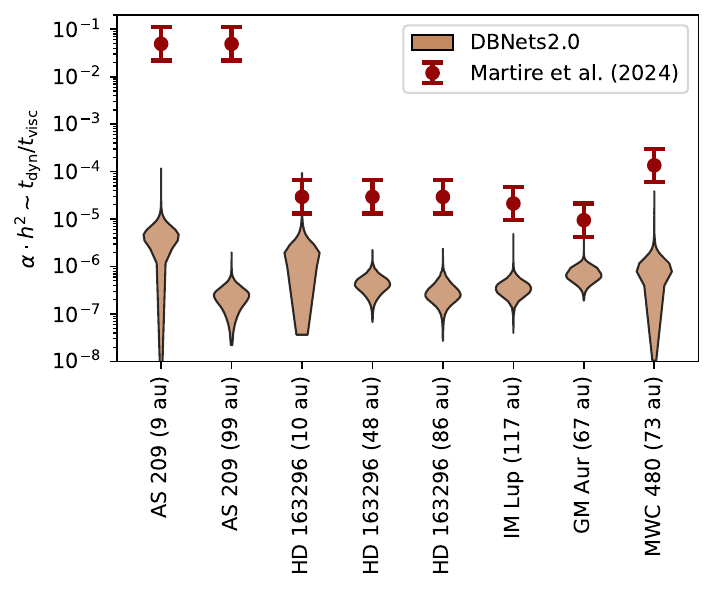}
    \caption{Comparison of DBNets2.0 and literature estimates of discs' viscous timescales for a subset of the analysed discs. Violin plots show the distribution $p(\alpha h^2 | x)$ inferred with DBNets2.0. Red and green points correspond to literature estimates obtained through $\alpha h^2 \sim \dot M/(M_{\text d} \Omega)$. We use dynmical measures of disc masses $M_\text d$ from \cite{Martire2024RotationDisks}. Measures of the star accretion rate $\dot M$ are from \cite{Manara2014GasWinds} and \cite{Oberg2021MoleculesHighlights}, provided with an uncertainty of 0.35 dex which dominates the $\alpha h^2$ uncertainties shown in this plot.}
    \label{fig:ah2_paper}
\end{figure}

\begin{figure}
    \centering
    \includegraphics[width=\linewidth]{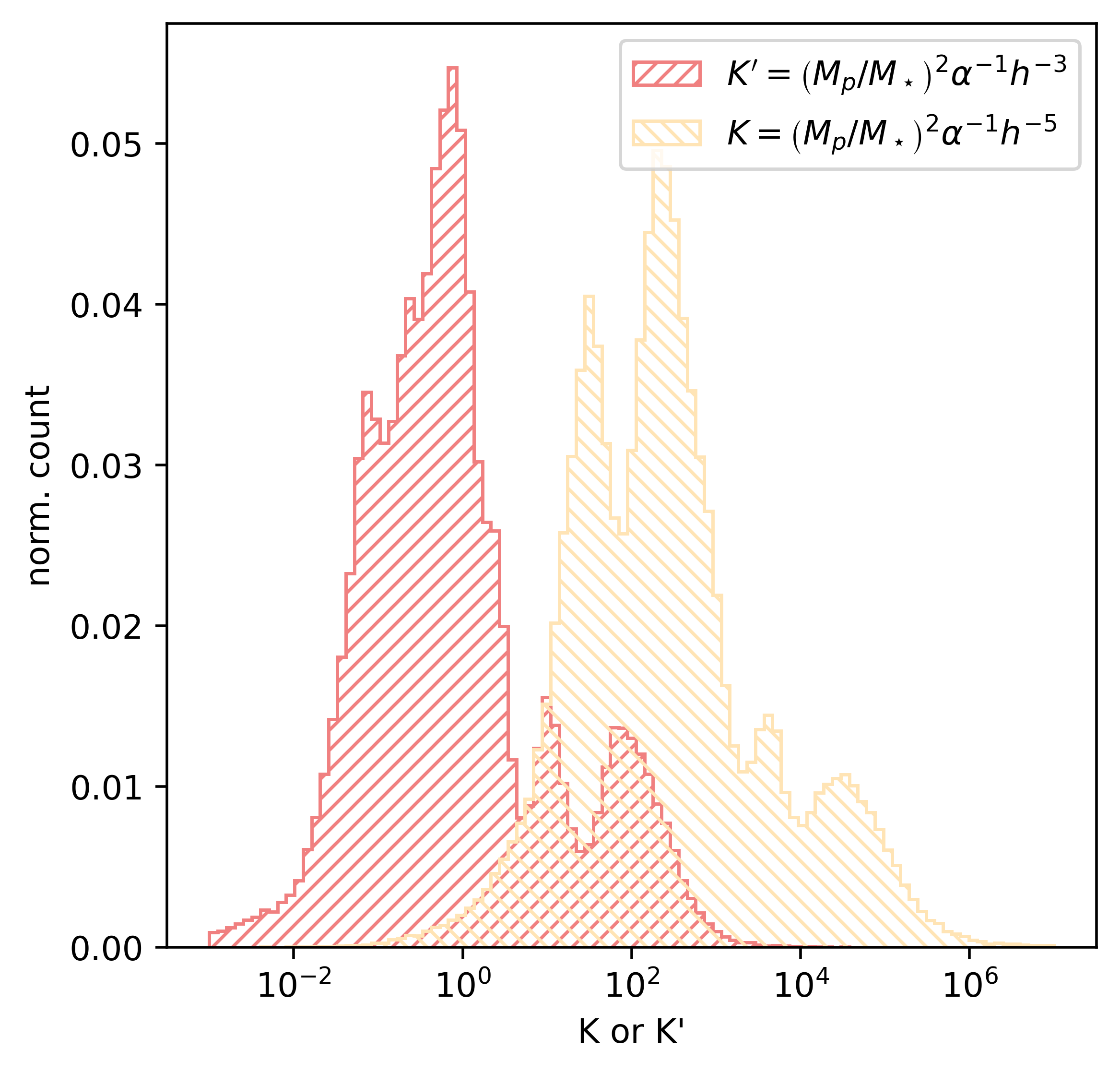}
    \caption{Inferred distributions of the K and K' coefficients for the population of proposed planets within dust substructures. The figure combines K and K' values obtained from 5000 samples of the posterior $p(\alpha, h, \text{St}, M_p|x)$  for each of the 49 analysed gaps. }
    \label{fig:kanagawa}
\end{figure}

\begin{figure*}
    \centering
    \includegraphics[width=0.431\linewidth]{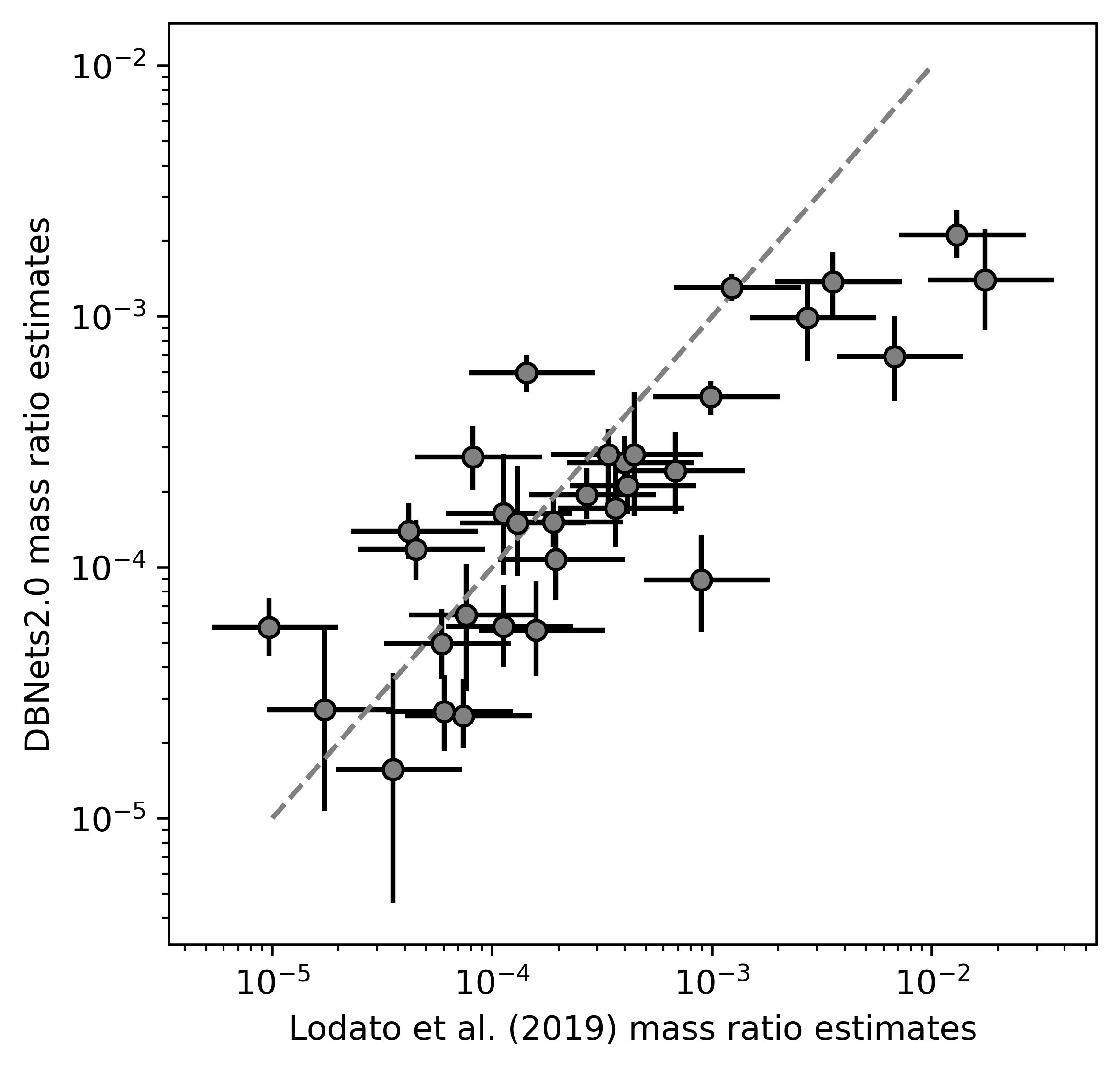}
    \includegraphics[width=0.51\linewidth]{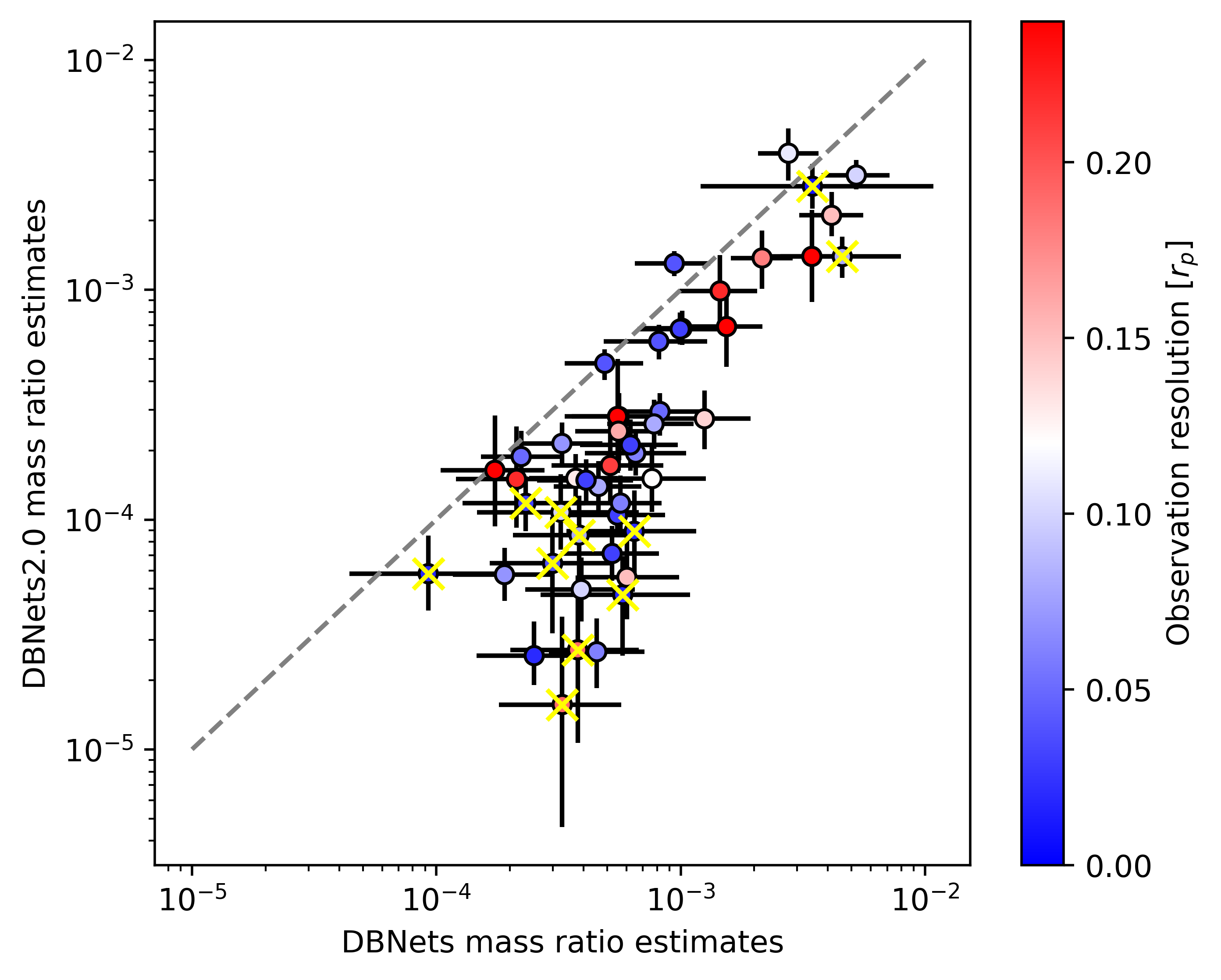}
    \caption{Comparison between DBNets2.0 and literature estimates for the mass of the proposed planets in the actual 49 observations analysed. The left plots shows a comparison with \cite{Lodato2019TheDiscs} who assumed the gap width to scale as the planet Hill radius. The right plot presents a comparison with the estimates of our tool's previous version DBNets \citep{Ruzza2024DBNets:Discs}. The color map refers to the resolution of the considered observations measured as the beam size over the assumed planet location. The diverging colour bar is centred on the only resolution used in \cite{Ruzza2024DBNets:Discs} to both train and test. Yellow crosses mark DBNets estimates that were considered unreliable by \cite{Ruzza2024DBNets:Discs} according to their rejection criterion.}
    \label{fig:real_lit_comp}
\end{figure*}

\begin{figure*}
\centering
\includegraphics[width=0.95\textwidth]{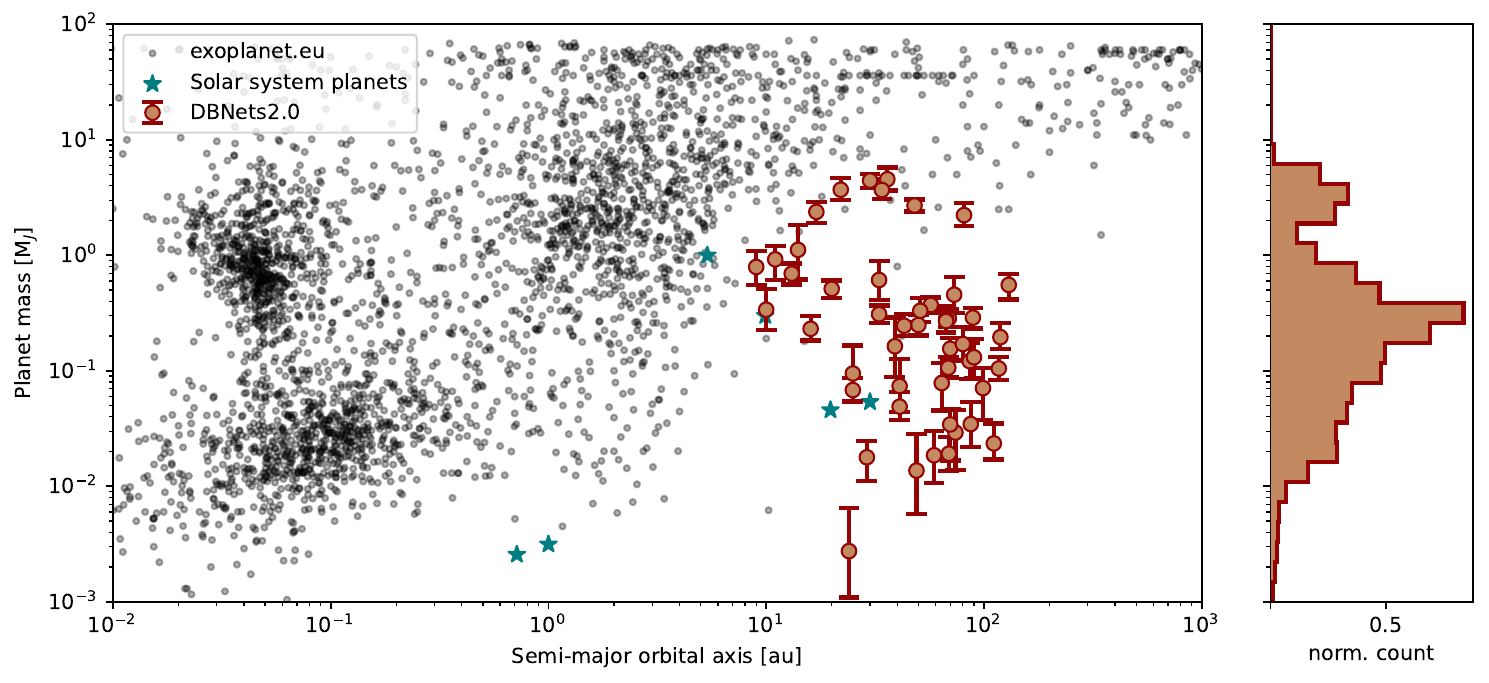}
\caption{Mass and semi-major orbital axis of the over 7000 exoplanets’ confirmed detections (grey points, data from \href{exoplanet.eu}{exoplanet.eu}). The red points are the proposed planets in protoplanetary discs characterised by DBNets2.0 with error bars marking the $16^{\text{th}}$ and $84^{\text{th}}$ percentiles of the inferred distributions.}
\label{fig:ma_plot}
\end{figure*}

We applied the newly developed pipeline to the same sample of discs, with ALMA Band 6 or 7 continuum observations showing observable axisymmetric substructures, that we collated in \cite{Ruzza2024DBNets:Discs}. Names and properties of these objects are reported in Appendix \ref{app:real_obs}. In the cases where multiple substructures are present, we applied the tool once per gap.

Results are presented in aggregated form in Fig. \ref{fig:all_samples_real}, and in detail for each disc in Fig. \ref{fig:all_real}. For each given input our tool provides an estimate of the four-dimensional posterior for the planet mass, $\alpha$-viscosity, disc scale height and dust Stokes number.  For their visualisation and analysis, we sample from each of them 5000 points $\theta \in \mathcal{R}^4 \sim p(\theta|x)$.  We show in Fig. \ref{fig:all_samples_real} the distributions of the inferred properties, for all 49 inputs, within the parameter space where we trained our tool. In Fig. \ref{fig:all_real} we report the one-dimensional marginalized posteriors for each analysed substructure and target property. The violin plot shows their shape within the 16$^\text{th}$ and $84^\text{th}$ percentiles. A few cases exhibit very elongated distributions covering the entire tool's training scope. We interpret these results as failures to extract information on that property from the observed dust substructure.

According to our analysis, the observed substructures imply relatively low values of $\alpha$-viscosity. The inferred distribution for $\log \alpha$ has mean -3.4 and standard deviation 0.45. These results are in agreement with other literature constraints obtained modelling planet-disc interaction (e.g. \citealt{Zhang2018TheInterpretation}, Sect. 4.4 of \citealt{Rosotti2023EmpiricalDiscs}), and with direct measures of line broadening (e.g. \citealt{ Flaherty2015WEAKOBSERVATIONS, Teague2016MeasuringLimitations,Flaherty2020MeasuringSgr}), which \citep[with some exceptions, e.g.][]{Flaherty2024EvidenceLup} yield strong upper limits.

We similarly observe that the distribution of inferred $h_0$ is slightly skewed towards lower values.
Since $\alpha h^2 \sim t_\text{dyn}/t_{visc}$ these results would suggest generally long viscous timescales $\sim 10^{5}-10^{7} \Omega^{-1}$ corresponding to $~10^6-10^9$~yr when rescaled for the dynamical timescales ($t_\text{dyn}\sim \Omega^{-1}$) at the location of the gaps for which they have been inferred. Plots can be found in Appendix \ref{app:visc_time}. 
    Assuming a self-similar disc model the product $\alpha h^2$ can be related to the star accretion rate and disc mass. Figure \ref{fig:ah2_paper} shows, for a small sample of common objects, the viscous timescales computed in this way using literature \citep{Martire2024RotationDisks} estimates of the discs masses and mass accretion rates \citep{Manara2014GasWinds, Oberg2021MoleculesHighlights}. In the same plot, we also report the local estimates of $\alpha h^2$ inferred with our tool from dust substructures. We observe that these are all systematically lower than the viscous timescales globally probed more directly from measures of the star accretion rate. Variability might explain, to some extent, the measure of higher-than-expected accretion rates. Another explanation, especially given the systematic nature of this difference, could be that the local mechanisms of angular momentum transport which, modelled as an effective viscosity affect the outcome of planet-disc interaction, are not sufficient to reproduce the global disc accretion timescales. This hints towards the presence of an additional global mechanism for dispersing angular momentum. There is considerable evidence for wind-driven angular momentum loss (see e.g. the reviews \citealt{Pascucci2023TheDisks, Manara2023DemographicsFormation}) and the observed offset would be broadly consistent with that picture. However, a word of caution is necessary as we use dynamical estimates for the disc masses  \citep{Lodato2022DynamicalDiscs, Longarini2025ExoALMA.Modelling} which can only be obtained for high mass discs possibly introducing a selection bias in our sample.

The inferred Stokes numbers appear to be more uniformly distributed across the parameter space. However, the inferred distribution shown in \ref{fig:all_samples_real} still exhibits a noticeable peak around $\text{St} \sim 0.03$.

The dependence of the main features of planet-induced gas substructures on disc and planet properties is often expressed only as a function of the coefficients $K$ and $K'$ defined as: 
\begin{equation}
K' = \left(\frac{M_p}{M_\star}\right)^2h^{-3}\alpha^{-1}
\end{equation}
and
\begin{equation}
K =  \left(\frac{M_p}{M_\star}\right)^2h^{-5}\alpha^{-1}.
\end{equation}
Specifically, several studies (e.g. \citealt{Duffell2013GAPDISK, Kanagawa2015FormationRotation}) using both analytical arguments and hydrodynamical simulations have shown the gas gap depth ($\Sigma_\text{min}/\Sigma_0$) to scale approximately as $1/(1+0.04 K)$, while \cite{Kanagawa2016MassWidth} found an empirical relation for the gap width as $\Delta_\text{gap} \propto K'^{1/4}$.
We report in Fig. \ref{fig:kanagawa} the overall distribution for these parameters found by analyzing the 49 gaps of our sample of discs' dust observations. 
Among others, the K distribution shows a significant peak around $K\sim10$ which, according to \cite{Kanagawa2015FormationRotation}, would correspond to planets weakly perturbing the disc's gas component.
We compute and report these distributions because, arising from the planet-disc interaction theory, these coefficients are commonly used to study the dependence of different phenomena in this context on the disc and planet properties. \cite{Scardoni2022InwardRadius}, for example, studied the relation between the sign of planet migration and the parameter $K$, finding a limit value ($K \sim 1.4\cdot 10^4$) where the planet switches from inward to outward migration. It is interesting to observe how, according to this criterion, the proposed planet population would mainly experience inner migration even though a significantly non-null tail of the distribution falls beyond the critical $K$ value. This last observation should however be taken with caution as planet migration, and thus its effects on gap morphologies \citep{Nazari2019RevealingObservations}, was not allowed in our simulations.

In Fig. \ref{fig:real_lit_comp}, we compare planet masses inferred using DBNets2.0 with those estimated with the previous version DBNets (\citealt{Ruzza2024DBNets:Discs}; right panel) and with those proposed by \cite{Lodato2019TheDiscs} (left panel) assuming that the gap width scales as the planet Hill radius. 
In the latter case, we observe a good agreement between the two works suggesting that our tool is indeed focusing on morphological features of the observed substructures. We interestingly observe that at the higher end of the mass spectrum, DBNets2.0 tends to return lower estimates than \cite{Lodato2019TheDiscs}. Compared to our previous work, we observe a good agreement of the new estimates for planet masses above, approximately, $1 \text{M}_\text{J}$. Lower masses present, instead, a less clear correlation between the two mass estimates with the values inferred by DBNets2.0 systematically lower than those inferred in \cite{Ruzza2024DBNets:Discs}. This disagreement may be due to two reasons. First, in \cite{Ruzza2024DBNets:Discs} tests on synthetic observations showed that low mass planets were usually slightly overestimated. Additionally, observational effects, like statistical noise or an image resolution different to the one used in the training dataset, were observed to affect the mass estimates leading, usually to their overestimation. Nevertheless, \cite{Ruzza2024DBNets:Discs} provided a rejection criterion to help identify these cases where their pipeline was failing. In Fig. \ref{fig:real_lit_comp} we mark with a yellow cross the estimates rejected according to this criterion observing that many of the discrepant estimates would be rejected.

We finally report, in Fig. \ref{fig:ma_plot}, the masses of the proposed planets in dust substructures and their radial distance from the host star together with all the confirmed detections of exoplanets as of December 2024 (\href{exoplanet.eu}{exoplanet.eu}). We highlight, with this plot, how the population of young planets that would emerge from dust substructures covers a region of the $M_p-a$ space where no exoplanets have been detected with traditional techniques due to their limitations. As also observed in \cite{Ruzza2024DBNets:Discs} we confirm that most (83\%) of the proposed exoplanets would be sub-Jupiters ($M_p < 1 M_J$) possibly explaining why more direct signatures of these objects have not yet been detected \citep{Reggiani2016TheOrbits, Nielsen2019TheAu, Vigan2021TheSHINE, Asensio-Torres2021Perturbers:Disks, Wallack202410Castella}.

\section{Discussion}
\label{sec:discussion}
\subsection{(In)dependence of the results on the observation resolution}

\begin{figure}
    \centering
    \includegraphics[width=\linewidth]{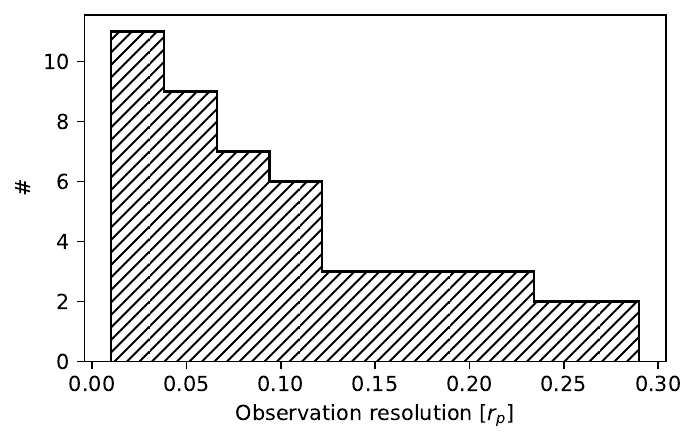}
    \caption{Distribution of resolutions of the set of dust continuum observations with substructures considered in this work.}
    \label{fig:res_hist}
\end{figure}

Unlike in our previous work \cite{Ruzza2024DBNets:Discs}, we trained this new pipeline with synthetic observations of virtually any resolution corresponding to a synthetic beam size between 0 and $0.2 r_p$. We have shown, in Sect. \ref{sec:results}, that, as a result, the accuracy of the estimated posteriors is not influenced by the resolution of the input observation. This means that our tool can be safely applied to any actual observation whose beam size lies within this range of values. The histogram in Fig. \ref{fig:res_hist} shows the distribution of beam sizes of the set of actual observations with dust substructures that we analysed with DBNets2.0 in Sect. \ref{sec:real_obs}. All of them lie within the scope of our tool with only a few exceptions that present a slightly worse resolution.

\subsection{Integration of independent constraints}
 Access to the full joint posterior for the target disc and planet properties enables the integration of our tool's estimates with independent constraints from other studies. For instance, we have shown the existence of a strong degeneracy between the planet's mass and the disc effective viscosity. Hence, if we were able to constrain the value of $\alpha$ we could break this degeneracy and improve our estimate of the planet mass. The same also applies to the other inferred disc properties.

\begin{figure}
    \centering
    \includegraphics[width=0.9\linewidth]{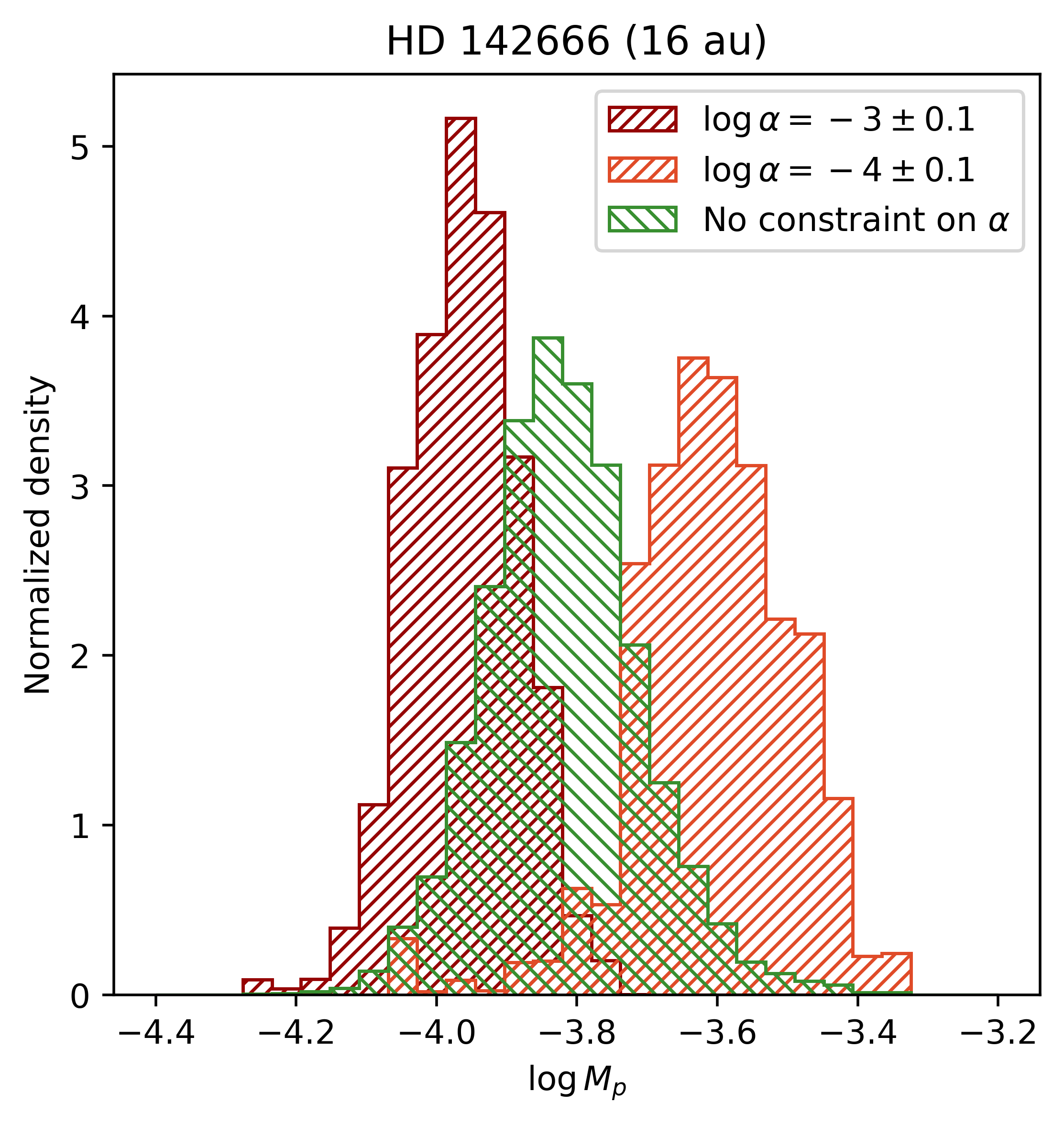}
    \caption{Marginalized posteriors $p(M_p|x)$ for the planet mass given the 16 au gap in HD142666 with no additional constraint on $\alpha$ (green histogram) and setting Gaussian priors for $\log \alpha$ centered at -3 (dark red histogram) or -4 (orange histogram) with standard deviation 0.1. Means $mu$ and standard deviations $\sigma$ of the shown distributions are $\mu=-3.82, \sigma=0.11$ (no constraint on $\alpha$), $\mu=-3.60, \sigma=0.11$ ($\alpha=10^{-3}$) and $\mu=-3.96, \sigma=0.08$ ($\alpha=10^{-4}$).}
    \label{fig:example_esthd142666}
\end{figure}

We showcase this capability considering the results obtained on the 16 au gap in HD142666 whose inferred posterior $p(\alpha, M_p | x)$, shown in Fig. \ref{fig:ex_corr}, presents a marked degeneracy between the two properties with the possible values for the viscosity spanning an entire order of magnitude. In this case, an independent measure of $\alpha$ could imply different estimates of the planet mass. This is shown in Fig. \ref{fig:example_esthd142666} where we plot, for this specific example, the inferred marginalized posterior for the planet mass with no constraint on $\alpha$ (in green) and those obtained assuming Gaussian priors with means $\alpha=10^{-3}$ and $\alpha=10^{-4}$ and standard deviations corresponding to a 10\% error on $\log \alpha$. The choice of the error magnitude is further discussed in the following paragraphs. We observe that different priors on the disc viscosity cause a shift of the planet mass posterior. Additionally, we also observe a slight improvement in precision with the $\alpha=10^{-4}$ constraint, measured by a $\sim 30 \%$ reduction of the distribution's standard deviation. 

Combining independently-constrained priors can thus shift the results and improve the estimates' precision. We demonstrated this with the results obtained on a real observation constituting an ideal case because of the marked $\alpha-M_p$ degeneracy in the posterior and high uncertainties for both these properties. We are now interested in studying systematically the consequences of combining DBNets2.0 estimates with independent constraints in terms of both accuracy and precision.
 To address this question, we conducted the following test on the test set. First, we consider an inferred property $\theta_c$ and assume that another method could provide an estimate with uncertainty $\sigma_{c, \theta}$. We thus represent these "fake" measures as Gaussian priors $p(\theta_c) = \mathcal N (\hat \theta_c, \sigma_{c, \theta})$ of standard deviation $\sigma_{c, \theta}$ and with means $\hat \theta_{c}$ that are themselves randomly sampled from a normal distribution centred on the true value of $\theta_c$ (which we know because we are using synthetic observations) with standard deviation $\sigma_{c, \theta}$. We then evaluate how the constraint on $\theta_c$ affects our estimate of the target property $\theta_t$. To do that, we marginalize the 4D posteriors obtained on the test set into the 1D posteriors $p(\theta_t|x)$ using the prior constraints $p(\theta_c)$. We then compute the best estimates for $\theta_t$ as the medians of the 1D marginalized posteriors and evaluate their distance to the target values. 

 A delicate point of this procedure is the choice of the errors $\sigma_{c, \theta}$ with which we assume to be able to constrain the disc properties. If the errors were larger than the typical uncertainties of our tool's estimates we would expect little or no improvement from additional constraints. On the other hand, we want to perform a realistic proof of concept, and all available methods for constraining the properties targeted in this work usually have high or difficult-to-quantify uncertainties.  With this premises, we assume an uncertainty of approximately $10\%$ on each disc (or planet) property which translates to 0.1 in normalized units. We point out that these values are lower than the typical uncertainties of DBNets2.0 on the test set, which can be read in Fig.~\ref{fig:metrics_sing}.

 The complete results of this test are reported in Fig. \ref{fig:const} as the mean absolute error (mae), root mean squared error (rmse) and mean of the posteriors' standard deviations ($\sigma$). To compute these metrics, we consider the median of the inferred posteriors as our best estimate.  The mae can be compared to understand global under- or overestimations, the rmse measures the tool's accuracy and $\sigma$ its precision. As explained in Sect. \ref{sec:evmethods}, we remind that the rmse computed on the normalized units can also be interpreted as the mean squared relative error with respect to the mean value of the range of values sampled in our training dataset for that property (these ranges are listed in Table \ref{tab:par_table}).
 From these values, we observe little or no improvement in both the accuracy and precision of our estimates when coupled with additional constraints on another disc property. There are, however, a few interesting exceptions. First, both $\alpha$ and \emph{St} benefit from a constraint on the other showing a slight improvement in both accuracy and precision. This result was expected from the high correlation between these properties observed in the results obtained on the test set (see Sect. \ref{sec:results_corr}).
 
\begin{figure}
    \centering
    \includegraphics[width=\linewidth]{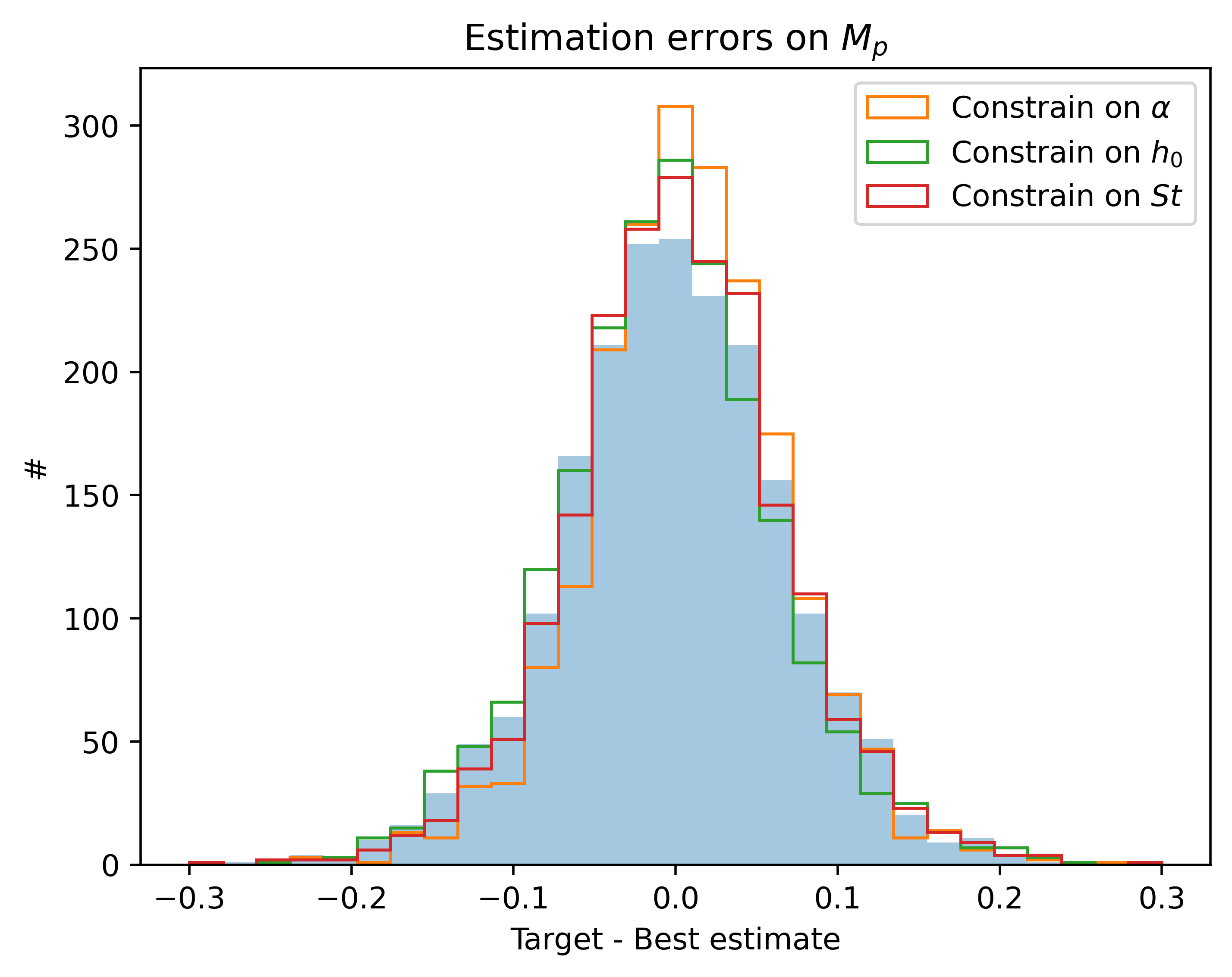}
    \caption{Distributions of errors of DBNets best estimates (median of the inferred posteriors) for the planet mass obtained with and without prior constraints on the other target properties. The test was performed on the test set.}
    \label{fig:histconstraints}
\end{figure}

 The main case that is worth discussing separately regards the planet's mass estimates which would be the main target of our tool. Figure \ref{fig:histconstraints} shows the distribution of errors for this property in all the cases when a constraint on another property is (or is not) given. We observe that, on average, an independent estimate of the disc $\alpha$-viscosity would be the most helpful to reduce the uncertainty of DBNets2.0 planet mass estimates. Specifically, in the test presented here, we achieved a 15\% reduction of both rmse and $\sigma$.
 
\begin{figure}
    \centering
    \includegraphics[width=\linewidth]{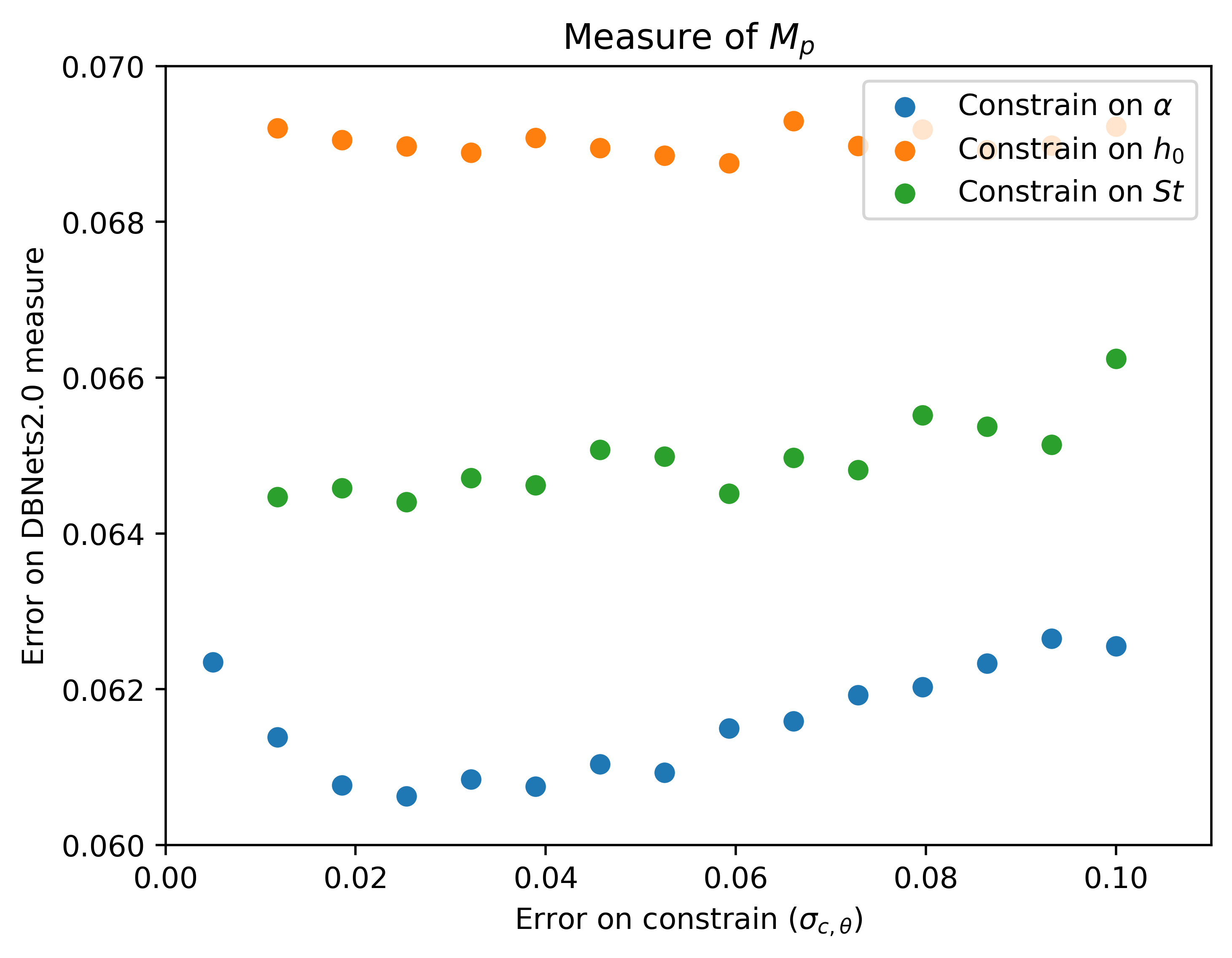}
    \caption{The rmse of the planet mass estimates, obtained by integrating the tool's results with external constraints on one of the other inferred properties, as a function of the assumed uncertainty of the independent constraint.}
    \label{fig:dim_errconst_pmass}
\end{figure}

\begin{figure}
    \centering
    \includegraphics[width=0.95\linewidth]{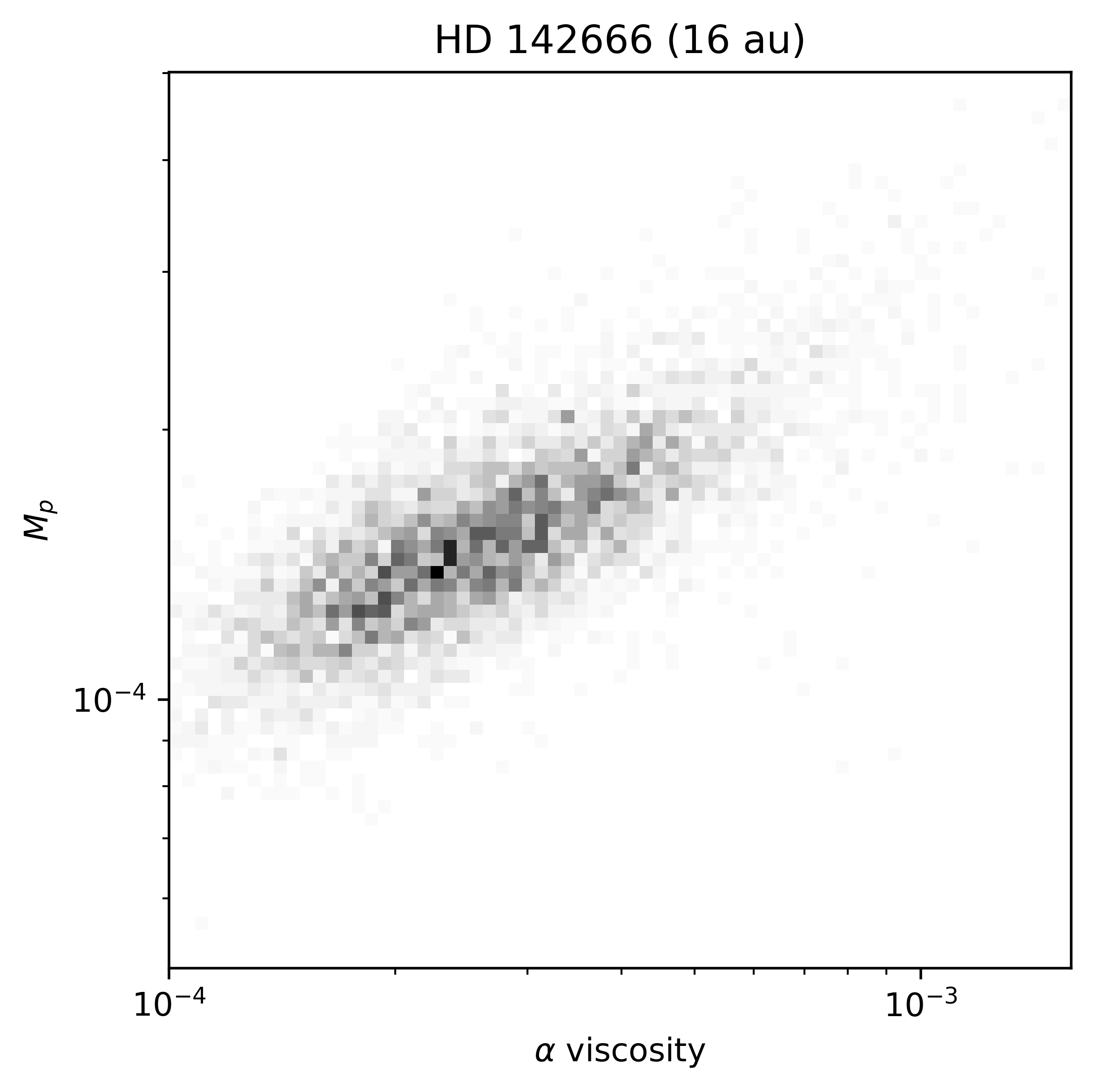}
    \caption{DBNets2.0 inferred posterior for the $\alpha$-viscosity and planet mass given the gap at 16 au in HD14266.  }
    \label{fig:ex_corr}
\end{figure}

\begin{figure}
    \centering
    \includegraphics[width=\linewidth]{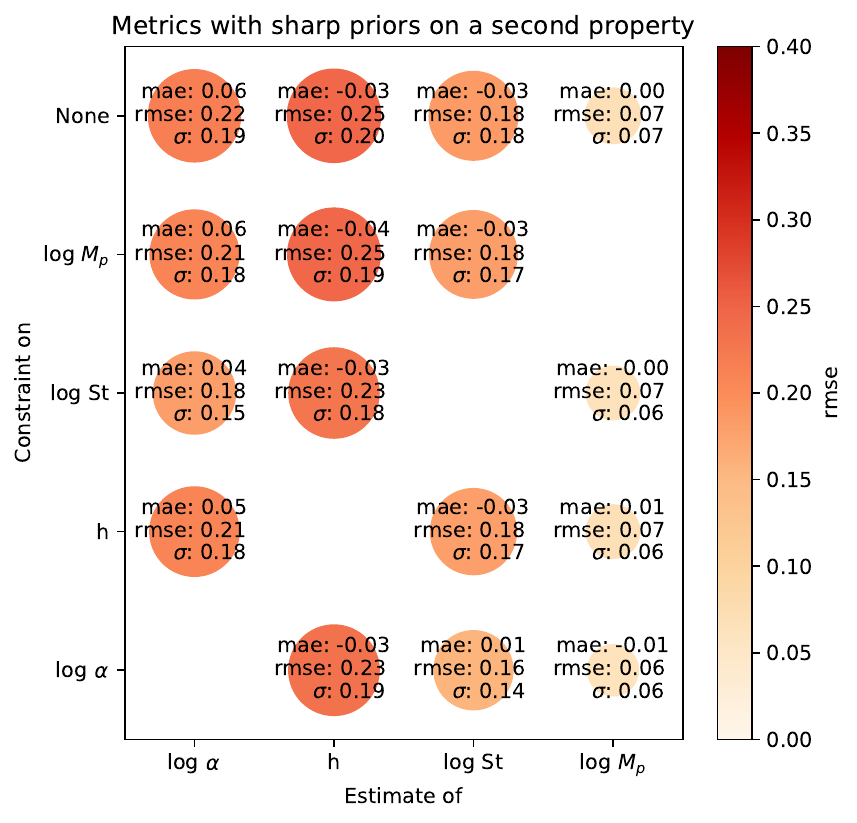}
    \caption{Evaluation metrics for each inferred property with (and without) setting a sharp Gaussian prior on a different property with standard deviation corresponding to a 10\% uncertainty. In each case, we report the mean absolute errors (mae), indicative of overall biases, the root mean squared error (rmse, indicative of the tool's accuracy) and the mean standard deviation of the inferred posteriors ($\sigma$, indicative of the tool's precision). The size of the markers is proportional to $\sigma$. }
    \label{fig:const}
\end{figure}

 All the results presented so far were obtained using constraints for the other properties with uncertainties corresponding to a relative error of approximately 10\%. In Fig. \ref{fig:dim_errconst_pmass} we show how the rmse of planet mass estimates would decrease if the uncertainties on the other disc properties were lower. We observe that in all cases the rmse would gradually decrease slightly to then reach a plateau or restart increasing. 
 These results show that, on average, external constraints would provide only a small increase in accuracy on the planet's mass estimates. We think that this is because generally, the errors and uncertainties of our tools' estimates are already low requiring a very tight constraint on other properties to provide a significant improvement. Although we proved a satisfactory overall accuracy, very tight priors highlight detailed regions of the inferred posteriors requiring an always higher accuracy as the uncertainties of the external constraints get lower. Our posterior estimates are probably not accurate enough at this level explaining the plateau or worsening of the rmse observed in Fig. \ref{fig:dim_errconst_pmass}. However, the stage at which we observe the rmse to remain constant (or increase) corresponds to unrealistically low errors on the other properties. 
 
 We remark that these results are statistical averages over the full test set. As shown at the beginning of this section, for single cases where we get large errors on one (or more) of the target properties, external constraints might indeed improve or change our estimates.
 

\subsection{Observed degeneracies and comparison with literature results}
\label{sec:deg_study}
\begin{figure*}
    \centering
    \includegraphics[width=0.48\linewidth]{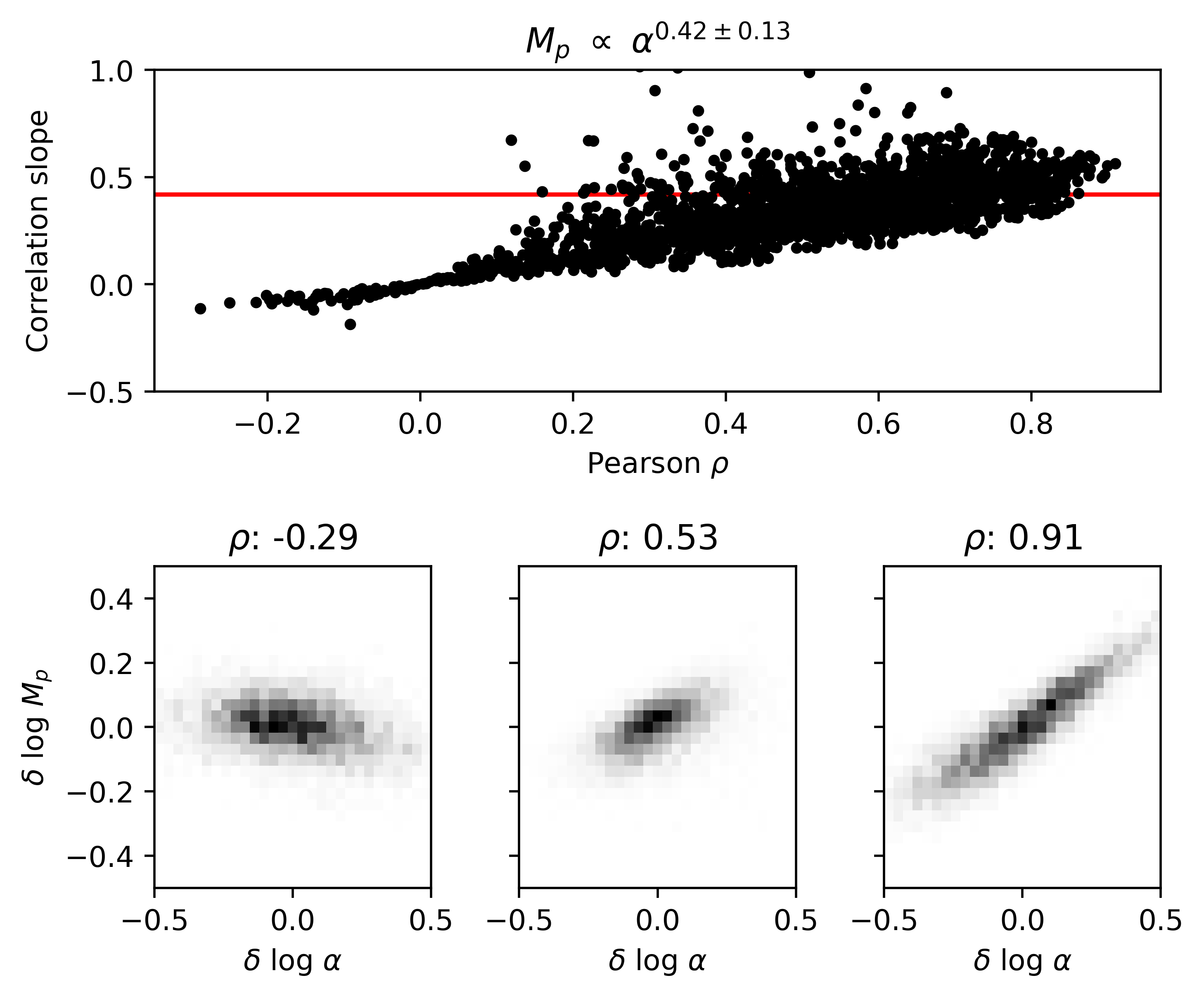}
    \includegraphics[width=0.48\linewidth]{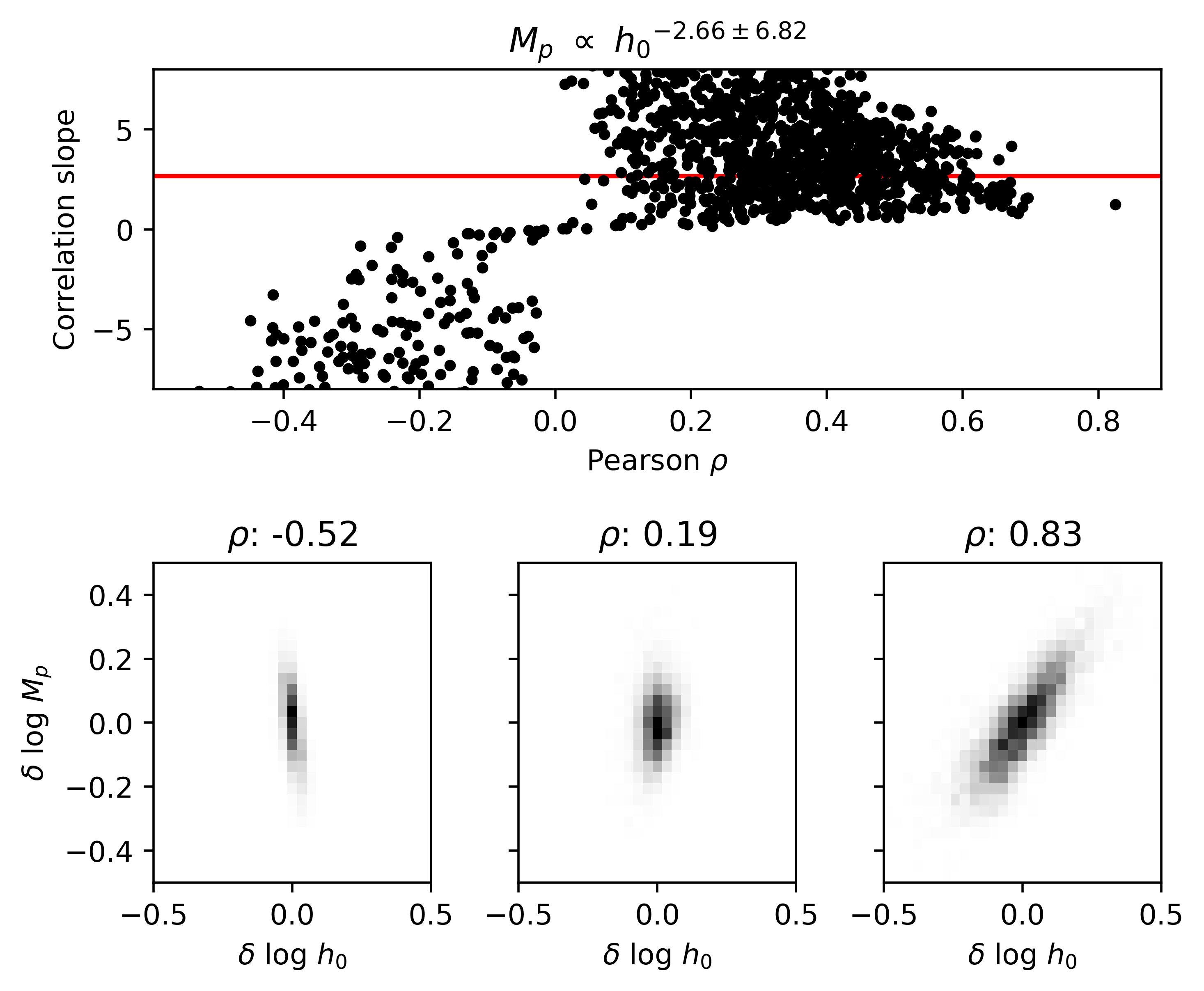}
    \includegraphics[width=0.48\linewidth]{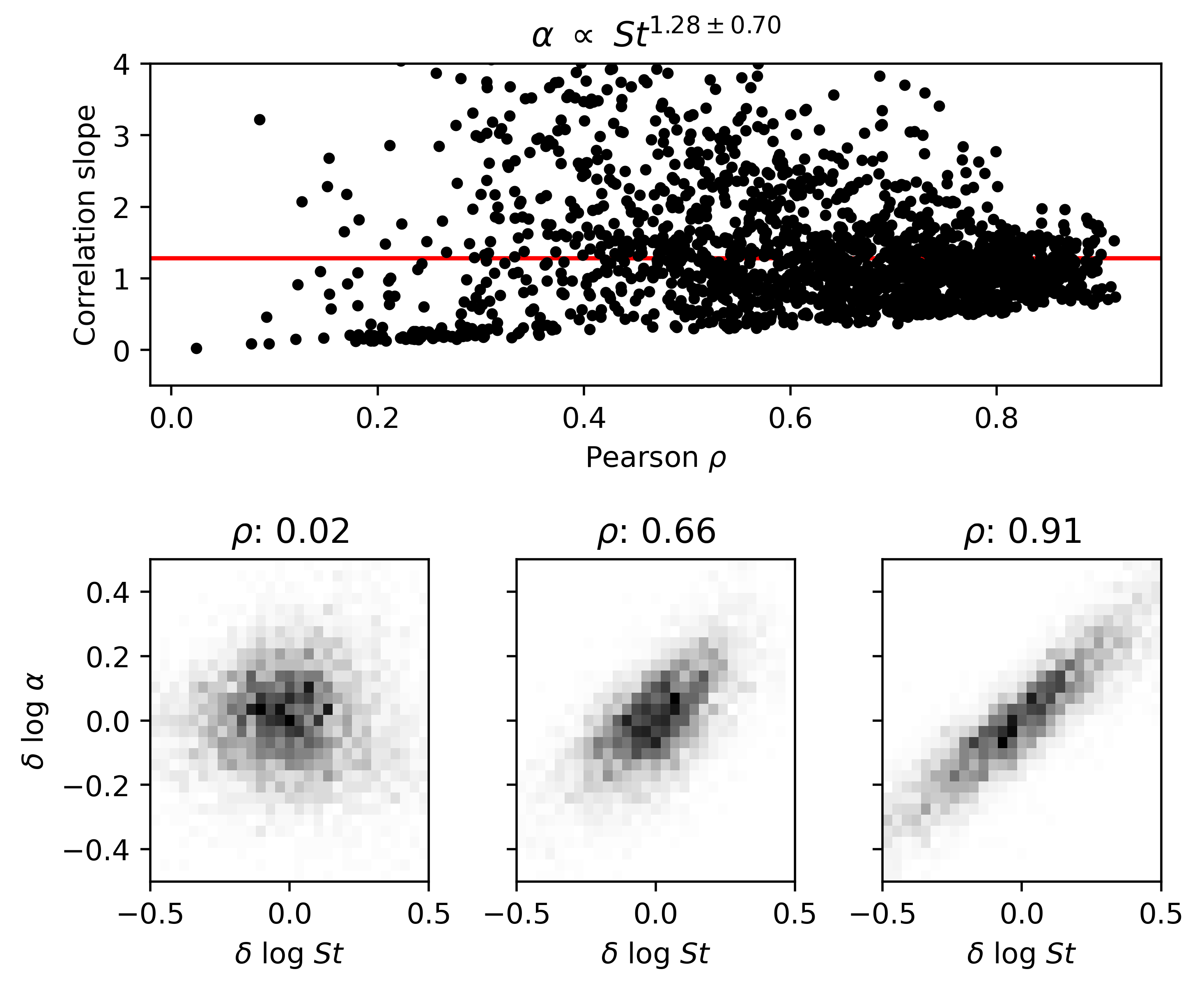}
    \includegraphics[width=0.48\linewidth]{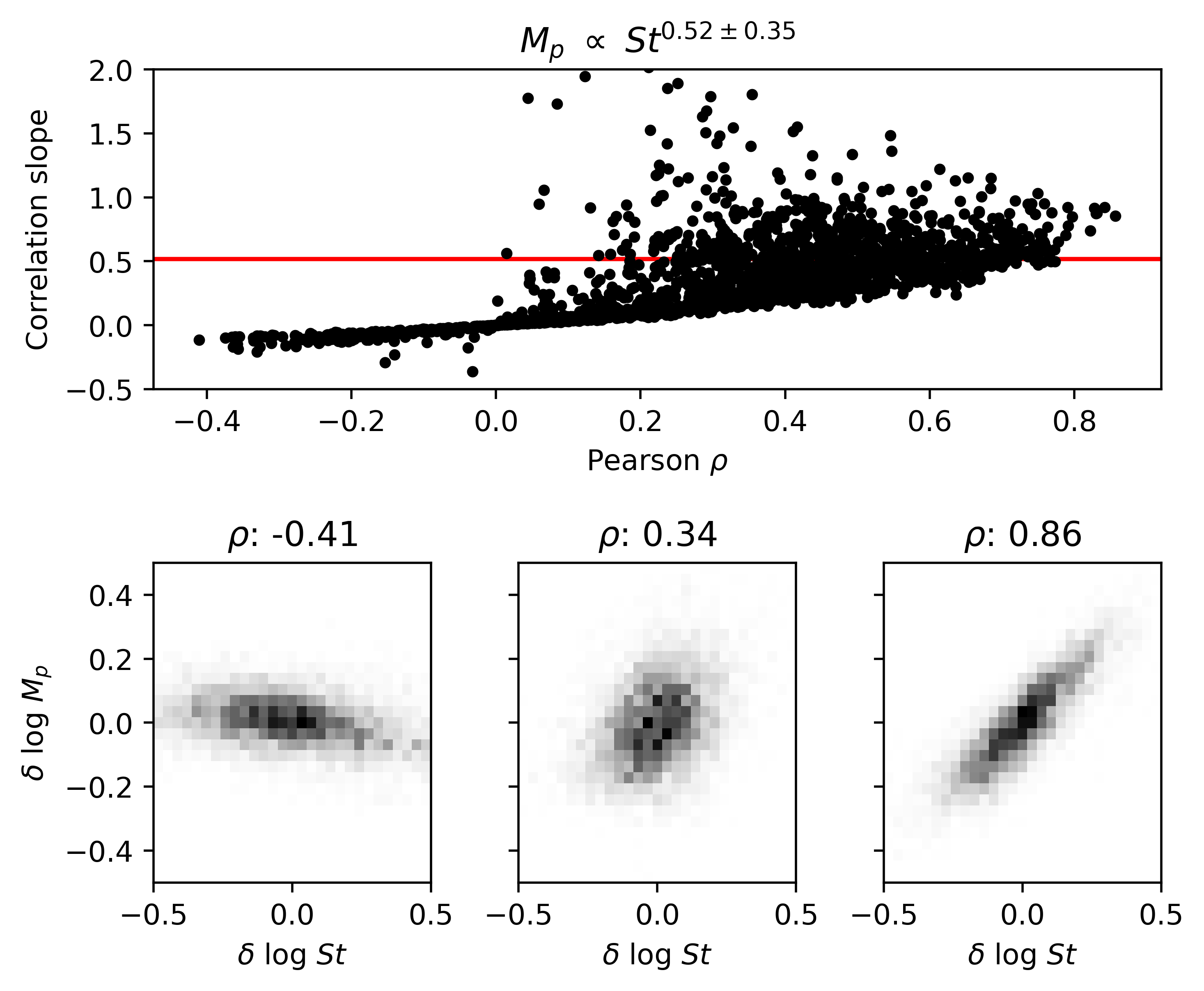}
    \caption{Degeneracies between pairs of properties highlighted by the inferred posteriors on the test set. For each simulation in the test set, these plots show the slope of the major axis of the 2D Gaussian that best fits the inferred posterior as a function of the Pearson correlation coefficients. Higher correlation coefficients mark sharper distributions towards the major axis of the Gaussian ellipses as shown by the exemplificative examples plotted. As indicated by the titles of these plots, the inferred slopes are indicative of possible degeneracies between the pairs of properties in forming the observed substructures. The red lines mark the weighted average slope using the Pearson coefficients as weights.}
    \label{fig:slopes}
\end{figure*}

We discussed in Sect. \ref{sec:results_corr} the correlation between disc and planet properties captured by the posteriors inferred on the test set. These can be interpreted considering that the same morphological features of dust substructures can be obtained with different combinations of properties. To provide a general but quantitative picture, we consider each pair of target properties (e.g. $M_p$ and $\alpha$) and assume that the degeneracy between the two can be described by a power law $M_p \propto \alpha^{\gamma}$ whose meaning is that, within a certain domain, systems with individually different values of $M_p$ and $\alpha$ can present the same morphology if the power-law relation holds with the same multiplicative coefficient. We thus consider each pair of properties and measure the possible indexes of these degeneracies by fitting the marginalized posteriors in the logged parameter space with a bivariate Gaussian distribution and computing the slope of its semi-major axis.  This is indeed the line where all points would lie in the limit of $\rho \rightarrow 1$, where $\rho$ indicates the Pearson correlation coefficient. In general, $\rho$ is the cosine of the angle between the two separate best linear predictors of one variable with respect to the other. Hence, the most reliable dependence between the two variables is given by the distributions with the highest $\rho$.
We show in Fig. \ref{fig:slopes} the slopes obtained in this way as a function of $\rho$ for the respective distributions. We show, in the same plot, three examples of the marginalised two-dimensional posteriors relative to the lower, median and highest correlation coefficients to visually assess the meaning of these values in terms of the distribution's shape.
For each pair of properties, we provide a single estimate, with uncertainties, for the index of the possible power law degeneracy averaging the slopes computed on each element of the test set weighted by the square of their Pearson correlation coefficient. 
It is interesting to check whether these relations correspond to those captured by the $K$ and $K'$ coefficients (introduced in Sect. \ref{sec:real_obs}) which have been found to capture the main dependence of gap morphologies from the properties of the disc and planet. 

The slope we measure for the typical correlation between $M_p$ and $\alpha$ in our test set  ($M_p \propto \alpha^{0.42}$) is in good agreement with the degeneracy captured by both the $K$ and $K'$ coefficients which, fixing the other properties, would result in $M_p\propto \alpha^{1/2}$. 
The measured, almost linear, relation between $\alpha$ and St is compatible with the substructures' morphologies being primarily determined by the ratio St$/\alpha$ rather than by the two properties separately and, analogously, the $M_p \propto \text{St}^{0.52\pm0.35}$ relation agrees with this picture having an index compatible with that between $M_p$ and $\alpha$. This might suggest that the main morphological features from which the inference is performed are those related to dust trapping, which are those relative to the rings.

Considering the $M_p-h$ degeneracy, the mean power law index that we measure in this case (2.66) is in fair agreement with that expected from the $K$ coefficient. However, we observe here a larger scatter and typically lower correlation coefficients indicating that this degeneracy is less strong and coherent across different regions of the parameter space.
Some scatter may be explained by the different dependence of the $K$ and $K'$ coefficients from the disc aspect ratio. These are related respectively to the gap depth and width, which should be simultaneously taken into account by the CNN in our pipeline. 

Despite being less marked, a certain scatter in both the correlation coefficients and the measured power-law indexes is also observed in the other cases. We point out that the empirical relations based on the $K$ and $K'$ are derived for morphological features of gas gaps while, in this work, we are only considering dust substructures. We expect these formulae to still hold at some level due to the low Stokes numbers of our simulations, but it may also explain the imperfect agreement with the analytical expectations and the lack of a unique parameter degeneracy.
Furthermore, we estimate the expected degeneracies between pairs of parameters starting from the empirical formulae by fixing all the other properties but we measure them on the two-dimensional marginalised posterior distributions. Hence, if the integrated properties are not completely independent an uncertainty on them could hide or change the relation between the two properties under exam. 
A more systematic study of how these relations change within the parameter space and considering multidimensional dependences is beyond the scope of this paper.

\section{Conclusion}
\label{sec:conclusions}

In this work, we developed a simulation-based inference pipeline for the analysis of protoplanetary discs' observations of the dust thermal emission. From the morphology of observed substructures, we estimate the posterior for the disc $\alpha$-viscosity, disc scale height and dust Stokes number, as well as the mass of a putative planet. This is the first publicly available and fully automated tool with this capability. We performed TARP tests \citep{Lemos2023Sampling-BasedInference} and computed standard metrics (rmse and r2-score) to quantify the accuracy and precision of the inferred posteriors.

With respect to our previous work \cite{Ruzza2024DBNets:Discs}, the most significant improvement lies in the inference of the full posterior for the planet mass and additional disc properties but we also solved two minor issues providing a new pipeline whose results' accuracy is not affected by the resolution of the input observation or by the position of the outer disc edge. Additionally, DBNets could only approximate the inferred posterior as a Gaussian mixture while, in this work, we were able to remove this limitation through the use of normalising flows.  
On the test set, we achieve a rmse for the planet mass of 0.07 which, keeping into account the different normalisation, is in line with the results obtained in our previous work \citep{Ruzza2024DBNets:Discs}. On actual observations, DBNets and DBNets2.0 planet mass estimates are in good agreement towards the higher part of the mass spectrum. For lower masses, we instead typically infer lower values with the new pipeline. This result is consistent with the tests that we performed in  \cite{Ruzza2024DBNets:Discs} as we observed, on synthetic observations, a systematic overestimation of low planet masses, especially after the injection of observational errors (e.g. noise, deprojection effects, change of resolution).

Evaluating our tool's precision on the test set, we conclude that the morphology of planet-induced substructures is primarily determined by the planet mass as this is always the best-constrained property. Nevertheless, the other properties are also inferred successfully with errors significantly lower than the parameter space dimension suggesting that they also imprint detectable signatures on the substructures' morphologies. We also studied, on the same results, the degeneracies captured by the inferred posteriors between each pair of parameters finding $\alpha$ and St to be the properties with the most marked degeneracy with the planet mass. A strong correlation between $\alpha$ and St themselves, consistent with $\alpha \propto \text{St}$, suggests that, in addition to the planet mass, the morphology of dust substructures is mainly controlled by the ratio between these two properties.

We used the developed pipeline on a set of 49 gaps across 34 discs assuming independently the presence of a planet in each gap. The results present generally low $\alpha$-viscosities and disc scale heights which would imply long viscous timescales ($\sim 10^{5}-10^7 \Omega^{-1}$). The population of proposed planets also presents generally low masses with approximately 83\% of them being below 1M$_J$, consistent with the lack of direct detections of these objects.

The pipeline presented here demonstrates an approach to fitting a (simulation-based) model to data with techniques that improve both speed and accuracy compared to alternative procedures. Inferred posteriors should be interpreted with caution: they represent results conditioned on the model and assumptions underlying the synthetic observations used in the training dataset. As such, they do not account for situations or phenomena not included in these simulations. Future work could improve the underlying models by introducing additional effects, such as planet migration, dust feedback, or planet multiplicity. In this scenario, the access to the full posteriors for the disc and planet properties, rather than single value best estimates, would allow us to perform a proper statistical comparison of the different models to determine which one best fits the data.

\section{Data availability}
The entire pipeline is publicly available as a Python package at \href{https://github.com/dust-busters/DBNets}{https://github.com/dust-busters/DBNets}.
DBNets2.0 requires a disc observation of the dust continuum, the disc geometrical 
properties (inclination, position angle, and location of the disc
centre within the image), and a guess on the alleged planet location either provided as the angular separation from the disc
centre or in physical units with the disc distance. The tool thus
returns the best fit posterior distribution for the planet mass, disc viscosity, aspect ratio and dust Stokes number.
Utility functions are also provided to interpret and analyse the results. Comprehensive documentation with commented examples
is available via the same link.

\begin{acknowledgements}
We thank the anonymous referee for the insightful comments and suggestions. Computational resources have been provided by the INDACO Core facility, which is a
project of High Performance Computing at the Università degli Studi di Milano
(https://www.unimi.it). We also acknowledge ISCRA for awarding this project access to the LEONARDO supercomputer, owned by the EuroHPC Joint Undertaking, hosted by CINECA (Italy). This work has been supported by Fondazione Cariplo,
grant n° 2022-1217, from the European Union’s Horizon Europe Research \&
Innovation Programme under the Marie Sklodowska-Curie grant agreement No.
823823 (DUSTBUSTERS) and from the European Research Council (ERC) under grant agreement no. 101039651 (DiscEvol). GL acknowledges support from PRIN-MUR 20228JPA3A and from the European Union Next Generation EU, CUP:G53D23000870006. Views and opinions expressed
are however those of the author(s) only, and do not necessarily reflect those of
the European Union or the European Research Council Executive Agency. Neither the European Union nor the granting authority can be held responsible for
them.     
\end{acknowledgements}

\bibliographystyle{aa}
\bibliography{references}

\begin{appendix}

\section{Confidence score}

\label{app:conf_score}
DBNets2.0 is a simulation-based inference tool. As such, it infers the posterior distribution for some target properties, fitting data with a model that is intrinsically defined by the training synthetic data. This means that the results are dependent on all the assumptions made in running the simulations used to generate these data and on the structure of the explored parameter space. Data that satisfy these assumptions and therefore are within the scope of the tool are typically referred to as in-distribution (ID) data in opposition to out-of-distribution (OOD) data, which are instead out of the tool's scope. Deep learning methods, like those we used in our SBI pipeline, are designed to interpolate within ID data but are typically not reliable when applied to OOD data. While directly detecting OOD data is challenging, as most of the assumptions made in generating the training dataset cannot be directly and reliably assessed, we can instead check the similarity of the data that we are targeting with the training data. Although this approach does not eliminate the possibility of degeneracies with OOD scenarios, it provides a practical means to estimate the reliability of the model’s outputs. Following this idea, we designed and equipped our tool with the ``confidence score'' (CS), a new metric that helps the user to assess the reliability of the inferred posterior.

The idea underlying this metric is to quantify the difference between the input observation and a synthetic observation corresponding to DBNets2.0 best estimates for the disc and planet properties. In other words, this metric summarises in a faster and automated procedure the PPC tests presented in Appendix \ref{app:ppc}.
Given an observation $x$ and DBNets2.0 estimate $p(\theta|x)$, CS is computed as follows: 1) we sample ten $\theta_i$ from $p(\theta|x)$, 2) we linearly interpolate the synthetic images in the training dataset at all $\theta_i$, 3) we remap the interpolated images $b_i$ and the input image $x$ to polar coordinates, 4) we compute the FFT of the remapped $b_i$ and $x$, 5) we compute CS as 
\begin{equation}
    \text{CS} = 1 - \frac {1} {10} \sum_i \frac{(|F(x)| - |F(b_i)|)^2 }{(|F(x)|^2 + |F(b_i)|^2)}.
\end{equation}
This pipeline implements the idea previously exposed. We use linear interpolation of the training dataset as a cheap surrogate of hydrodynamical simulations, but we sample 10 points from the inferred posterior to average out artefacts and errors. Additionally, sampling 10 points also makes the CS definition sensitive to the uncertainty of the inferred properties, lowering its value in the case of broad inferred posteriors. We remap the images to polar coordinates and compute the Fourier decomposition, using only the modulus of the complex Fourier coefficients, to remove any azimuthal dependence. For instance, if the input data presents a strong azimuthal asymmetry the image is found similar to the training data if a similar asymmetry is present, regardless of its azimuthal position. We finally included a normalisation factor in our definition of CS.

\subsection{Validation and calibration}
\begin{figure}
    \centering
    \scalebox{0.6}{\includegraphics{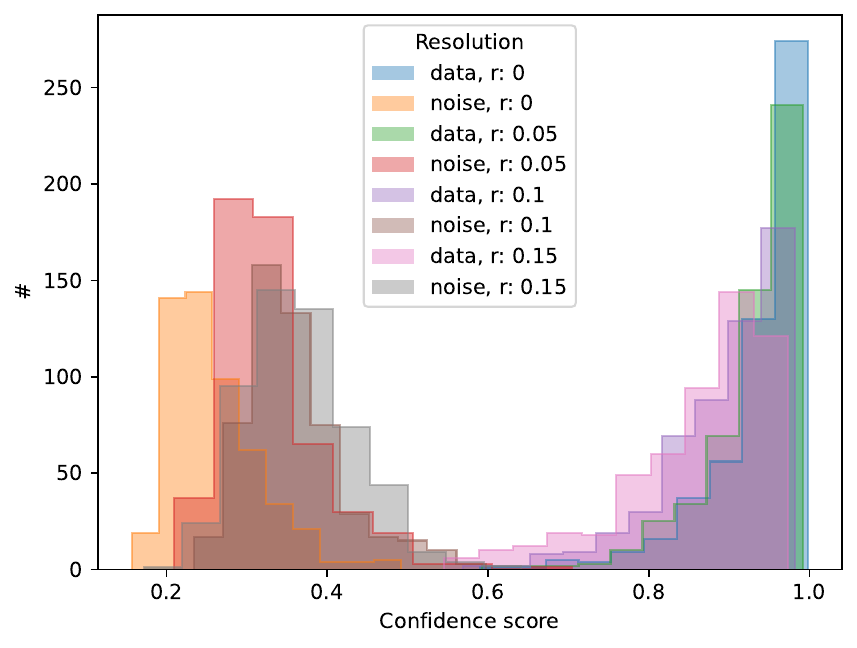}}
    \caption{Confidence score distributions for DBNets2.0 estimates on the synthetic observations of the test set and on white noise images. In both cases, we show the results for the same images convolved with a Gaussian beam of different sizes (standard deviation reported in the legend).}
    \label{fig:ood_fig1}
\end{figure}

\begin{figure}
        \centering
        \includegraphics[width=\linewidth]{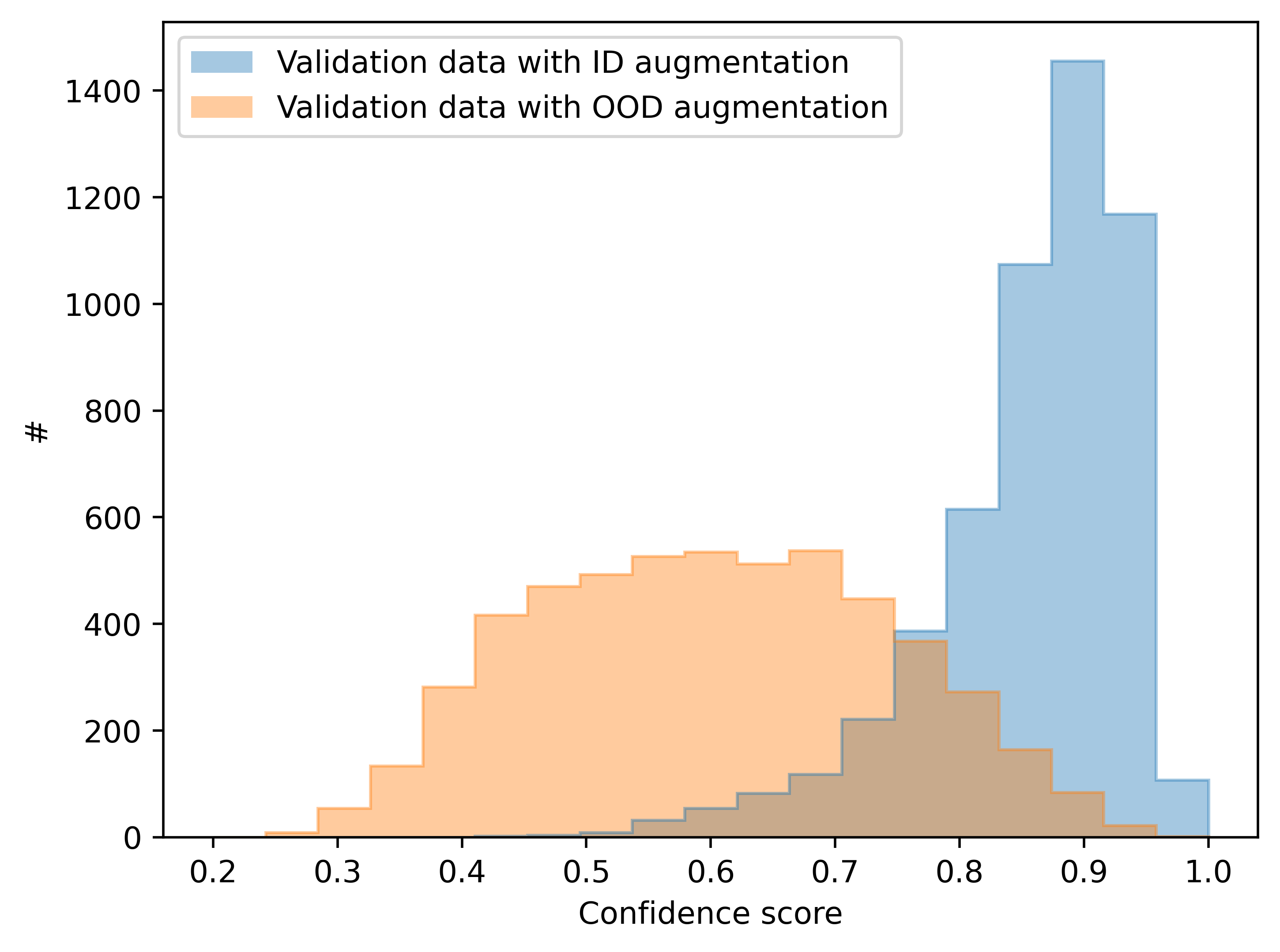}
        \caption{Confidence score distributions for DBNets2.0 estimates on the synthetic observations of the test set randomly augmented with the same augmentation layers used during training: random rotation, random translation, random cut of the outer disc, convolution with a Gaussian beam of random size and addition of white noise. The blue histogram corresponds to in-distribution (ID) augmentation, i.e. using the same hyperparameters used during training. The orange histogram corresponds to values obtained for images on which we performed OOD augmentation increasing the maximum translation factor from 0.01 to 0.1, the noise standard deviation from 0.1 to 0.2 and the maximum synthetic beam size from 0.2 to 0.4 (in code units).}
        \label{fig:ood_fig2}
    \end{figure}

To calibrate this confidence score and assess its correct functioning we performed several tests. First, we checked CS values for input data completely out of scope, i.e. white noise, observing, as shown in Fig. \ref{fig:ood_fig1}, that their distribution is well separated from that of CS values computed for our validation set. Furthermore, in Fig. \ref{fig:ood_fig2} we compare the CS values computed for the test set images augmented with the same augmentation layers used in the DBNets2.0 CNN (see Sect. \ref{sec:CNN}) using the same (ID) and wider (OOD) ranges for the augmentation hyperparameters (e.g. translation factor, noise level, synthetic resolution, etc.). We observe a shift of CS values towards lower values in the case of inputs on which OOD augmentation was performed. Note that while in the case of OOD augmentation we increased the ranges of allowed values for the augmentation's hyperparameters, these are randomly sampled for each new observation and might thus, in specific cases, still fall within ID values. Additionally, particularly simple morphologies might still appear similar to the training data even when OOD augmentation is performed. This explains the overlap in the CS values distributions.

\begin{figure}
    \centering
    \includegraphics[width=\linewidth]{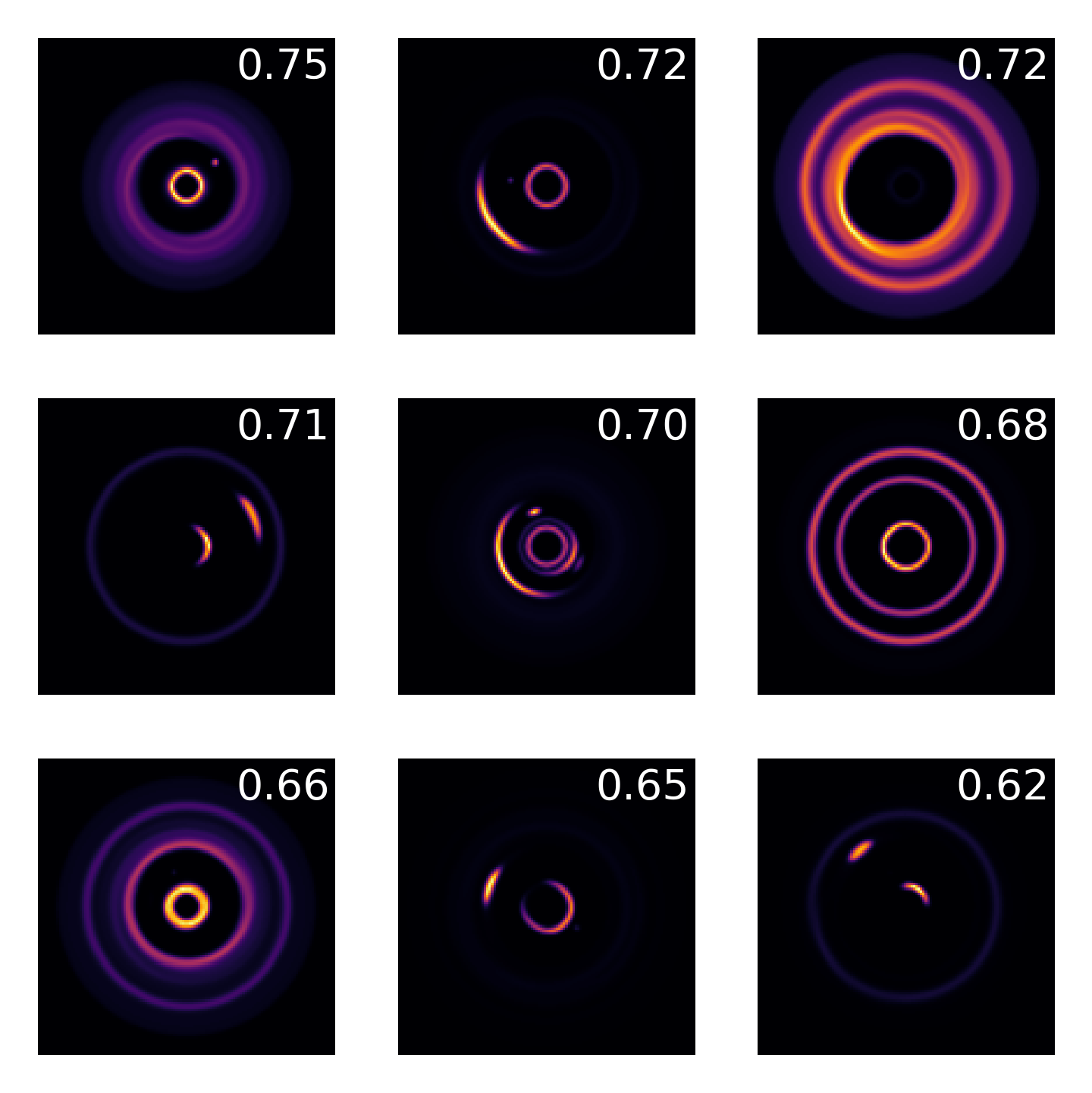}
    \caption{Morphology of the substructures in the test set synthetic observations corresponding to the lowest confidence scores (reported in the upper right corner of each image).}
    \label{fig:ood_fig3}
\end{figure}

From the results shown in Fig.~\ref{fig:ood_fig1} and Fig~.\ref{fig:ood_fig2} we can suggest a threshold of 0.6 for CS under which DBNets2.0 estimates should be rejected. However, we strongly encourage future users to consider the actual CS value, treating it exactly as a continuous confidence score whose value must be considered together with all the information available about the disc under exam. Indeed, individually inspecting the disc morphologies, we observe that they become qualitatively more peculiar as the CS decreases. For instance, Fig. \ref{fig:ood_fig3} shows the 9 elements of the validation set corresponding to the lowest CS. As shown, these exhibit untypical morphologies with strong asymmetries or multiple rings and gaps.

\begin{figure}
    \centering
    \includegraphics[width=\linewidth]{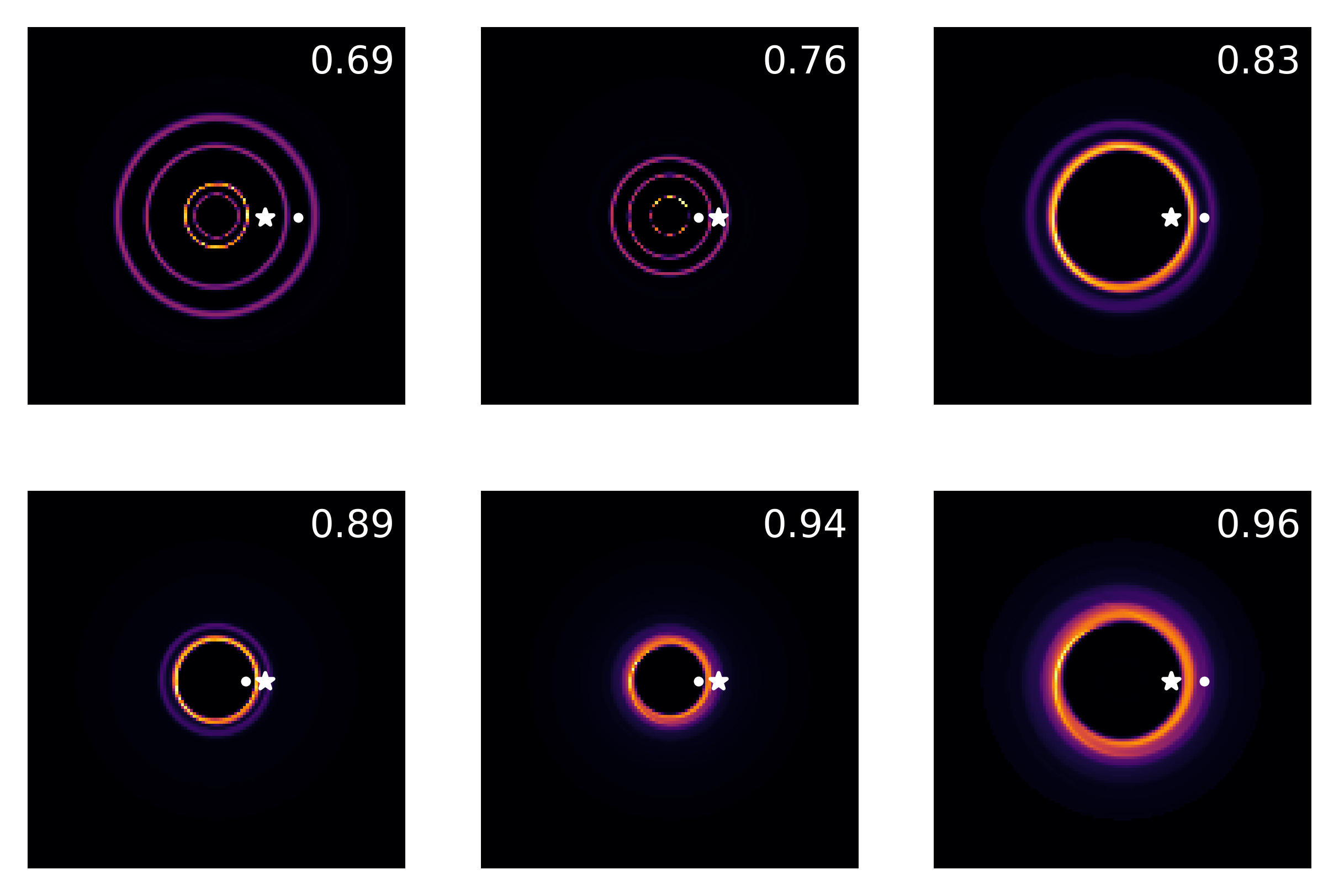}
    \caption{Synthetic observations obtained from hydrodynamical simulations with two planets. The confidence score of DBNets2.0 estimates on these images is reported in the upper right corners. In each image, the star marks the position of the planet which is also assumed for the analysis of DBNets2.0, while the point marks the position of the second planet in the simulation.}
    \label{fig:ood_fig4}
\end{figure}

As a final test, we computed the CS on synthetic observations generated from simulations with two planets. This is meant to benchmark the CS in a case where the physics assumptions in the model that DBNets2.0 is fitting are not met. Figure \ref{fig:ood_fig4} shows these synthetic images ordered with respect to the CS of DBNets2.0 estimate. We observe that even though none of these CS values fall below 0.6, the images corresponding to the lowest CS exhibit the most peculiar morphologies with multiple gaps and rings that are probably difficult to explain with only one planet. It is, however, important to keep in mind that this metric cannot point out degeneracies with OOD systems, and therefore, generally, a high CS is still expected if the observed morphology could also be explained with only one planet.

\subsection{Confidence score of DBNets2.0 application to actual observations}

\begin{figure}
    \centering
    \scalebox{0.65}{\includegraphics{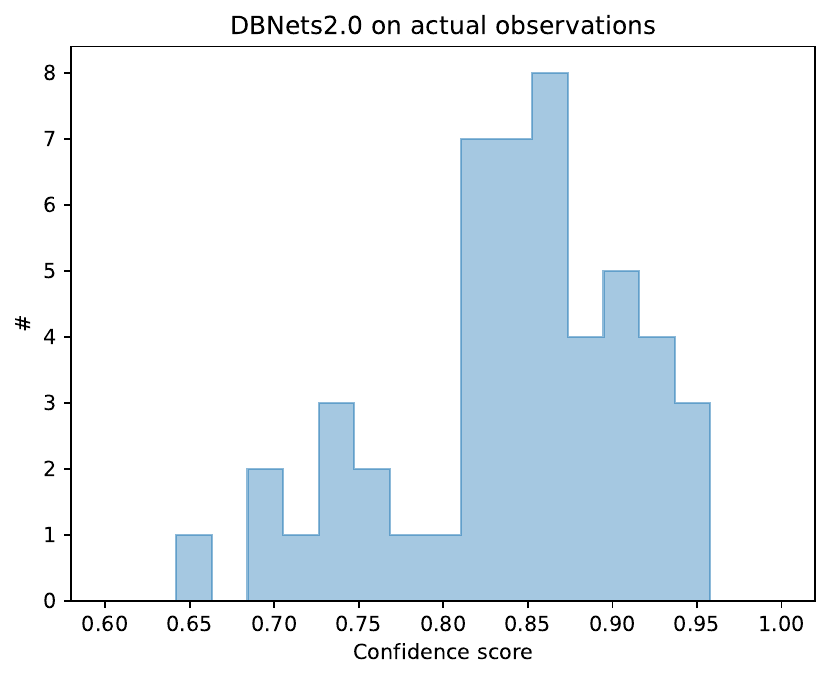}}
    \caption{Distribution of confidence scores obtained for DBNets2.0 estimates on the actual observations analysed in this paper and listed in Table \ref{tab:prop}.}
    \label{fig:ood_fig5}
\end{figure}

\begin{figure*}
    \centering
    \includegraphics[width=\linewidth]{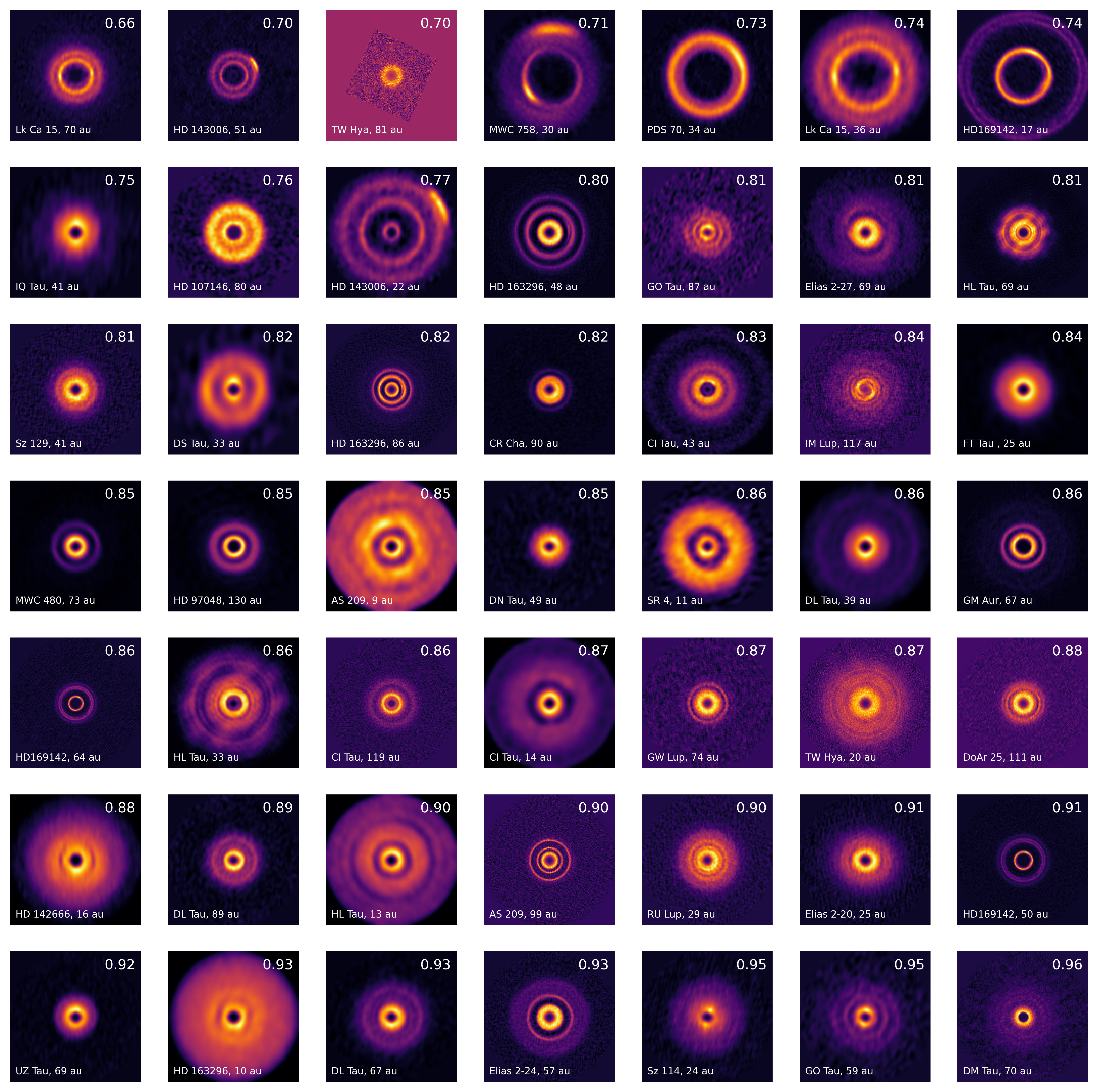}
    \caption{Deprojected continuum observations of actual discs analysed with DBNets2.0 in this paper. Images are ordered according to the confidence score of DBNets2.0 estimate and are meant to show how this metric relates to the observations' features.}
    \label{fig:ood_fig6}
\end{figure*}

We show in Fig. \ref{fig:ood_fig5} the distribution of CS obtained for the actual observations on which we applied DBNets2.0 in this paper. Figure \ref{fig:ood_fig6} shows the morphology of the observed substructures ordered with respect to the obtained CS. We also report CS values for each DBNets2.0 estimate in Table \ref{tab:prop}.

\newpage
\section{Evaluation of the feature extracting CNN (first step of the pipeline)}
\label{app:appa}

\begin{figure}
    \centering
    \includegraphics[width=\linewidth]{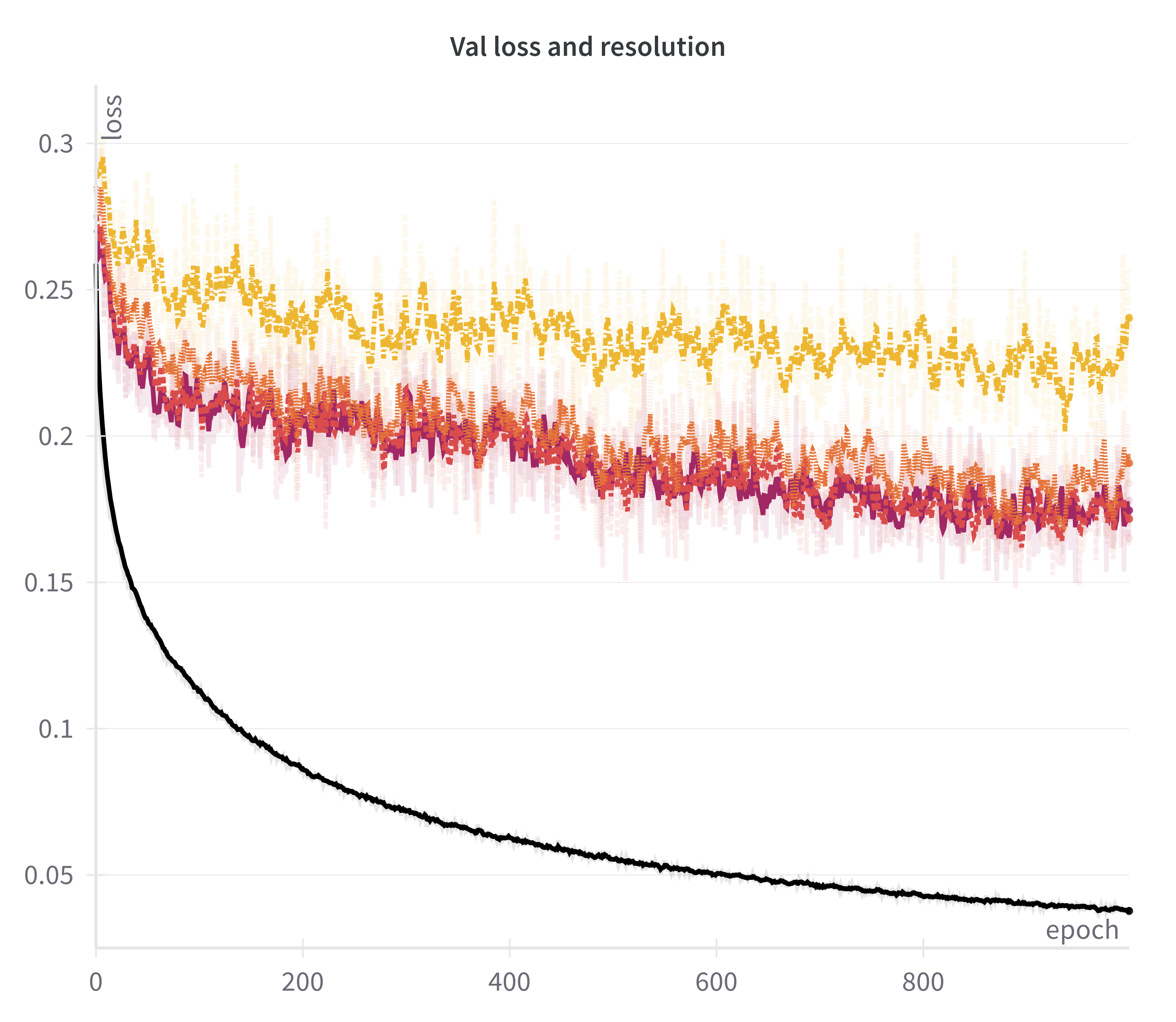}
    \caption{Loss function (mse averaged over all 4 inferred parameters) computed during training for both the training and test sets. The black line refers to the metric computed on the training dataset. Coloured lines refer to the metric computed on the test set where each image has been convolved with a Gaussian beam of fixed size among [0, 0.05, 0.1, 0.15]~$r_p$. Darker colours correspond to smaller beam sizes. Plots are shown for one of the 5 folds trained but the others exhibit a similar behaviour.}
    \label{fig:learn_curves}
\end{figure}

\begin{figure}
    \centering
    \includegraphics[width=\linewidth]{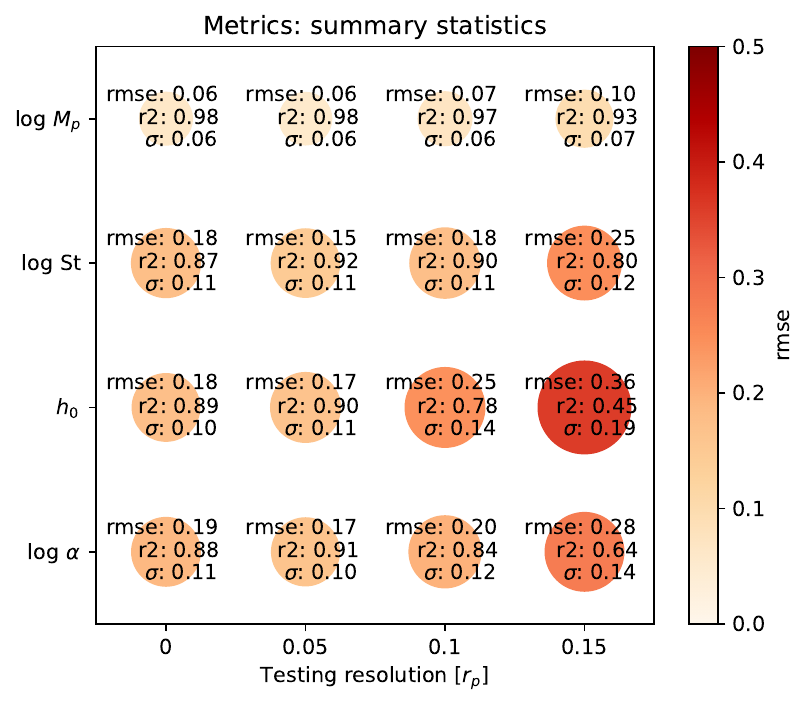}
    \caption{Different metrics summarising the CNN results on the test set for each inferred parameter and for different synthetic resolutions. The colours ascend from brighter to darker as the mse. Sizes scale as $\sigma$. Definitions of the plotted metrics can be found in the text (Appendix~\ref{app:appa}).}
    \label{fig:metrics_cnn}
\end{figure}

Figure \ref{fig:learn_curves} shows the learning curve (i.e. loss function as a function of the training epoch) of the CNN trained in this work. The black line shows the loss function computed on the training dataset while the others refer to the loss function computed on the test set with synthetic images convolved with Gaussian beams of different dimensions where darker colours refer to smaller beams. We observe that the CNN is trained effectively exhibiting a significant lowering of the loss function evaluated on the training set. The test set curves show that the CNN is not overtrained. Beyond the stochastic behaviour, a small but systematic improvement of these curves can be observed up to the last training epochs.
We remind that the CNN output, for each input observation, is a set of 4 real values interpreted as estimates of $\log \alpha, h, \log \text{St}$ and $\log \text{M}_p$. Enabling dropout layers during inference, we collect for each input a set of 1500 estimates for each target property which, in our pipeline, are used as input to the normalising flows. A different solution could have been to directly interpret these values as samples extracted from the target posterior distribution $p(\alpha, h, \text{St}, M_p|x)$. We already show in Sect.  \ref{sec:results4d}, performing the TARP test, that this method would provide inaccurate estimates. We additionally report here in Fig. \ref{fig:metrics_cnn} some metrics analogously to Fig. \ref{fig:metrics_sing} but computed using only the CNN outputs. The rmse values are all systematically higher than the $\sigma$ values, which correspond to the standard deviations of the distributions obtained from the CNN outputs, indicating that these systematically underestimate the uncertainties.

\newpage
\section{Posterior predictive check}
\label{app:ppc}
In this appendix, we present the results of posterior predictive checks (PPC; e.g. \citealt{Gabry2019VisualizationWorkflow}) that we used to further test our pipeline. In a PPC, one considers the observed data $x_0$ and compares them with those generated from the pipeline's best estimates of the target properties. We call "best estimates" the medians of the inferred marginalised posterior distributions corresponding also, in all these cases, to the maximum a posteriori estimate (MAP). The general idea is that the generated data should look similar to the observed one if the inference is correct. We performed this test using synthetic observations and selected examples with a large difference between the target properties and their inferred best estimates. These are the most interesting to perform this test as they can show if the inferred posteriors are correctly tracing degeneracies between parameters. Instead, if the best estimates were close to the actual target values, the test would be comparing synthetic observations that are expected to be similar because obtained from simulations with small differences in their disc and planet properties.

\begin{figure*}
    \centering
    \includegraphics[width=0.4\linewidth]{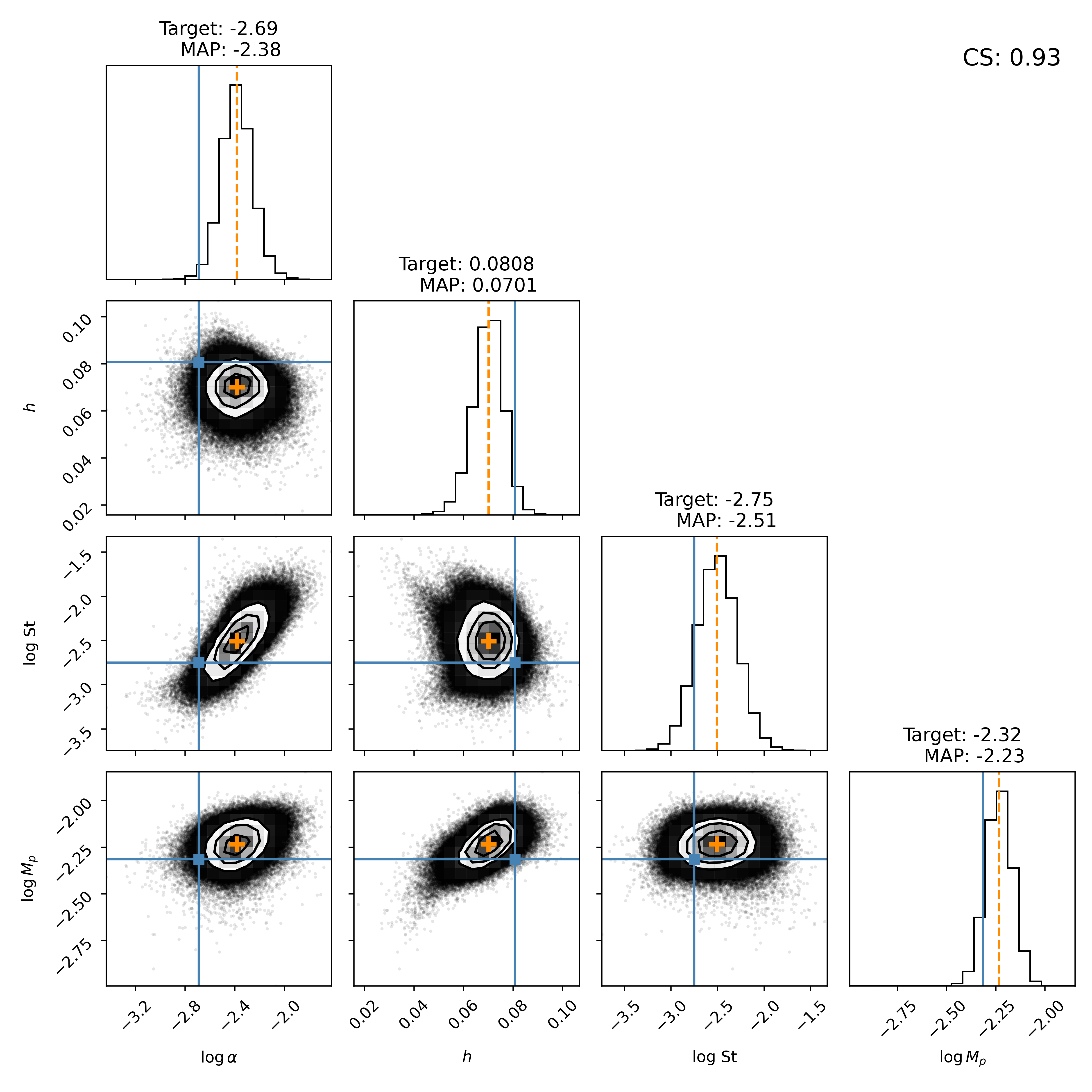}
    \includegraphics[width=0.58\linewidth]{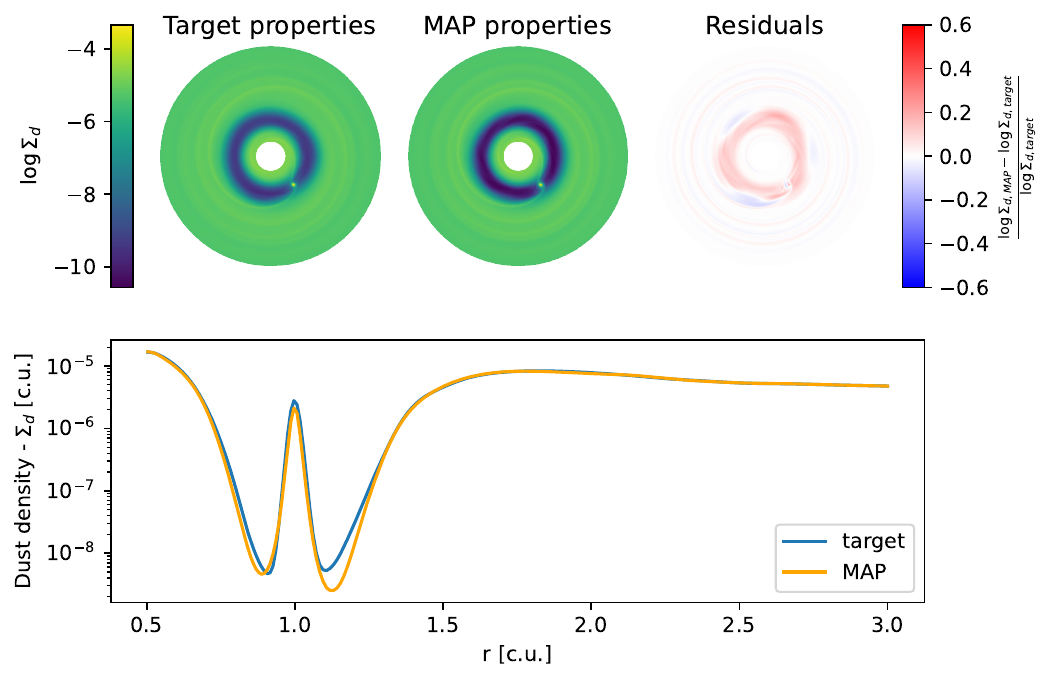}
        \includegraphics[width=0.4\linewidth]{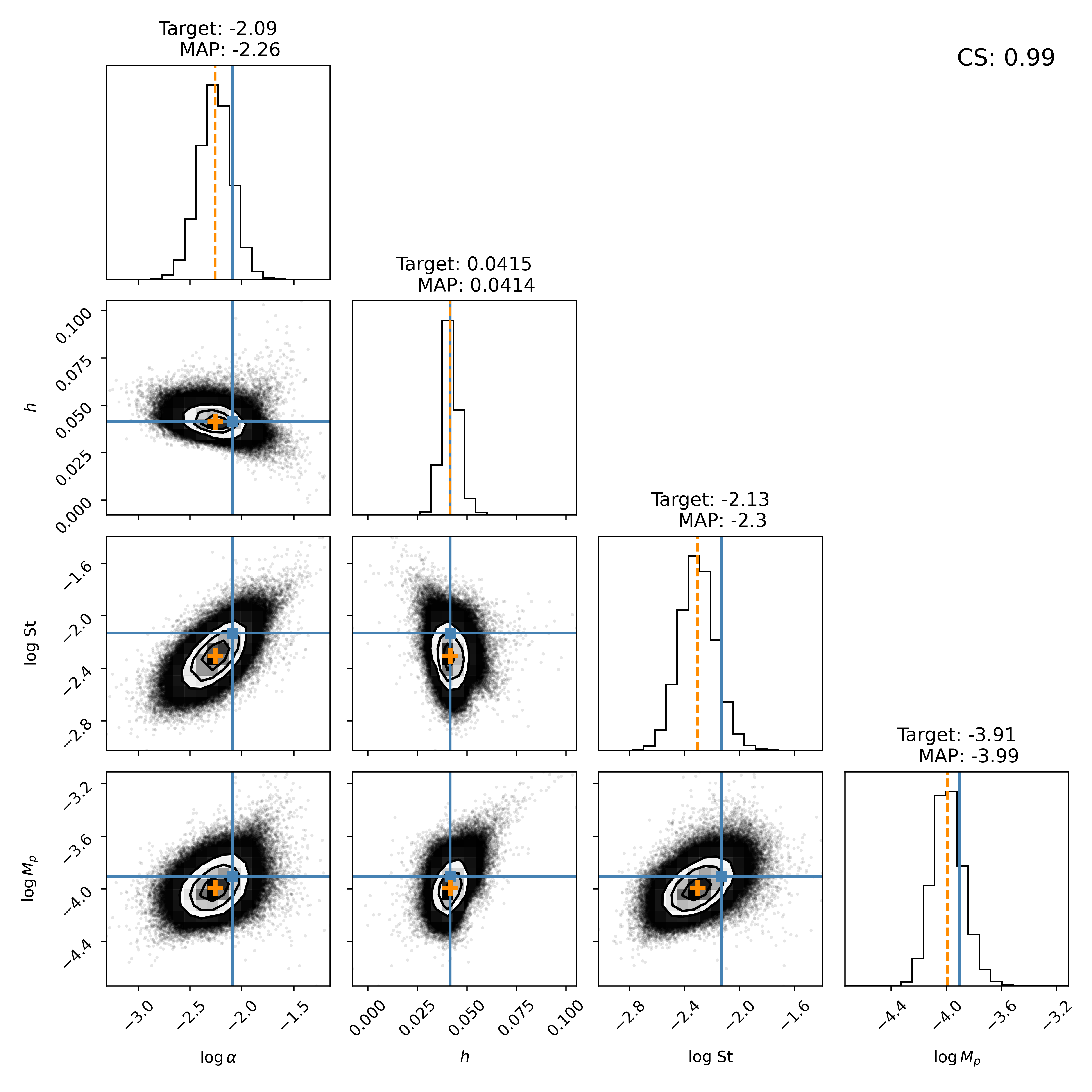}
    \includegraphics[width=0.58\linewidth]{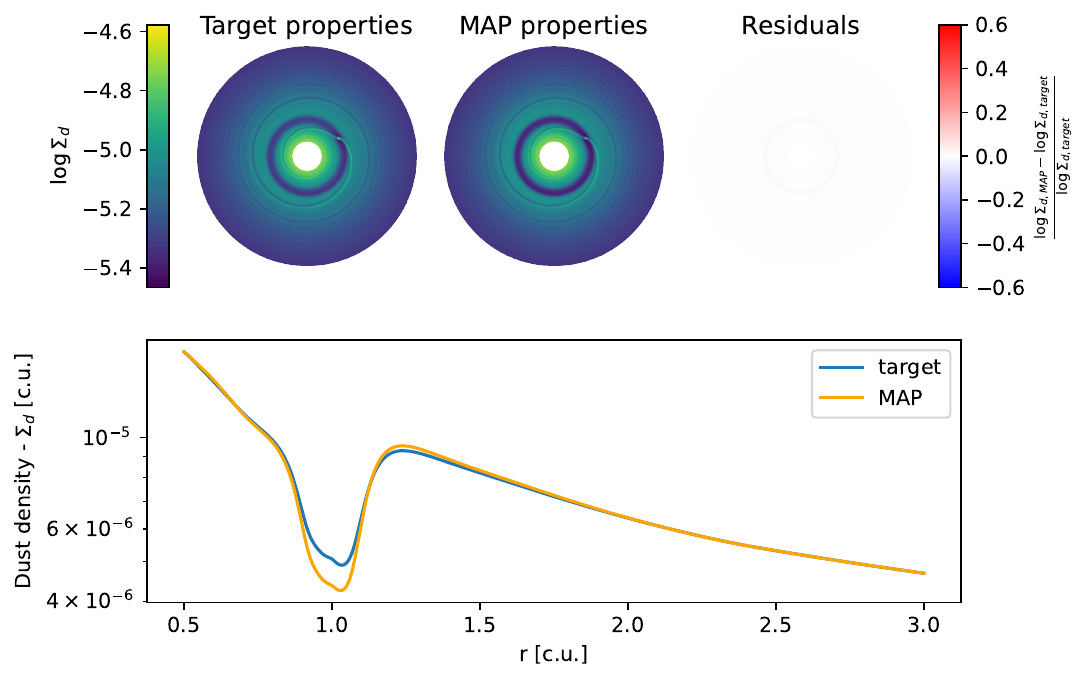}
        \includegraphics[width=0.4\linewidth]{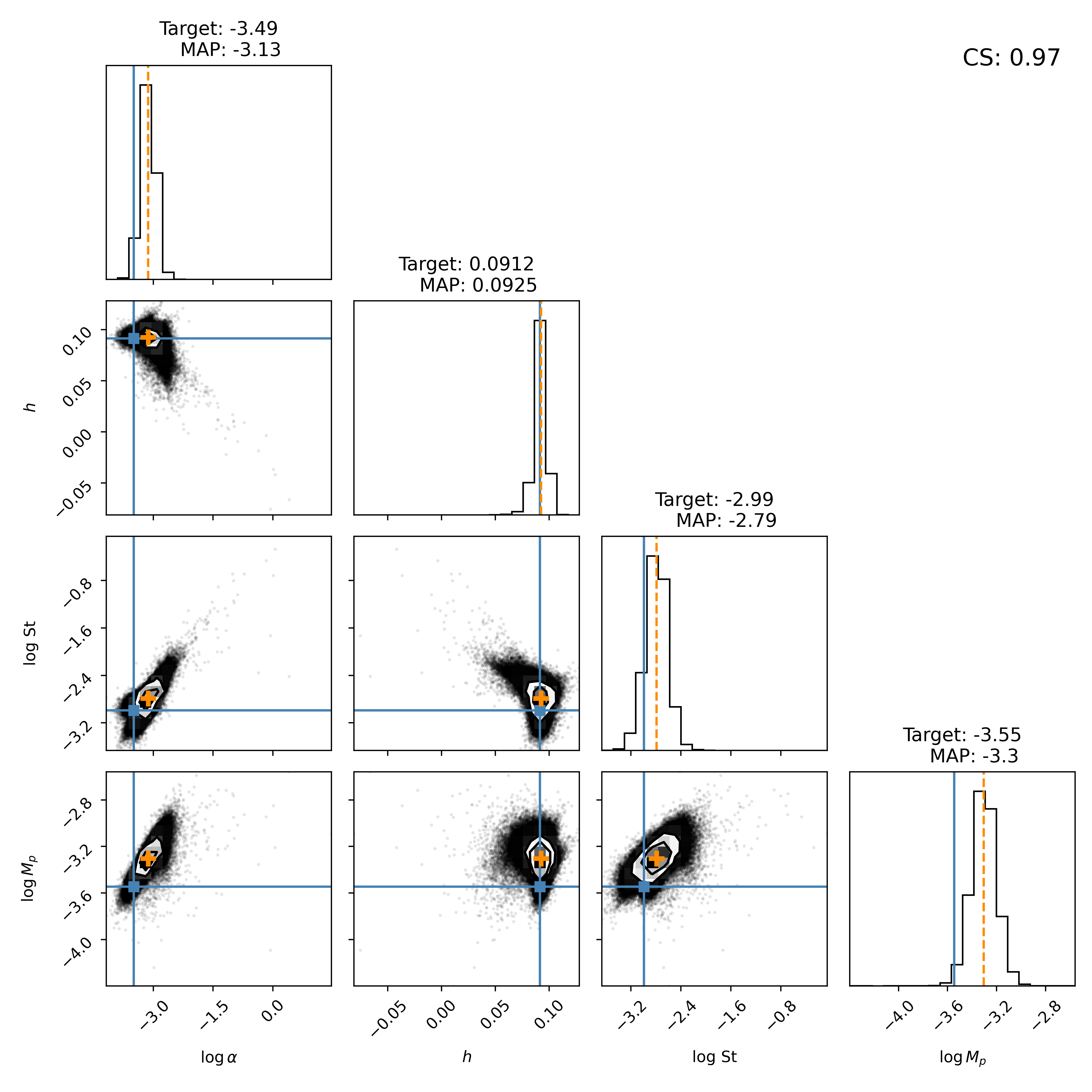}
    \includegraphics[width=0.58\linewidth]{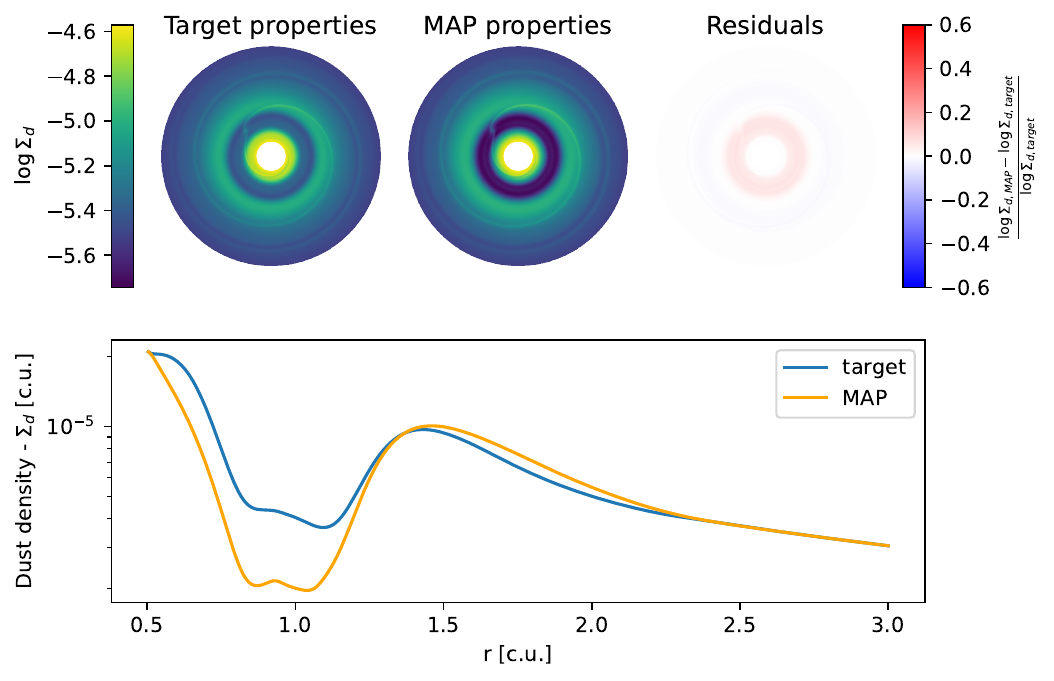}
    \caption{Results of three posterior predictive checks. Each row corresponds to a PPC considering a different synthetic observation in the test set. \emph{Left panel:} corner plots showing the inferred posterior distributions for the target properties. The orange lines mark the medians of the marginalised posteriors for each target property which are then assumed as the respective best estimates. The blue lines mark the target ``true" values. Confidence scores (CS) of DBNets2.0 estimates (see appendix \ref{app:conf_score}) are reported in the upper right corners. \emph{Right panel:} comparison between the simulated dust distributions with the ``true" and ``best estimates" values of the systems' properties. Both the 2D distribution (upper panels) and the azimuthally averaged radial profile of the dust density (lower panels) are compared. }
    \label{fig:ppc}
\end{figure*}

Figure \ref{fig:ppc} shows the results of three iterations of this test (one per row) for three different synthetic observations. The corner plots in the left panel show the inferred posterior distributions (black histograms with medians in orange) compared to the "true" values of the target properties (blue lines). Note that these mostly lie in the tail of the inferred posteriors. 
In the right panel we compare the simulated dust distributions with the ``true" values and best estimates of the system's properties.
We observe a good agreement both in the 2D map and in the azimuthally averaged radial profile. The main noticeable difference is in the value of the dust density in the gap minimum which, in all three cases, appears lower in the simulation with the inferred best estimates of the disc and planet properties. This difference, however, being orders of magnitude lower than the gap depth, is practically negligible.
The third case presents the most significant difference of the synthetic observation generated with the MAP properties with respect to the original. This is probably due to the low target viscosity close to the lower end of the explored parameter space. Therefore, although the target value is still plausible according to the inferred posterior, its best estimate is slightly higher causing also, as a consequence, a shift of the best estimates of the other properties.

\newpage

\section{TARP on single parameters marginalised posteriors}
\label{app:tarp_single}
\begin{figure}[h]
    \centering
    \includegraphics[width=\linewidth]{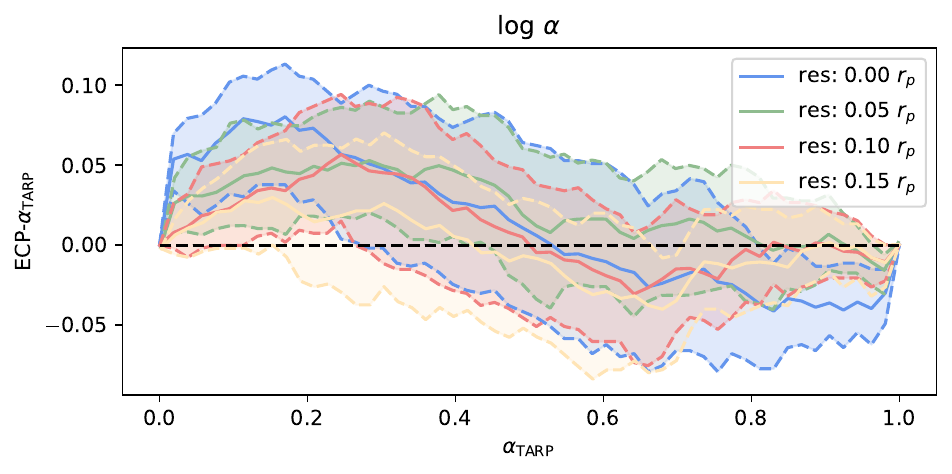}
    \includegraphics[width=\linewidth]{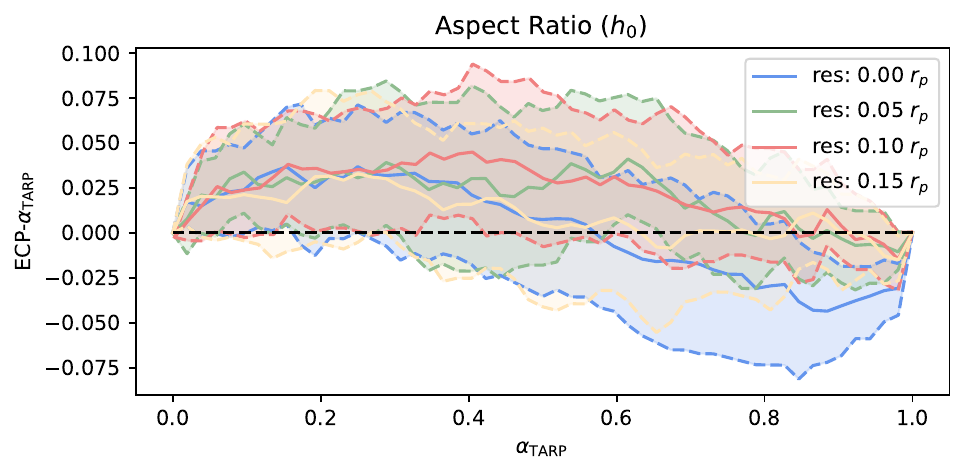}
    \includegraphics[width=\linewidth]{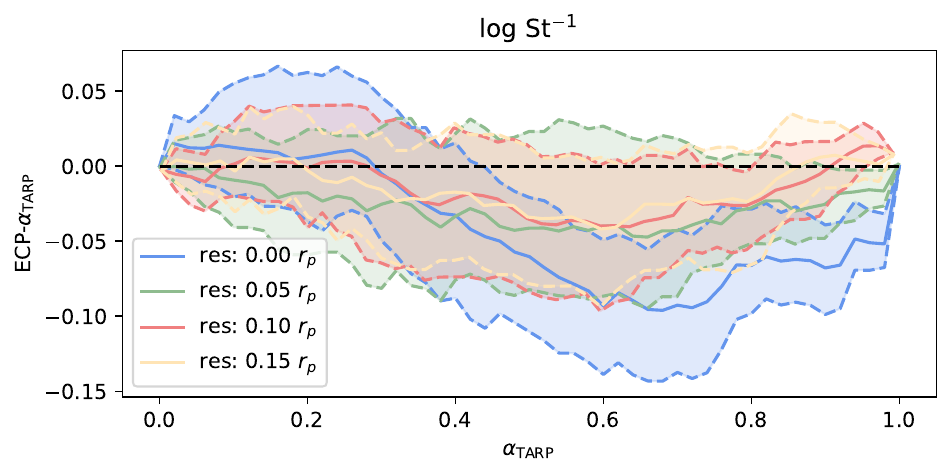}
    \includegraphics[width=\linewidth]{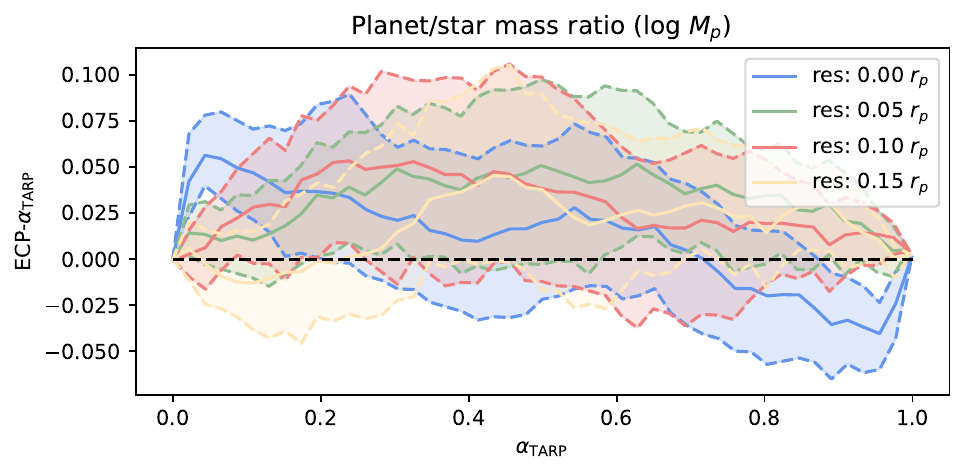}
    \caption{Results of TARP tests performed for each target property on the marginalised distributions over the other parameters. The different colours refer to the results obtained on synthetic observations convolved with different beam sizes. The dashed black lines mark the target curves that would be obtained if the inferred posteriors were perfectly accurate.}
    \label{fig:tarp_single}
\end{figure}

We report in Fig. \ref{fig:tarp_single} the results of TARP tests performed, for the test set, on the inferred posteriors marginalised for all but one target property. 
\FloatBarrier
\newpage

\section{Set of continuum observations considered and DBNets2.0 inference results}
\label{app:real_obs}
\begin{figure*}
    \centering
    \includegraphics[width=0.85\linewidth]{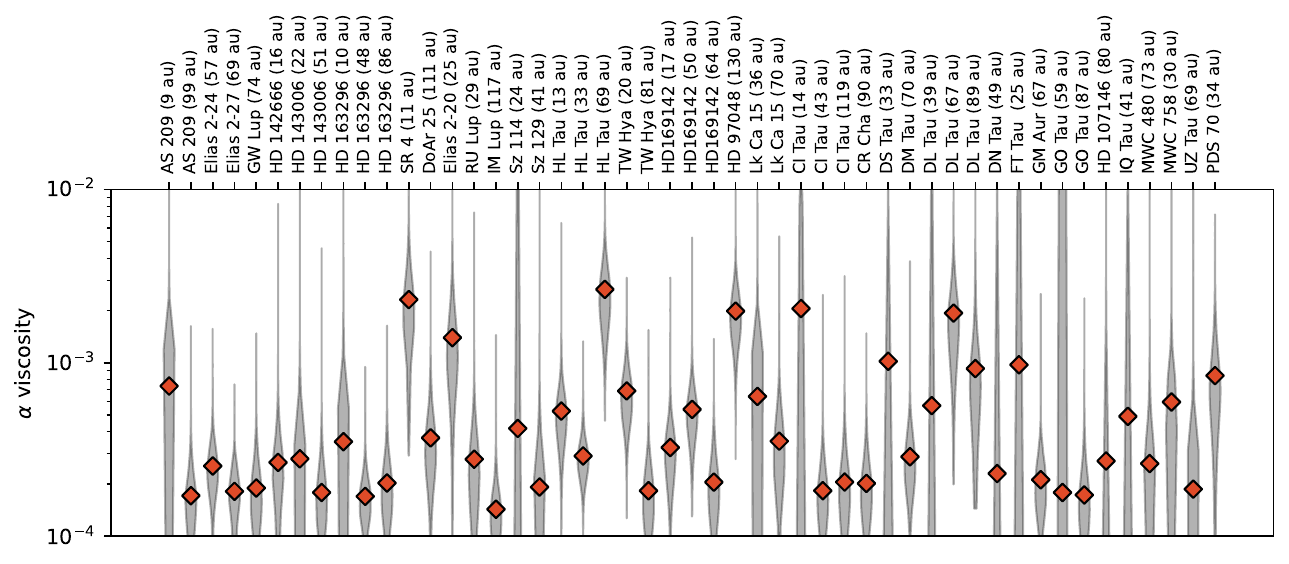}
    \includegraphics[width=0.85\linewidth]{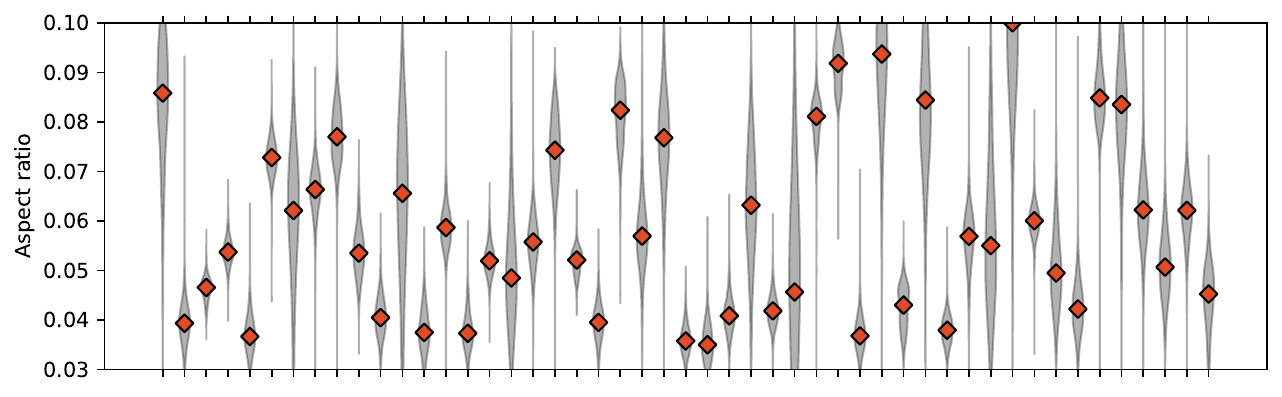}

    \includegraphics[width=0.85\linewidth]{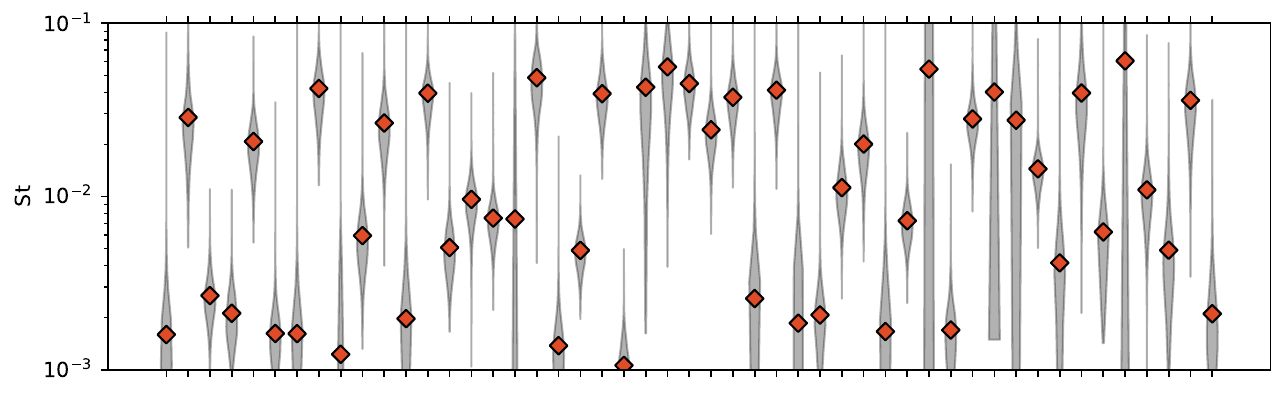}

    \includegraphics[width=0.85\linewidth]{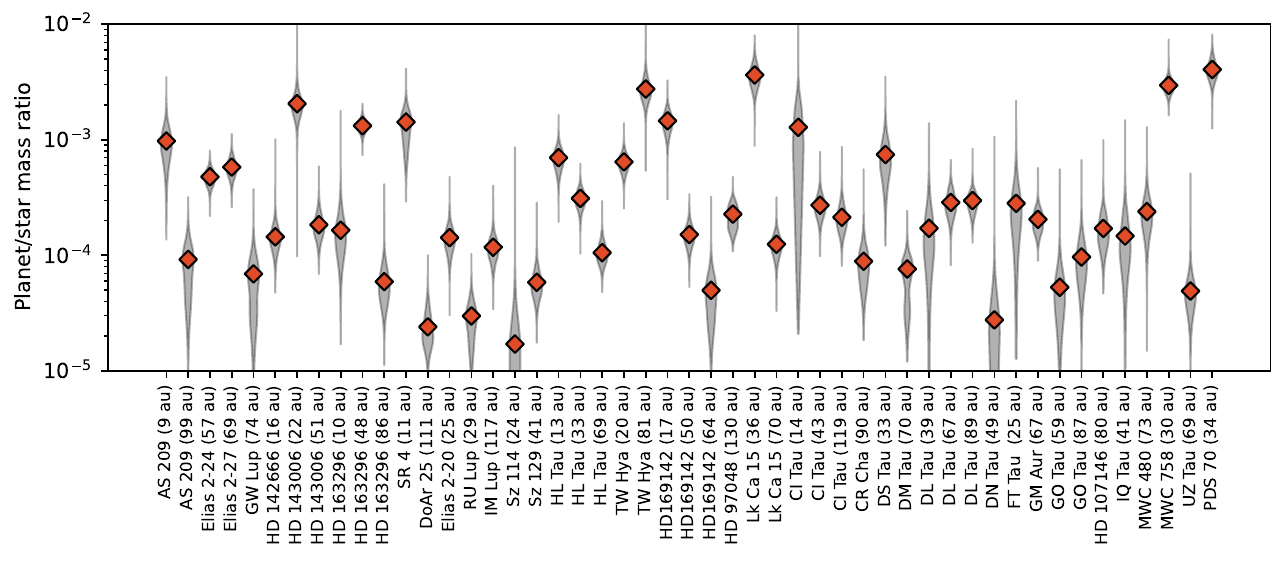}
    \caption{Overview of the best estimates of each target property for all the 49 dust substructures considered. The grey violin plots represent the inferred posterior distributions marginalised over all other parameters. The red squares mark the medians of these distributions.}
    \label{fig:all_real}
\end{figure*}

In Table \ref{tab:prop} we list the name and properties of the 34 protoplanetary disc observations on which we applied our tool in Sect. \ref{sec:real_obs}. We also report in the same table DBNets2.0 estimates of the targeted properties with uncertainties marking the 16$^\text{th}$ and 84$^\text{th}$ percentiles of the respective marginalised distributions. DBNets2.0 estimates for each disc are also shown with violin plots in Fig. \ref{fig:all_real}.

\begin{table*}[h!]
\vspace{0.3cm}
\caption{Catalogue of ALMA observations on which we applied DBNets2.0. }
\label{tab:prop}
\resizebox{\linewidth}{!}{%
\begin{tabular}{lrrrrrrrrrrrrrrr}
\toprule
\\
\multicolumn{1}{c}{\textbf{\begin{tabular}[c]{@{}c@{}}Disc\\  name\end{tabular}}} & \multicolumn{1}{c}{\textbf{\begin{tabular}[c]{@{}c@{}}$M_\star$\tablefootmark{1}\\  {[}M$_{\sun}$ {]}\end{tabular}}} & \multicolumn{1}{c}{\textbf{\begin{tabular}[c]{@{}c@{}}d\tablefootmark{2}\\  {[}pc{]}\end{tabular}}} & \multicolumn{1}{c}{\textbf{\begin{tabular}[c]{@{}c@{}}i\tablefootmark{3} \\ {[}°{]}\end{tabular}}} & \multicolumn{1}{c}{\textbf{\begin{tabular}[c]{@{}c@{}}PA\tablefootmark{4} \\ {[}°{]}\end{tabular}}} & \multicolumn{1}{c}{\textbf{\begin{tabular}[c]{@{}c@{}}res\tablefootmark{5} \\ {[}a{]}\end{tabular}}} & \multicolumn{1}{c}{\textbf{\begin{tabular}[c]{@{}c@{}}band\end{tabular}}} & \multicolumn{1}{c}{\textbf{\begin{tabular}[c]{@{}c@{}} a  {[}au{]} \tablefootmark{6}\end{tabular}}} & \multicolumn{1}{c}{\textbf{\begin{tabular}[c]{@{}c@{}}Data \\ ref.\tablefootmark{7}\end{tabular}}} & \multicolumn{1}{c}{\textbf{\begin{tabular}[c]{@{}c@{}}Prop. \\ ref.\tablefootmark{8}\end{tabular}}} 
&  
\multicolumn{1}{c}{\textbf{\begin{tabular}[c]{@{}c@{}}DBNets2.0 \\ $\log \alpha$ \end{tabular}}}
&
\multicolumn{1}{c}{\textbf{\begin{tabular}[c]{@{}c@{}}DBNets2.0 \\ $h$\end{tabular}}}
&
\multicolumn{1}{c}{\textbf{\begin{tabular}[c]{@{}c@{}}DBNets2.0 \\ $\log \text{St}$\end{tabular}}} 
&
\multicolumn{1}{c}{\textbf{\begin{tabular}[c]{@{}c@{}}DBNets2.0 \\ M$_p$ {[}$\text{M}_\text{J}${]}\end{tabular}}}
&
\multicolumn{1}{c}{\textbf{\begin{tabular}[c]{@{}c@{}}Confidence \\ score \end{tabular}}}
\\   \\
\midrule 
AS 209 & 0.83 & 121 & 35 & 86 & 0.22, 0.02 & 6 & 9, 99 & a &  1 & $-3.13^{+0.24}_{-0.22}$, $-3.77^{+0.18}_{-0.18}$ & $0.09^{+0.01}_{-0.02}$, $0.04^{+0.00}_{-0.00}$ & $-2.80^{+0.25}_{-0.23}$, $-1.54^{+0.20}_{-0.15}$ & $0.85^{+0.31}_{-0.25}$, $0.08^{+0.04}_{-0.03}$ & 0.85, 0.90\\
CI Tau & 0.90 & 159 & 50 & 11 & 0.24, 0.08, 0.03 & 6 & 14, 43, 119 & g &  10 & $-2.69^{+0.27}_{-0.27}$, $-3.74^{+0.19}_{-0.17}$, $-3.69^{+0.22}_{-0.20}$ & $0.05^{+0.02}_{-0.03}$, $0.08^{+0.00}_{-0.01}$, $0.09^{+0.00}_{-0.01}$ & $-2.73^{+0.36}_{-0.38}$, $-2.68^{+0.16}_{-0.17}$, $-1.95^{+0.14}_{-0.13}$ & $1.21^{+0.73}_{-0.52}$, $0.26^{+0.07}_{-0.05}$, $0.20^{+0.07}_{-0.04}$ & 0.87, 0.83, 0.86\\
CR Cha & 1.50 & 187 & 31 & 36 & 0.08 & 6 & 90 & h &  11 & $-3.70^{+0.20}_{-0.19}$ & $0.04^{+0.00}_{-0.00}$ & $-1.70^{+0.13}_{-0.14}$ & $0.14^{+0.06}_{-0.05}$ & 0.82\\
DL Tau & 0.98 & 159 & 45 & 52 & 0.24, 0.14, 0.11 & 6 & 39, 67, 89 & i  &  14 & $-3.25^{+0.60}_{-0.47}$, $-2.71^{+0.20}_{-0.22}$, $-3.03^{+0.31}_{-0.23}$ & $0.08^{+0.02}_{-0.01}$, $0.04^{+0.00}_{-0.00}$, $0.06^{+0.00}_{-0.01}$ & $-1.26^{+0.34}_{-0.35}$, $-2.77^{+0.17}_{-0.19}$, $-1.55^{+0.12}_{-0.11}$ & $0.18^{+0.13}_{-0.08}$, $0.29^{+0.09}_{-0.07}$, $0.31^{+0.06}_{-0.06}$ & 0.86, 0.93, 0.89\\
DM Tau & 0.50 & 145 & 33 & -25 & 0.03 & 3 & 70 & l &  13 & $-3.54^{+0.17}_{-0.16}$ & $0.04^{+0.00}_{-0.01}$ & $-2.14^{+0.11}_{-0.11}$ & $0.04^{+0.01}_{-0.02}$ & 0.96\\
DN Tau & 0.52 & 128 & 35 & 79 & 0.19 & 6 & 49 & i  &  14 & $-3.64^{+0.53}_{-0.37}$ & $0.06^{+0.02}_{-0.01}$ & $-1.40^{+0.25}_{-0.27}$ & $0.02^{+0.02}_{-0.01}$ & 0.85\\
DS Tau & 0.83 & 159 & 65 & -19 & 0.29 & 6 & 33 & i  &  12 & $-2.99^{+0.31}_{-0.31}$ & $0.09^{+0.01}_{-0.01}$ & $-2.78^{+0.33}_{-0.32}$ & $0.65^{+0.30}_{-0.21}$ & 0.82\\
DoAr 25 & 0.95 & 138 & 67 & 111 & 0.02 & 6 & 111 & a &  1 & $-3.43^{+0.21}_{-0.19}$ & $0.04^{+0.00}_{-0.00}$ & $-1.40^{+0.10}_{-0.11}$ & $0.02^{+0.01}_{-0.01}$ & 0.88\\
Elias 2-20 & 0.48 & 138 & 49 & 153 & 0.08 & 6 & 25 & a &  1 & $-2.85^{+0.24}_{-0.23}$ & $0.06^{+0.00}_{-0.00}$ & $-2.29^{+0.14}_{-0.14}$ & $0.07^{+0.02}_{-0.02}$ & 0.91\\
Elias 2-24 & 0.78 & 136 & 29 & 46 & 0.04 & 6 & 57 & a &  1 & $-3.60^{+0.15}_{-0.14}$ & $0.05^{+0.00}_{-0.00}$ & $-2.57^{+0.14}_{-0.12}$ & $0.39^{+0.06}_{-0.06}$ & 0.93\\
Elias 2-27 & 0.49 & 140 & 56 & 117 & 0.04 & 6 & 69 & a &  1 & $-3.74^{+0.14}_{-0.18}$ & $0.05^{+0.00}_{-0.00}$ & $-2.67^{+0.15}_{-0.15}$ & $0.30^{+0.06}_{-0.05}$ & 0.81\\
FT Tau  & 0.34 & 127 & 35 & 122 & 0.28 & 6 & 25 & i  &  14 & $-3.01^{+0.60}_{-0.49}$ & $0.10^{+0.02}_{-0.02}$ & $-1.56^{+0.34}_{-0.37}$ & $0.10^{+0.07}_{-0.04}$ & 0.84\\
GM Aur & 1.32 & 159 & 53 & 57 & 0.05 & 6 & 67 & m &  15 & $-3.68^{+0.16}_{-0.17}$ & $0.06^{+0.00}_{-0.00}$ & $-1.84^{+0.09}_{-0.10}$ & $0.28^{+0.07}_{-0.06}$ & 0.86\\
GO Tau & 0.36 & 144 & 54 & 21 & 0.15, 0.10 & 6 & 59, 87 & i  &  14 & $-3.75^{+0.43}_{-0.29}$, $-3.76^{+0.22}_{-0.20}$ & $0.05^{+0.01}_{-0.01}$, $0.04^{+0.01}_{-0.01}$ & $-2.38^{+0.24}_{-0.24}$, $-1.40^{+0.21}_{-0.19}$ & $0.02^{+0.01}_{-0.01}$, $0.04^{+0.02}_{-0.01}$ & 0.95, 0.81\\
GW Lup & 0.46 & 155 & 39 & 38 & 0.04 & 6 & 74 & a &  1 & $-3.72^{+0.20}_{-0.19}$ & $0.04^{+0.00}_{-0.00}$ & $-1.68^{+0.12}_{-0.13}$ & $0.03^{+0.02}_{-0.02}$ & 0.87\\
HD 107146\tablefootmark{9} & 1.00 & 28 & 19 & 153 & 0.12 & 6 & 80 & n &  16 & $-3.57^{+0.27}_{-0.27}$ & $0.08^{+0.01}_{-0.01}$ & $-2.20^{+0.14}_{-0.17}$ & $0.18^{+0.07}_{-0.05}$ & 0.76\\
HD 142666 & 1.58 & 148 & 62 & 162 & 0.13 & 6 & 16 & a &  1 & $-3.57^{+0.21}_{-0.17}$ & $0.07^{+0.00}_{-0.00}$ & $-2.79^{+0.15}_{-0.17}$ & $0.24^{+0.07}_{-0.05}$ & 0.88\\
HD 143006 & 1.78 & 165 & 19 & 169 & 0.15, 0.06 & 6 & 22, 51 & a &  1 & $-3.55^{+0.23}_{-0.20}$, $-3.75^{+0.19}_{-0.17}$ & $0.06^{+0.01}_{-0.02}$, $0.07^{+0.00}_{-0.00}$ & $-2.79^{+0.25}_{-0.19}$, $-1.38^{+0.13}_{-0.14}$ & $3.83^{+1.06}_{-0.74}$, $0.34^{+0.10}_{-0.07}$ & 0.77, 0.70\\
HD 163296 & 2.04 & 101 & 48 & 133 & 0.21, 0.04, 0.02 & 6 & 10, 48, 86 & a &  1 & $-3.46^{+0.30}_{-0.26}$, $-3.77^{+0.13}_{-0.14}$, $-3.69^{+0.19}_{-0.18}$ & $0.08^{+0.01}_{-0.01}$, $0.05^{+0.00}_{-0.00}$, $0.04^{+0.00}_{-0.00}$ & $-2.91^{+0.23}_{-0.22}$, $-2.23^{+0.15}_{-0.15}$, $-1.58^{+0.14}_{-0.15}$ & $0.35^{+0.18}_{-0.12}$, $2.83^{+0.36}_{-0.32}$, $0.13^{+0.06}_{-0.04}$ & 0.93, 0.80, 0.82\\
HD 97048 & 2.50 & 183 & 40 & 3 & 0.07 & 7 & 130 & c &  7,8 & $-2.70^{+0.15}_{-0.15}$ & $0.04^{+0.00}_{-0.00}$ & $-1.43^{+0.12}_{-0.12}$ & $0.59^{+0.14}_{-0.15}$ & 0.85\\
HD169142 & 1.65 & 117 & 5 & 13 & 0.08, 0.03, 0.02 & 6 & 17, 50, 64 & d &  4,5,6,7,8 & $-3.49^{+0.15}_{-0.17}$, $-3.27^{+0.14}_{-0.15}$, $-3.69^{+0.19}_{-0.19}$ & $0.08^{+0.01}_{-0.01}$, $0.04^{+0.00}_{-0.00}$, $0.03^{+0.00}_{-0.00}$ & $-1.25^{+0.18}_{-0.18}$, $-1.35^{+0.11}_{-0.12}$, $-1.61^{+0.12}_{-0.12}$ & $2.52^{+0.54}_{-0.50}$, $0.26^{+0.06}_{-0.05}$, $0.09^{+0.04}_{-0.03}$ & 0.74, 0.91, 0.86\\
HL Tau & 1.00 & 140 & 47 & 138 & 0.14, 0.05, 0.03 & 7 & 13, 33, 69 & b &  2 & $-3.28^{+0.15}_{-0.15}$, $-3.54^{+0.12}_{-0.11}$, $-2.58^{+0.17}_{-0.18}$ & $0.07^{+0.01}_{-0.01}$, $0.05^{+0.00}_{-0.00}$, $0.04^{+0.00}_{-0.00}$ & $-2.86^{+0.18}_{-0.18}$, $-2.31^{+0.09}_{-0.10}$, $-1.41^{+0.13}_{-0.12}$ & $0.73^{+0.16}_{-0.14}$, $0.33^{+0.06}_{-0.05}$, $0.11^{+0.03}_{-0.02}$ & 0.90, 0.86, 0.81\\
IM Lup & 0.89 & 158 & 48 & 143 & 0.03 & 6 & 117 & a &  1 & $-3.85^{+0.16}_{-0.15}$ & $0.05^{+0.00}_{-0.00}$ & $-2.12^{+0.10}_{-0.11}$ & $0.11^{+0.03}_{-0.02}$ & 0.84\\
IQ Tau & 0.50 & 131 & 62 & 42 & 0.22 & 6 & 41 & i  &  14 & $-3.31^{+0.58}_{-0.46}$ & $0.08^{+0.01}_{-0.01}$ & $-1.22^{+0.30}_{-0.32}$ & $0.08^{+0.06}_{-0.03}$ & 0.75\\
Lk Ca 15 & 1.25 & 159 & 50 & 62 & 0.11, 0.06 & 6 & 36, 70 & e &  3,9 & $-3.19^{+0.24}_{-0.23}$, $-3.45^{+0.21}_{-0.22}$ & $0.06^{+0.01}_{-0.01}$, $0.04^{+0.00}_{-0.00}$ & $-2.59^{+0.30}_{-0.21}$, $-1.39^{+0.11}_{-0.12}$ & $4.77^{+1.25}_{-0.96}$, $0.16^{+0.04}_{-0.03}$ & 0.74, 0.66\\
MWC 480 & 1.91 & 161 & 36 & 148 & 0.16 & 6 & 73 & i  &  14 & $-3.58^{+0.27}_{-0.25}$ & $0.06^{+0.01}_{-0.01}$ & $-1.96^{+0.15}_{-0.15}$ & $0.48^{+0.21}_{-0.15}$ & 0.85\\
MWC 758 & 1.50 & 160 & 21 & 62 & 0.10 & 7 & 30 & o &  17 & $-3.23^{+0.22}_{-0.21}$ & $0.05^{+0.01}_{-0.01}$ & $-2.31^{+0.26}_{-0.22}$ & $4.65^{+0.68}_{-0.61}$ & 0.71\\
PDS 70 & 0.90 & 112 & 52 & 160 & 0.07 & 7 & 34 & p &  18 & $-3.07^{+0.25}_{-0.24}$ & $0.05^{+0.01}_{-0.01}$ & $-2.68^{+0.28}_{-0.26}$ & $3.83^{+0.90}_{-0.66}$ & 0.73\\
RU Lup & 0.63 & 154 & 19 & 121 & 0.06 & 6 & 29 & a &  1 & $-3.56^{+0.22}_{-0.23}$ & $0.04^{+0.00}_{-0.00}$ & $-2.02^{+0.11}_{-0.12}$ & $0.02^{+0.01}_{-0.01}$ & 0.90\\
SR 4 & 0.68 & 134 & 22 & 18 & 0.18 & 6 & 11 & a &  1 & $-2.63^{+0.22}_{-0.19}$ & $0.07^{+0.02}_{-0.02}$ & $-2.71^{+0.26}_{-0.23}$ & $1.01^{+0.30}_{-0.32}$ & 0.86\\
Sz 114 & 0.17 & 162 & 21 & 165 & 0.19 & 6 & 24 & a &  1 & $-3.38^{+0.86}_{-0.58}$ & $0.05^{+0.01}_{-0.01}$ & $-2.13^{+0.27}_{-0.28}$ & $0.00^{+0.00}_{-0.00}$ & 0.95\\
Sz 129 & 0.83 & 161 & 34 & 151 & 0.07 & 6 & 41 & a &  1 & $-3.72^{+0.28}_{-0.24}$ & $0.06^{+0.01}_{-0.01}$ & $-1.31^{+0.13}_{-0.14}$ & $0.05^{+0.02}_{-0.01}$ & 0.81\\
TW Hya & 0.80 & 56 & 7 & 155 & 0.03, 0.01 & 7 & 20, 81 & f &  3 & $-3.16^{+0.15}_{-0.16}$, $-3.74^{+0.18}_{-0.19}$ & $0.08^{+0.01}_{-0.01}$, $0.06^{+0.01}_{-0.01}$ & $-2.97^{+0.15}_{-0.15}$, $-1.37^{+0.25}_{-0.23}$ & $0.54^{+0.10}_{-0.09}$, $2.30^{+0.65}_{-0.45}$ & 0.87, 0.70\\
UZ Tau & 0.39 & 131 & 56 & 90 & 0.10 & 6 & 69 & i  &  14 & $-3.73^{+0.31}_{-0.26}$ & $0.06^{+0.01}_{-0.01}$ & $-1.44^{+0.15}_{-0.16}$ & $0.02^{+0.01}_{-0.01}$ & 0.92\\
\bottomrule
\end{tabular}
} 
\tablefoottext{1}{Star mass}
\tablefoottext{2}{Distance}
\tablefoottext{3}{Inclination}
\tablefoottext{4}{Position Angle}
\tablefoottext{5}{Resolution expressed as the major standard~deviation of the two-dimensional Gaussians that approximate the observational beams. Values are expressed in units of the putative planet radial location $a$.}
\tablefoottext{6}{Putative planet location(s)}
\tablefoottext{7}{\textit{Data references:} (a) \href{https://almascience.eso.org/almadata/lp/DSHARP/}{DSHARP Data Release}, (b) \href{https://almascience.eso.org/alma-data/science-verification}{ALMA Science Verification Data}, (c) \cite{Pinte2019HD13CO}, (d) \cite{Perez2019DustRing}, (e) \cite{Facchini2020AnnularJ1610}, (f) \cite{Andrews2016RINGEDDISK}, (g) Zagaria et al. (in prep.), (h) \cite{Kim2020TheDisk}, (i) \cite{Long2018GapsRegion}, (l) \cite{Hashimoto2021ALMATau}, (m) \cite{Huang2020ADisk}, (n) \cite{Marino2021ConstrainingEdges}, (o) \cite{Baruteau2019DustPlanets}, (p) \cite{Benisty2021APDS70c}.}
\tablefoottext{8}{\textit{Properties references:} (1) \cite{Zhang2018TheInterpretation}, (2) \cite{Lodato2019TheDiscs}, (3) \cite{Dong2017WhatPlanet}, (4) \cite{Toci2019Long-lived169142}, (5) \cite{Perez2019DustRing}, (6) \cite{Gratton2019Blobs169142}, (7) \cite{Pinte2019KinematicDisk}, (8) \href{https://exoplanets.nasa.gov/}{NASA exoplanets catalogue}, (9) \cite{Facchini2020AnnularJ1610}, (10) \cite{Clarke2018High-resolutionAu}, (11) \cite{Kim2020TheDisk}, (12) \cite{Veronesi2020IsPlanet}, (13) \cite{Wang2021ArchitectureGap}, (14) \cite{Long2018GapsRegion}, (15) \cite{Huang2020Large-scaleLup}, (16) \cite{Marino2021ConstrainingEdges}, (17) \cite{Dong2018MultipleSystems}, (18) \cite{Benisty2021APDS70c}.}
\tablefoottext{9}{ Note that, unlike the others, this is a debris disc.} 
\vspace{0.4cm}
\end{table*}

\newpage
\section{Inferred viscous timescales}
\label{app:visc_time}
\begin{figure}[h]
    \centering
    \includegraphics[width=0.9\linewidth]{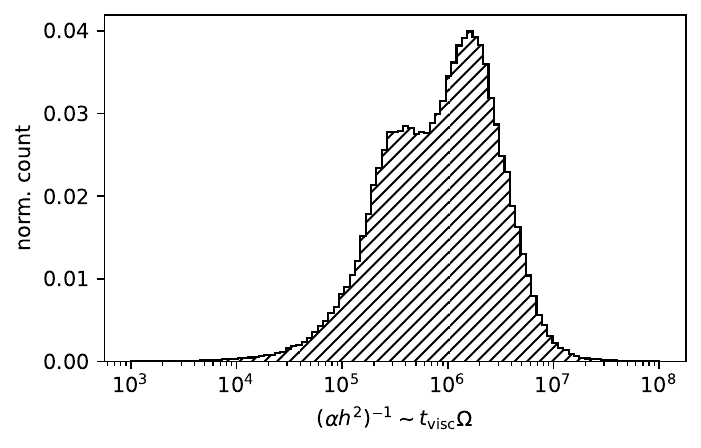}
    \caption{Inferred distribution of $(\alpha h^2)^{-1} \sim t_\text{visc}\Omega$ for the population of proposed planets within dust substructures. The plot is made extracting 5000 samples for $\alpha$ and $h$ from the inferred $p(\alpha, h, St, M_p|x)$ for each of the 49 analysed gaps.}
    \label{fig:ah2}
\end{figure}

\begin{figure}[h]
    \centering
    \includegraphics[width=0.9\linewidth]{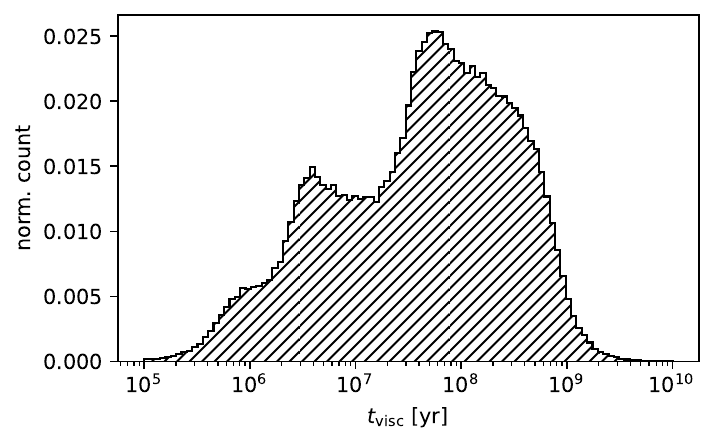}
    \caption{Inferred distribution of $t_\text{visc}$ for the population of proposed planets within dust substructures. The plot is made extracting 5000 samples for $\alpha$ and $h$ from the inferred $p(\alpha, h, St, M_p|x)$ for each of the 49 analysed gaps and computing $t_\text{visc}$ as $t_\text{visc} = 1/(\Omega \alpha h^2)$.}
    \label{fig:visct}
\end{figure}

We show in Fig. \ref{fig:ah2} the distribution of $(\alpha h^2)^{-1}$ inferred with DBNets2.0 from the set of dust observations considered in this work. As explained in the main text, this quantity can be interpreted as the ratio between the dynamical and viscous timescales measured locally at the gap location. In Fig. \ref{fig:visct} we provide the same distribution for the viscous timescale in years rescaling for the appropriate dynamical timescales.

\FloatBarrier

\end{appendix}
\end{document}